\documentclass{aa}  
\usepackage{graphicx}
\usepackage{txfonts}
\usepackage{natbib}
\bibpunct{(}{)}{;}{a}{}{,}%
\begin{document}
%

\newcommand{\dmom}{\ensuremath{D_{{\mathrm m}{\mathrm o}{\mathrm m}}}}
\newcommand{\finf}{\ensuremath{f_{\infty}}}
\newcommand{\ha}{H\ensuremath{\alpha }}
\newcommand{\hb}{H\ensuremath{\beta }}
\newcommand{\hg}{H\ensuremath{\gamma }}
\newcommand{\hd}{H\ensuremath{\delta }}
\newcommand{\he}{H\ensuremath{\epsilon }}
\newcommand{\hi}{H\,{\sc i}}
\newcommand{\hii}{H\,{\sc ii}}
\newcommand{\htwo}{\ensuremath{H_{2}}}
\newcommand{\hei}{He\,{\sc i}}
\newcommand{\heii}{He\,{\sc ii}}
\newcommand{\heff}{\ensuremath{h_{\rm eff}}}
\newcommand{\kms}{\ensuremath{\rm{km~s}^{-1}}}
\newcommand{\logg}{\ensuremath{\log g}}
\newcommand{\lsolar}{\ensuremath{L_{\odot}}}
\newcommand{\lstar}{\ensuremath{L_{*}}}
\newcommand{\llsolar}{\ensuremath{L/L_{\odot}}}
\newcommand{\logl}{\ensuremath{\log L}}
\newcommand{\loglstar}{\ensuremath{\log L_{*}}}
\newcommand{\loglbol}{\ensuremath{\log L_{{\mathrm b}{\mathrm
o}{\mathrm l}}}}
\newcommand{\logllsolar}{\ensuremath{\log\ (L/\lsolar)}}
\newcommand{\loglx}{\ensuremath{\log L_{{\mathrm x}}}}
\newcommand{\loglxlbol}{\ensuremath{\log\,(L_{{\mathrm x}}/L_{{\mathrm b}{\mathrm
o}{\mathrm l}})}}
\newcommand{\logmdot}{\ensuremath{\log \dot{M}}}
\newcommand{\logmdotradio}{\ensuremath{\log\,(\mdot_{{\mathrm r}{\mathrm a}{\mathrm d}{\mathrm i}{\mathrm o})}}}
\newcommand{\mm}{\ensuremath{\mu {\rm m}}}
\newcommand{\mdot}{\ensuremath{\dot{M}}}
\newcommand{\mdotrho}{\ensuremath{\dot{M}(\rho^{2})}}
\newcommand{\mdothalpha}{\ensuremath{\mdot_{\mathrm \ha}}}
\newcommand{\mdotradio}{\ensuremath{\mdot_{{\mathrm r}{\mathrm a}{\mathrm d}{\mathrm i}{\mathrm o}}}}
\newcommand{\mdotradiohalpha}{\ensuremath{\mdot_{{\mathrm r}{\mathrm a}{\mathrm d}{\mathrm i}{\mathrm o}}/\mdot_{\mathrm \halpha}}}
\newcommand{\mbol}{\ensuremath{M_{\rm bol}}}
\newcommand{\mevol}{\ensuremath{M_{\rm evol}}}
\newcommand{\mh}{\ensuremath{m_{\rm H}}}
\newcommand{\minit}{\ensuremath{M_{\rm init}}}
\newcommand{\msolar}{\ensuremath{M_{\odot}}}
\newcommand{\msolaryr}{\ensuremath{M_{\odot}~\rm{yr}^{-1}}}
\newcommand{\msolaryrsquared}{\ensuremath{M_{\odot}^{2}~\rm{yr}^{-2}}}
\newcommand{\mspec}{\ensuremath{M_{\rm spec}}}
\newcommand{\mstar}{\ensuremath{M_{*}}}
\newcommand{\mv}{\ensuremath{M_{V}}}
\newcommand{\rphot}{\ensuremath{R_{phot}}}
\newcommand{\rsolar}{\ensuremath{R_{\odot}}}
\newcommand{\rstar}{\ensuremath{R_{*}}}
\newcommand{\teff}{\ensuremath{T_{\rm eff}}}
\newcommand{\vzero}{\ensuremath{v_{0}}}
\newcommand{\vcl}{\ensuremath{v_{cl}}}
\newcommand{\vcore}{\ensuremath{v_{core}}}
\newcommand{\vphot}{\ensuremath{v_{phot}}}
\newcommand{\vesc}{\ensuremath{v_{\rm esc}}}
\newcommand{\vinf}{\ensuremath{v_{\rm \infty}}}
\newcommand{\vmax}{\ensuremath{v_{\rm max}}}
\newcommand{\vmin}{\ensuremath{v_{\rm min}}}
\newcommand{\vrot}{\ensuremath{v_{\rm rot}}}
\newcommand{\vsini}{\ensuremath{v\sin i}}
\newcommand{\vtherm}{\ensuremath{v_{\rm th}}}
\newcommand{\vturb}{\ensuremath{v_{\rm turb}}}
\newcommand{\wl}{\ensuremath{W}_{\lambda}}
\newcommand{\yhe}{\ensuremath{Y(\rm He)}}
\newcommand{\zsolar}{\ensuremath{Z_{\odot}}}


\title{Quantitative studies of the optical and UV spectra \\
of Galactic early B supergiants}
\subtitle{I. Fundamental parameters}   

   \author{Samantha C. Searle\inst{1}, Raman K. Prinja\inst{1}, Derck Massa\inst{2}, \and Robert Ryans\inst{3}}

   \offprints{S. C. Searle, \email{scs@star.ucl.ac.uk}}

   \institute{Department of Physics \& Astronomy, University College London, 
Gower Street, London WC1E 6BT England, UK \\
              \email{scs@star.ucl.ac.uk, rkp@star.ucl.ac.uk}
         \and
             SGT, Inc., Code 665.0, Nasa Goddard Space Flight Center, Greenbelt, MD 20771, USA \\
             \email{massa@taotaomona.gsfc.nasa.gov}
         \and
             Department of Physics \& Astronomy, The Queen's University of Belfast, BT7, 1NN, Northern Ireland, UK \\
             \email{r.ryans@qub.ac.uk}
             }

   \date{Accepted January 3, 2008}

 
  \abstract 
{} 
  {We undertake an optical and ultraviolet spectroscopic analysis of a
    sample of 20 Galactic B0 -- B5 supergiants of luminosity classes
    Ia, Ib, Iab, and II. Fundamental stellar parameters are obtained
    from optical diagnostics and a critical comparison of the model
    predictions to observed UV spectral features is made.}
  {Fundamental parameters (e.g., \teff, \loglstar, mass-loss rates and
    CNO abundances) are derived for individual stars using CMFGEN, a
    nLTE, line-blanketed model atmosphere code. The impact of these newly derived parameters on the Galactic
    B supergiant \teff\ scale, mass discrepancy, and wind-momentum
    luminosity relation is examined.}
  {The B supergiant temperature scale derived here shows a reduction
    of about 1\,000 -- 3\,000 K compared to previous results
    using unblanketed codes. Mass-loss rate estimates are in good
    agreement with predicted theoretical values, and all of the 20 B0~
    -- B5 supergiants analysed show evidence of CNO processing. A
    mass discrepancy still exists between spectroscopic and
    evolutionary masses, with the largest discrepancy occurring at
    \logllsolar\ $\sim$ 5.4. The observed WLR values calculated for B0
    -- B0.7 supergiants are higher than predicted values, whereas the
    reverse is true for B1 -- B5 supergiants. This means that the
    discrepancy between observed and theoretical values cannot be
    resolved by adopting clumped (i.e., lower) mass-loss rates as for
    O stars. The most surprising result is that, although CMFGEN
    succeeds in reproducing the optical stellar spectrum accurately,
    it fails to precisely reproduce key UV diagnostics, such as the \ion{N}{v}
    and \ion{C}{iv} P Cygni profiles. This problem arises
    because the models are not ionised enough and fail to reproduce
    the full extent of the observed absorption trough of the P Cygni
    profiles. }
  {Newly-derived fundamental parameters for early B supergiants are in
    good agreement with similar work in the field. The most
    significant discovery, however, is the failure of CMFGEN to predict
    the correct ionisation fraction for some ions. Such findings
    add further support to revising the current standard model of
    massive star winds, as our understanding of these winds is
    incomplete without a precise knowledge of the ionisation structure
    and distribution of clumping in the wind. }

   \keywords{stars: early type -- stars: supergiants
                galaxies: Milky Way -- stars: atmospheres
                stellar winds -- stellar evolution
                -- mass loss -- abundances}

\titlerunning{Fundamental parameters of Galactic B supergiants}
\authorrunning{S. C. Searle et al.}
\maketitle

%

\section{Introduction}

The study of luminous, massive stars is fundamental to improving our
understanding of galactic evolution, since the radiatively driven
winds of these stars have a tremendous impact on their host galaxies.
This huge input of mechanical energy is responsible for creating H II
regions, making a significant contribution to the integrated light of
starburst galaxies and providing star formation diagnostics at both
low and high redshifts. They substantially enrich the local ISM with
the products of nucleosynthesis via their stellar winds and supernovae
explosions and are a possible source of gamma ray bursts. It is
therefore imperative to obtain accurate fundamental parameters for
luminous massive stars since they contribute to many currently active
areas of astrophysical research.\\

\noindent
However there are still some uncertainties regarding the post-main
sequence evolution of massive stars since their evolution is
controlled by variable mass loss from the star as well as rotation,
binarity and convective processes, the latter leading to surface
enrichment as the products of nuclear burning are brought to the
surface. Until recently, stellar evolution models failed to predict
the correct amount of CNO processing in massive stars. However, new
evolutionary tracks that account for the effects of rotation
\citep{meynet2001} show better agreement between predicted and
observed amounts of CNO enrichment in massive stars. A far greater
problem in stellar astrophysics is the determination of {\em accurate}
observed mass loss rates. Recent research (e.g.,
\citealt{fullerton2006,prinja2005,massa2003,puls2006,repolust2004}),
has shown that current OB star mass loss rates might be over-estimated
by at least a factor of 10. Such large uncertainties in the mass loss
rates of massive stars suggest that our understanding of their winds
is incomplete. It is now widely accepted that the winds of both O and
B stars are highly structured and clumped and therefore our
assumptions that they are smooth and homogeneous are invalid. Evidence
to support this claim has come from hydrodynamical, time-dependent
simulations of stellar winds (e.g. \citealt{owocki1988, owocki2002}),
the latter of which proposed the idea that instabilites in the
line-driving of the wind can produce small-scale, stochastic structure
in the wind. Further evidence for the inhomogeneity of stellar winds
comes in the form of various observational studies
\citep[e.g.,][]{puls2006,bouret2005,massa2003,
  prinja2002,bianchi2002}. Time-series analyses of both Balmer and
metal spectral lines \citep[e.g.,][]{prinja2004} in OB stars highlight
clear, periodic patterns of variability that correspond to the
evolution of structure in the wind. \cite{ puls2006} demonstrated that
the discrepancy between values of \mdothalpha\ and \mdotradio\ implies
the presence of different amounts of clumping at the base of and
further out in the wind.  \cite{massa2003} showed that for a sample of
O stars in the LMC, the empirical ionisation fractions derived were
several orders of magnitude lower than expected, indicating a lack of
dominant ions in the wind. Similar results were found by Prinja et al.
(2005) for early B supergiants. More recently, \cite{fullerton2006}
demonstrated that the ionisation fraction of \ion{P}{v}, which is
dominant over a given range in \teff\ in the O star spectral range,
never approaches a value of unity. They subsequently showed that a
reduction in mass loss rate of at least a factor of 100 is required to
resolve the situation. We intend to re-address the issue of the
ionisation structure of early B supergiants, following on from
\cite{prinja2005}, in a forthcoming paper (Searle et al., 2007b, in
preparation; hereafter Paper II). Such drastic reductions in OB star
mass loss rates would have severe consequences for the
post-main-sequence evolution of these stars; in particular it would
affect the numbers of Wolf-Rayet stars produced and the ratio of
neutron stars to black
holes produced in the  final stages of massive star evolution. \\

\noindent
Early type B supergiants are particularly important since they are the
most numerous massive luminous stars and are ideal candidates for
extra-galactic distance indicators, essential for calibrating the
Wind-Momentum-Luminosity Relation (WLR)
\citep[e.g.,][]{kudritzki1999}.  Research into this WLR calibration
has highlighted a spectral type dependence for Galactic OBA type stars
\cite[e.g.,][]{kudritzki1999,repolust2004,markova2004}, whilst others
have explored the effect of metallicity on the WLR by studying OB
stars in the metal-poor environment of the Magellanic Clouds
\cite[e.g.,][]{kudritzki2000,trundle2004,evans2004a}. Accurately
derived mass loss rates are essential in calibrating the WLR, yet
discrepancies still exist between observed mass loss rates obtained
from different wavelength regions (i.e.  optical, UV or IR).
Furthermore acknowledged discrepancies of up to 30~\% have been found
between observational and theoretically predicted mass loss rates
\citep{vink2000}. Vink has remarked that these discrepancies for
the mass loss rates of B stars can be attributed to systematic errors
in the methods employed to derive the observed values. Good agreement
was found between observed and predicted mass
loss rates for O stars in \cite{vink2000}. Additionally, \cite{puls2006}
recently derived values of both \mdothalpha\ and \mdotradio,
highlighting a discrepancy of roughly a factor of two between
both values. \\

\noindent
The layout of this paper is as follows. \S \ref{obs} introduces the
sample of 20 Galactic B0 -- B5 supergiants upon which this analysis is
based. \S \ref{param} describes the methodology used in deriving
fundamental parameters for this sample. The results are presented in
\S \ref{results} and a critical examination of the {\sc CMFGEN} model
fit to the UV spectra of the 20 B supergiants is made in \S \ref{uv}.
Finally the conclusions are given in \S \ref{conc}.


\section{Observations}
\label{obs}

Optical and UV spectra have been collected for a sample of 20 Galactic
B supergiants, covering the spectral range of B0--B5 and including Ia,
Ib, Iab and II luminosity classes as well as a hypergiant. Stars were
only included in the sample if both optical and {\it IUE} data were
available for them. Where possible, stars were selected such that
there would be 2 different luminosity classes at each spectral
sub-type. The details of observational data for each star are given
in Table \ref{sample}.  Fourteen of the chosen B supergiants belong to
OB associations (\citealt{humphreys1978}), so for these stars, the absolute
magnitude given in Table \ref{sample} is based on the distance to the
relevant association; for the remaining six stars \mv\ and therefore
the
distance modulus is derived from photometry. \\

\noindent
For fifteen of the twenty B supergiants in our sample, the optical
spectra were taken from an existing data set (see \citealt{lennon1992}
for further details). The blue spectra were observed using the 1-m
Jacobus Kapteyn Telescope (JKT) at the Observatorio del Roque de los
Muchacos, La Palma in October 1990 with the Richardson-Brealey
Spectrograph and a R1200B grating. They have a wavelength coverage
of 3950 -- 4750 \AA, a spectral resolution of 0.8 \AA\ and a
signal-to-noise ratio $\sim$ 150. The red spectra were obtained with the
2.5-m Isaac Newton Telescope (INT) using the Intermediate Dispersion
Spectrograph (IDS) and cover a wavelength range of 6260 -- 6870
\AA\ (\citealt{lennon1992}). The spectral resolution is 0.7 \AA\ and the
signal-to-noise ratio is $>$ 100. Spectra for HD 192660, HD 185859, HD
190066 and HD 191243 were taken at the INT in July 2003, again using
the IDS. The R400B grating was used for the blue spectra, giving a
central wavelength $\lambda_c$ of 4300 \AA, whereas the R600R grating
was used for the red spectra, giving $\lambda_c$ $\sim$ 6550 \AA. The
signal-to-noise ratio was $>$ 100 and the spectral resolution was 0.7 \AA.
Finally high resolution time-averaged blue and red spectra of HD~64760
were provided by RKP (see \citealt{kaufer2002} for more
details). These spectra were taken in 1996 on the HEROS fiber-linked
echelle spectrograph, which was mounted on the ESO 50-cm telescope at
the La Silla, Chile. The blue spectra have a range of 3450 -- 5560
\AA\ whilst the red have 5820 -- 8620 \AA. Both had a resolving
power of 20\,000. The signal-to-noise ratio varied with lambda but
for a red spectrum with a 40-minute exposure it was typically $>$ 150.

\begin{table*}
\caption{Observational data for the sample of 20 Galactic B Supergiants. Spectral types and V magnitudes are taken from \cite{lennon1992} for all stars except HD 192660, HD 185859, HD 190066 and HD 64760. The references for the spectral types of these 4 stars are as follows: HD 192660 from \cite{walborn1971}; HD185859 from \cite{lesh1968}; HD 190066 from \cite{hiltner1956} and HD 64760 from \cite{hj1982} (from which the V magnitude of HD 64760 was also taken). V magnitudes for the remaining 3 stars were obtained from \cite{fernie1983}. Values of $(B-V)_0$ taken from \cite{fp1970}. Absolute visual magnitudes, distance moduli and cluster associations have been taken from: 1. \cite{brown1994}, 2.\cite{garmany1992} or 3. \cite{humphreys1978}. For stars not associated with a cluster, an absolute visual magnitude scale based on spectral type (\citealt{egret1978}) was used. L1992 refers to archive data obtained from \cite{lennon1992}, INT2003 denotes data taken on the 2.5m INT and RKP marks data supplied by R.K. Prinja.}
\label{tab1}
\centering
\label{tab1}
\begin{tabular}{rlllllllllll}
\hline\hline\\
HD no. & Alias & Sp. Type & V & B-V & \mv & Association &
m-M & Optical & {\it IUE} \\
\hline\\
\object{37128}  & $\epsilon$ Ori & B0 Ia & 1.70 & -0.19 & -6.95 & Ori OB1b & 7.8$^{1}$ & L1992 & SWP30272 \\
\object{192660} & - & B0 Ib & 7.38 & 0.67 & -7.0 & Cyg 0B8 & 11.8$^{3}$ & INT2003 & SWP44625 \\
\object{204172} & 69 Cyg & B0.2 Ia & 5.94 & -0.08 & -6.2 & Cyg OB4 & 6.2 & L1992 & SWP48900 \\
\object{38771}  & $\kappa$ Ori & B0.5 Ia & 2.04 & -0.18 & -6.51 & Ori OB1c & 8.0$^{1}$ & L1992 & SWP30267 \\
\object{185859} & - & B0.5 Ia & 6.48 & +0.44 & -6.6 & - & 13.08 & INT2003 & SWP47509 \\
\object{213087} & 26 Cep & B0.5 Ib & 5.46 & +0.37 & -6.2 & Cep OB1 & 12.7 & L1992 & SWP02735 \\
\object{64760}  & - & B0.5 Ib & 4.24 & -0.15 & -6.2 & - & 10.44 & RKP & SWP53781 \\
\object{2905}   & $\kappa$ Cas & BC0.7Ia & 4.16 & +0.14 & -7.09 & Cas OB14 & 10.2$^{3}$ & L1992 & SWP54038 \\
\object{13854}  & V551 Per & B1 Iab(e) & 6.47 & +0.28 & -6.73 & Per OB1 & 11.8$^{2}$ & L1992 & SWP02737 \\
\object{190066} & - &  B1 Iab(e) & 6.53 & +0.18 & -6.1 & - & 12.63 & INT2003 & SWP18310 \\
\object{190603} & V1768 Cyg & B1.5 Ia+ & 5.64 & +0.54 & -7.0 & - & 12.56 & L1992 & SWP4325 \\
\object{193183} & - & B1.5 Ib & 7.01 & +0.44 & -6.24 & Cyg OB1 & 11.3$^{3}$ & L1992 & SWP52716 \\
\object{14818}  & V554 Per & B2 Ia & 6.25 & +0.30 & -6.93 & Per OB1 & 11.8$^{3}$ & L1992 & SWP18658 \\
\object{206165} & V337 Cep & B2 Ib & 4.74 & +.30 & -6.44 & Cep OB2 & 13.2 & L1992 & SWP06336 \\ 
\object{198478} & 55 Cyg & B2.5 Ia & 4.84 & +0.40 & -6.43 & Cyg OB7 & 9.6$^{2}$ & L1992 & SWP38688 \\
\object{42087}  & 3 Gem & B2.5 Ib & 5.75 & +0.22 & -6.26 & Gem OB1 & 10.9$^{3}$ & L1992 & SWP08645\\
\object{53138}  & 24 CMa & B3 Ia & 3.01 & -0.11 & -7.1 & - & 10.11 & L1992 & SWP30271 \\
\object{58350}  & $\eta$ CMa & B5 Ia & 2.41 & -0.07 & -7.0 & - & 9.41 & L1992 & SWP30198 \\
\object{164353} & 67 Oph & B5 II(Ib) & 3.97 & +0.02 & -4.2 & Coll 359 & 6.5 & L1992 & SWP08560 \\
\object{191243} & - & B5 II(Ib) & 6.09 & +0.16 & -6.5 & Cyg OB3 & 11.8$^{3}$ & INT2003 & SWP07737 \\
\hline\hline\\
\end{tabular}
\end{table*}

\subsection{Sample details}
\label{sample}

As previously mentioned, our sample covers a range of B0--B5
supergiants with luminosity classes varying from Ia down to II in a
couple of cases. Spectral type classifications have been taken from
\cite{lennon1992}, which includes some recent revisions. HD 204172 has
been changed from B0 Ib to B0.2 Ia due to the strength of its
\ion{Si}{iv} lines and narrowness of its H lines.  Comparing its UV
resonance lines to those of the B0 Ib star HD 164402 supports the
change to a more luminous spectral type (\citealt{prinja2002}). Also both
HD 164353 and HD 191243 have been reclassified from B5 Ia to B5 II
stars. It is also worth noting that \cite{dezeeuw1999} has questioned
the membership of the stars HD 53138 and HD 58350 to Collinder 121 on
account of insignificant proper motion and small parallax
respectively. Seven of the twenty stars in our sample have been
examined for \ha\ variability by \cite{morel2004}, who obtained both
photometric and spectroscopic data on these objects in order to
ascertain the amount of variability present in their light-curves and
\ha\ profiles. \cite{morel2004} quantify the amount of spectral and
photometric variability observed as well as determining periods where
cyclic behaviour is observed.  The sample includes a rapid rotator, HD
64760 (discussed in detail in the following section) and a hypergiant,
the B1.5 Ia+ star HD 190603. Furthermore, there are several objects 
of interest in the sample for which a significant amount of research
has already been undertaken and  merit further discussion. \\

\begin{itemize}
\item{$\mathbf{\epsilon}$ \textbf{Ori}}: There are several intriguing aspects of
  this star that are worth mentioning. Firstly it has been noted as
  moderately nitrogen deficient by \cite{walborn1976}. Secondly it is
  known to undergo significant variations in H$\alpha$, with
  \cite{morel2004} reporting variations of 81.9 \%. Further studies by
  Prinja et al. (2004) have revealed variability in not only
  H$\alpha$ but also H$\beta$, He absorption and metal lines with a 1.9
  day period. A modulating S-wave pattern has been discerned in the
  weaker lines, which cannot be fully explained by current non-radial
  pulsation models (\citealt{townsend1997}). These results highlight a
  direct connection between photospheric activity and perturbations in
  the stellar wind. Finally this star is the only normal early B
  supergiant to have a measured thermal radio flux (Blomme et al.
  2002), from which
  a radio mass loss rate of log \.{M} $= -5.72$ was derived. \\

\item{$\mathbf{\kappa}$ \textbf{Ori}}: Like $\epsilon$ Ori, $\kappa$ Ori also
  exhibits spectral variability in H$\alpha$. Its \ha\ profile was
  studied in detail by \cite{rusconi1980} who described it as a
  double-peaked absorption profile with a central emission core and
  broad profile wings, with variations on long (of the order of
  days) and short (of the order of minutes) time scales. More recently
  \cite{morel2004} reported changes in the profile amplitude and
  morphology of 32.6 \%. \cite{walborn1976} noted it as an example of
  a morphologically normal B supergiant in terms of the relative
  strengths of its CNO spectral lines. \\
  
\item{\textbf{HD 192660}}: \cite{bidelmann1988} noted that this star
  showed a ``faint \ha\ emission with a slight P Cygni absorption'',
  based on observations taken at the Lick observatory in
  1957. Our spectrum of this star shows a very similar \ha\ profile
  that is also only weakly in emission. \cite{walborn1976} and
  \cite{schild1985} both noted that HD192660 displays evidence for
  nitrogen deficiency. \\
  
\item{\textbf{HD 64760}}: This star is classified as a rapid rotator, having
  a {\em v} sin {\em i} of 265 km/s. Many interesting studies have
  been carried out regarding the periodic and sinusoidal modulations
  of its optical and UV lines
  (\citealt{prinja1995,fullerton1997,kaufer2002}), which have turned HD
  64760 into a key object for improving our understanding of the
  spatial structure and variations of hot star winds. More recently,
  \cite{kaufer2006} found, for the first time, direct observational
  evidence for a connection between multi-periodic non-radial
  pulsations (NRPs) in the photosphere and spatially structured winds.
  More specifically, they can use the interference of multiple
  photospheric pulsation modes on hourly timescales with wind
  modulation periods on time scales of several days. A beat period of
  6.8 days seen in the photosphere and base of the wind does not
  match with the derived periods of 1.2 and 2.4 days for wind
  variability, being closer to the longer 5-11 day repetitive
  timescales observed for discrete absorption components (DACs) in the
  {\it IUE} data sets. Evidently the precise nature of the
  wind-photosphere connection in this star is a complex one. Using
  hydrodynamical simulations, \cite{cranmer1996} succeeded in
  confirming the existence of co-rotating interaction regions (CIRs).
  These are spiral structures in the wind that are produced through
  the collisions of fast and slow streams rooted in the stellar
  surface. \\
    
\item{$\mathbf{\kappa}$ \textbf{Cas}}: \cite{walborn1972a} classified this star as
  carbon rich, giving it a spectral type of BC0.7 Ia, when comparing
  its optical spectrum with that of HD 216411, a B0.7 Ia star.
  He found that the nitrogen lines in $\kappa$ Cas were barely
  detectable, whereas the O II - C III blends were very prominent.
  Since Walborn also recognised that HD 216411 possesses a well
  developed nitrogen spectrum, he described $\kappa$ Cas as
  carbon-rich, rather than nitrogen-weak, with respect to a
  morphologically-normal B supergiant. Walborn had previously
  suggested that all OB stars begin their post-main-sequence evolution
  with an enhancement of carbon (\citealt{walborn1971}), which then
  becomes depleted as the star evolves and produces nitrogen as a
  by-product of the CNO bi-cycle. Therefore the implication of
  classifying $\kappa$ Cas as a BC 0.7 Ia star is that it is less
  evolved than other stars in the sample. Please consult Section
  \ref{cno} for a discussion of the carbon rich status of this star
  based on the results presented in that section. \\
  
\item{\textbf{HD 13854}}: \cite{mcerlean1999} describe this star as `highly processed' i.e. displaying a large amount of CNO processing in its spectrum.  Its \ha\ profile is seen mostly in emission, assuming a P Cygni shape.  \cite{morel2004}
  found that not only does the \ha\ profile of this star vary by 47.3
  \%, but that these variations have a period of 1.047 days $\pm 0.01$. {\it Hipparcos} light curves also show evidence for
  periodic behaviour. \\
      
\item{\textbf{HD 14818}}: possesses an \ha\ profile with a P Cygni
  profile. \cite{morel2004} report variations of 34.8 \% in the \ha\ 
  profile and, like HD 13854, they find that this star also shows
  periodic behaviour in its {\it Hipparcos} light curve. \\
  
\item{\textbf{HD 42087}}: also has its \ha\ profile in emission but
  more importantly \cite{morel2004} reported significant \ha\ 
  variability of 91.2 \% ({\em greater} than the percentage variability that
  they found for $\epsilon$ Ori), for which they find strong evidence
  of cyclic behaviour on a periodicity of 25 days $\pm$ 1-4 days.
  Moreover, the \ha\ variability correlates with variability in the
  \ion{He}{i} 6678 \AA\ line, such that as \ha\ emission increases,
  \ion{He}{i} 6678 becomes weaker. \cite{morel2004} also find periodic
  variability in its {\it Hipparcos} light curve. \\
  
\item{\textbf{HD 53138}}: \cite{walborn1976} notes that this star shows
  a morphologically normal CNO spectrum, despite other authors
  associating this star with OBN/OBC groups. It undergoes 66.4 \% \ha\ 
  variability (\citealt{morel2004}). \\

\end{itemize}

\section{Derivation of fundamental parameters}
\label{param}

\begin{figure}
\centering
\resizebox{\hsize}{!}
{\includegraphics[width=\textwidth]{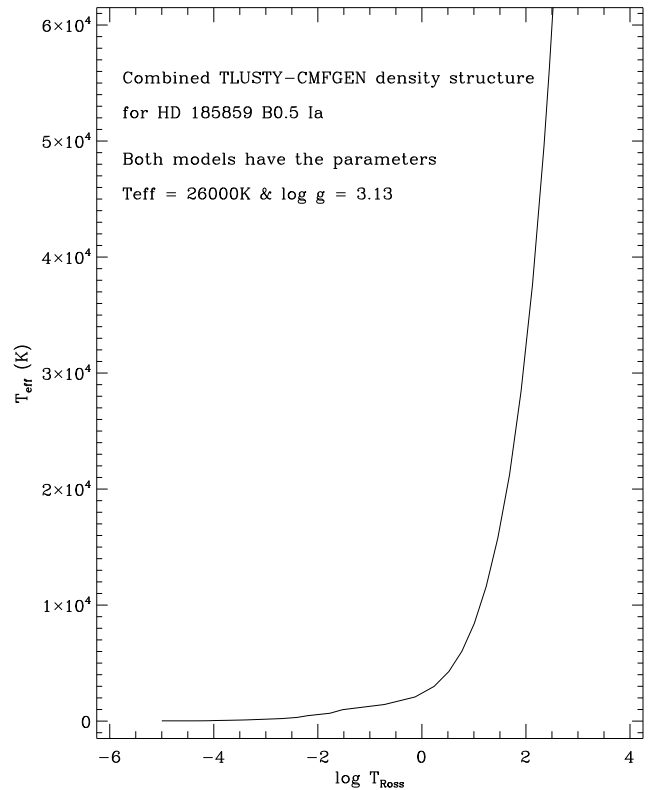}}
\caption{Temperature structure against Rosseland optical mean depth for the hydrostatic density structure produced from combining the subsonic TLUSTY velocity structure with the supersonic CMFGEN velocity structure, such that the velocity and velocity gradient are constant. The above example is for the B0.5 Ia star HD 185859 (\teff $=$ 26 000 K, log $g$ $=$ 3.13) }
\label{tvel}
\end{figure}

Fundamental parameters were derived for this sample of stars using the
nLTE stellar atmosphere codes {\sc TLUSTY}
(\citealt{hubeny1995,lanz2003}) and {\sc CMFGEN}
(\citealt{hillier1998}). The application of {\sc TLUSTY}, a
plane-parallel photospheric code that does not account for the
presence of a stellar wind, to modelling supergiants is valid as long
as it is used solely to model purely photospheric lines. An existing
grid of B star {\sc TLUSTY} models (\citealt{dufton2005}) was used as
a base for this work and the grid (originally incremented in steps of
$\sim$ 2\,000 K in \teff\ and 0.13 in log $g$) was refined further by
RR in the range 15\,000 K $\le$ \teff\ $\le$ 23\,000 K to cover the
same parameter space as the {\sc CMFGEN} grid. The {\sc TLUSTY} models
provide a hydrostatic structure that can be input into {\sc CMFGEN},
since the latter code does not solve for the momentum equation and
therefore requires a density/velocity structure (see e.g.,
\citealt{hillier2003,bouret2003,martins2005}). The {\sc TLUSTY} input
provides the subsonic velocity structure and the supersonic velocity
structure in the {\sc CMFGEN} model is described by a $\beta$ - type
law. The two structures are joined to a hydrostatic density structure
at depth, such that the velocity and velocity gradient are
consistent. The resulting structure is shown in
Fig. \ref{tvel}, which shows the change in Rosseland mean opacity,
$\tau_{Ross}$,with temperature and ensures that the model is
calculating deep enough into the photosphere to sample the regions
where the appropriate photospheric lines form 
(around -2 $\le$ log$\tau_{Ross}$ $\le$ 0). 
This structure is then input into the CMFGEN 
model, adopting a $\beta$-type velocity law of the form:   

\begin{equation}
v(r) = \frac{\vzero + (\vinf - \vzero){(1-\rstar/r)}^{\beta}}{1+ \frac{\vzero}{\vcore} e^{\frac{\rstar - r}{\heff}}} 
\end{equation}
  
\noindent
where \vzero\  is the photospheric velocity, \vcore\  is the core
velocity, \vinf\  is the terminal velocity, \heff\  is the scale height
expressed in terms of \rstar\ and $\beta$ is the acceleration
parameter. The value of $\beta$ is normally determined from fitting the
\ha\ profile, as discussed in \S \ref{swparam}, assuming typical values of 
$ 1.0 \le \beta \le 1.5$ for B supergiant. Additionally, values of 
\vcore\ = 0.002 km/s and \vzero\ = 0.1 km/s are adopted for B supergiants
as suggested by D.J. Hillier (priv. comm.).  All parameters except log $g$ were derived
using {\sc CMFGEN} and the precise details of the methods employed
will be discussed next. The general method employed was to produce a
grid of {\sc CMFGEN} models of varying temperatures and luminosities
(all other parameters were kept constant), incremented in steps of
1\,000 K in \teff\ and 5 in \logllsolar, and compare these synthetic
spectra to observed spectra in order to constrain these two parameters
(\S
\ref{teff}). Once satisfactory values had been derived for the
temperature and luminosity of a star, the mass loss rate, $\beta$
velocity law, turbulent velocity (\vturb) (\S \ref{swparam}) and CNO
abundances (\S \ref{cno}) could be constrained. These parameters are
all sensitive to changes in temperature and luminosity so can only be
derived once those values have been fixed.  Consistency checks are
made after each parameter is altered to ensure that the resulting
model output has not worsened the fit to the diagnostic lines and
overall spectrum. Given recent improvements in the treatment of metal
lines in nLTE stellar atmosphere codes such as {\sc CMFGEN} or {\sc
FASTWIND}, it is now possible to obtain accurate abundances by fitting
(by eye) appropriate diagnostic lines for each element. All models
were calculated to include the following elements; H, He, C, N, O, Al,
Mg, Si, S, Ca \& Fe, assuming solar abundances for silicon, magnesium,
aluminium, phosphorous, sulphur, calcium and iron and adopting a
relative number fraction of 5:1 for H:He (as used by
\citealt{hillier2003}). {\sc CMFGEN} adopts the superlevel
approximation (see \citealt{anderson1989} for more details), where a
superlevel can consist of several or even many energy levels grouped
together, such that all real levels {\it j} that form superlevel {\it
J} have the same nLTE departure coefficient (i.e., each component {\it
j} is in Boltzmann equilibrium with respect to the other
components). Details of the model atoms, including their full level
and superlevel groupings, are given in Table \ref{atomod}.

\begin{table}
\centering
\caption{{\sc CMFGEN} model atomic data, showing the number of full levels and superlevels treated as well as the number of bound-bound transitions considered for each ion included in a {\sc CMFGEN} model.}\label{atomod}
\begin{tabular}{lcccl}
\hline\noalign{\smallskip}
Ion & Full Levels & Superlevels & b-b transitions \\
\hline
H I & 30 & 20 & 435 \\
He I & 59 & 41 & 590 \\
He II & 30 & 20 & 435 \\
C II & 53 & 30 & 323 \\
C III & 54 & 29 & 268 \\
C IV* & 18 & 13 & 76 \\  
N I & 22 & 10 & 59 \\
N II & 41 & 21 & 144 \\
N III* & 70 & 34 & 430 \\
O I & 75 & 18 & 450 \\
O II & 63 & 22 & 444 \\
O III* & 45 & 25 & 182 \\
Mg II & 45 & 18 & 362 \\
Al II & 44 & 26 & 171 \\
Al III & 65 & 21 & 1452 \\
Si II & 62 & 23 & 365 \\
Si III & 45 & 25 & 172 \\
Si IV & 12 & 8 & 26 \\
S II & 87 & 27 & 786 \\
S III & 41 & 21 & 177 \\
S IV* & 92 & 37 & 708 \\
Ca II & 12 & 7 & 28 \\
Fe II & 510 & 100 & 7501 \\
Fe III & 607 & 65 & 5482 \\
Fe IV & 272 & 48 & 3113 \\
Fe V* & 182 & 46 & 1781 \\
\hline
\end{tabular}
\end{table}

\subsection{Derivation of \teff, \logllsolar, log $g$ and CNO abundances}
\label{teff}

In B stars, the silicon lines are used as the primary temperature
diagnostics, having the advantage that the abundance is well known as
silicon is unaffected by nuclear processing.  For B0 - B2 supergiants
the \ion{Si}{iv} 4089~\AA\ and \ion{Si}{iii} 4552, 4568 and 4575~\AA\
lines provide the main temperature diagnostics. \ion{Si}{iv}~4089~\AA\
decreases in strength as the temperature decreases until it is barely
detectable at a spectral type of B2.5, which corresponds to \teff\ $
\sim$ 18\,000 K.  At this point the \ion{Si}{ii}~4128-30 \AA\ doublet
is present and replaces
\ion{Si}{iv}~4089~\AA\ as the main temperature diagnostic for B2.5 - B9
stars, along with \ion{Si}{iii} 4552, 4568 and 4575 \AA. The \ion{He}{i} lines at
4144~\AA, 4387~\AA, 4471~\AA\ and 4713~\AA\ and \ion{Mg}{ii} line at 4481~
\AA\ can also be used as secondary criteria for both temperature and
luminosity, since they are sensitive to changes in both parameters.
The principal luminosity criteria used in spectral classification of B
stars is the ratio of \ion{Si}{iv}~4089~\AA\ to \ion{He}{i}~4026~\AA,
4121~\AA\ and/or 4144~\AA for B0 - B1 supergiants, whereas for stars
later than B1 the ratio of \ion{Si}{iii} 4552, 4568 and 4575~\AA\ to
\ion{He}{i} 4387~\AA\ is used.
The procedure adopted for deriving values of \teff, \logllsolar, log $g$
and CNO abundances is the same method adopted by 
\cite{hillier2003,crowther2006,bouret2003,martins2005} and is as follows:

\begin{enumerate}
  
\item An optical stellar spectrum of a chosen star is compared
   to a grid of CMFGEN models which differ in values of Teff and
   luminosity (all other parameters are kept constant). \\
  
\item A value of Teff is selected for the star by finding the
   model that provides the best fit to the temperature sensitive
   silicon lines. The diagnostic lines are \ion{Si}{iv} 4089 and
    \ion{Si}{iii} 4552, 4568 and 4575 for B0-B2 supergiants and
    \ion{Si} {ii} 4128-30 and \ion{Si}{iii} 4552, 4568 and 4575 for
    B2.5-B5 supergiants. This temperature would then be 
      confirmed by checking that the model also succeeded in fitting
    the helium lines listed earlier and \ion{Mg}{ii} (but these lines
    were not used to derive the initial value of \teff) to ensure it
    provided a reasonable match to all temperature-sensitive lines. \\
  
\item Once a value of \teff\ has been chosen, an inital
   estimate of \logllsolar\ is made by selecting the model from the
   grid (at the chosen value of \teff) whose value of \mv\ (output
   from the model) best matches the observed value of \mv\ for the
   star in question. The luminosity is then constrained further by
   taking observed values of \mv, the absolute visual magnitude and
   $V$, together with the estimate of $A(V)$, were used to obtain an
   initial estimate of the distance modulus. Optical photometry and
   ultraviolet spectroscopy were then de-reddened with respect to the
   model spectral energy distribution to obtain revised estimates of
   E(B-V) and \mv. This \mv\ derived from the model was then compared
   to an observed \mv, if the values matched then the model
   luminosity was correct. If not, the value of \mv\ was translated
   into a bolometric correction to obtain an estimate of the
   corrected luminosity for the model, which was then rerun with this
   value for the luminosity. This iterative process was continued
   until the observed and model values of \mv\ were in reasonable
   agreement and then the resulting model was checked against the
   \teff\ diagnostic lines (\ion{Si}{ii}, \ion{Si}{iii},
    \ion{Si} {iv}) and
    derived value of \teff\ was adjusted if necessary. \\
  
\item Estimates of log $g$ were then made from TLUSTY fits to
   the \hg\ lines of the observed spectra. \hg\ is normally chosen as
   the log $g$ diagnostic since \ha\ and \hb\ suffer from too much
   wind fill emission; \hd\ was used as a secondary diagnostic to
   check for consistency with values of log $g$ derived from \hg.
   Since \hg\ is affected by an \ion{O}{ii} blend around 4350 \AA\ 
    and \hd\ has a blend with \ion{N}{iii} 4097, neither line is ideal
    but both implied the same log $g$ values so we can have confidence
    that the derived log $g$ values are not affected by these blends.
    The adopted log g value was then incorporated into the {\sc
      CMFGEN} model and again the derived \teff\ and \logllsolar\ 
    values were
    revised if necessary. \\
  
\item Next CNO abundances were derived by varying the
   abundance of each element until the appropriate diagnostic lines
   were fitted by the model. For nitrogen, we use the \ion{N}{iii}
   4097 \AA\ line for B0 -- B2 stars (which is blended with H$\delta$)
   and the \ion{N}{ii} 3995 \AA, \ion{N}{ii} 4447 \AA\ and 4630 \AA\
   lines (for all B stars) as primary diagnostics. The main
   diagnostics for carbon and oxygen are the 4267 \AA\ and 4367 \AA\
   lines respectively; however their abundances are confirmed by
   checking for good fits to \ion{O}{ii} blends at 4070 \AA, 4317 --
   4319 \AA\ and 4650 \AA, the \ion{O}{ii} lines at 4590 \AA, 4596
   \AA\ and 4661 \AA\ and the \ion{C}{ii} doublet at 6578, 6582 \AA.
   The errors on constraining the CNO abundances using this method
   were typically up to $\sim$ 0.3 dex. \\

\end{enumerate}
 
\subsection{Determination of Stellar Wind Properties}
\label{swparam}

The stellar wind parameters \mdot, $\beta$ and \vturb were 
then constrained using the usual method outlined below (again the same
procedure used by
\citealt{hillier2003,crowther2006,bouret2003,martins2005}).
{\sc CMFGEN } allows for a treatment of turbulence in the stellar wind
by assuming a radially-dependent microturbulent velocity, defined as

\begin{equation}\label{vt}
  \vturb\ = \vmin + \frac{(\vmax - \vmin) v(r) }{\vinf}
\end{equation}

\noindent
where \vmin\ is the minimum turbulent velocity occurring in the
photosphere and \vmax\ is the maximum turbulent velocity.
\cite{hillier2003} found that varying the turbulent velocity had
little effect on the temperature structure calculated by {\sc CMFGEN}.
In this work, values of \vmin\ $=$ 10 km/s and \vmax\ $=$ 50 km/s are
adopted as limits. Typically $v$ reaches the value of \vturb\
around $r =$ 1.05\rstar.

\begin{enumerate} 
  
\item Values for the mass loss rate of each star were
   constrained by fits to the H$\alpha$ profile, with each fit aiming
    to reproduce the overall profile shape and amplitude. \\
  
\item Values of \vinf\ are best determined from the UV so
   values obtained through UV line synthesis modelling (see
    \citealt{prinja2005}; Paper II) are used here. \\
  
\item The value of $\beta$ is also varied in order to improve
   the shape of the model profile fit with respect to the observed
   profile, but this has no effect for \ha\ profiles in absorption,
    in which case values obtained from SEI analysis were used. \\
  
\item Estimates of the microturbulent velocity, \vturb, were
   then made by fitting the \ion{Si}{iii} 4552, 4568 and~ 4575
    lines. \\
  
\item The model fit to the temperature and luminosity
   sensitive lines was checked after altering these parameters to
    ensure consistency. \\

\end{enumerate}

\noindent
The sample of 20 B supergiants showed some variation in overall
profile shape, with four stars showing \ha\ in emission, six in
absorption and the remainder displaying a more complex morphology. In
the last case, the profiles are partly in absorption with some
emission component also detectable, implying that the profile has been
filled in by stellar wind emission.  The B2~-~B5 supergiants display
H$\alpha$ profiles with a P Cygni profile shape, which implies that
line scattering is playing a significant role in the line's formation,
though this is not normally observed until late B/A supergiants. The
majority of stars in this sample have no record of \ha\ variability so
that any changes in the line profile morphology and amplitude can be
considered negligible for the purpose of this analysis. However
$\epsilon$ Ori, HD 13854, HD 14818 and HD 42087 all display
significant \ha\ variability according to \cite{morel2004}. In view of
this problem, for $\epsilon$ Ori we have assumed a radio mass loss
rate, \mdotradio, of $1.9 \times 10^{-6}$ \msolaryr, as measured by
\cite{blomme2002}, thereby avoiding the inaccuracies involved in
deriving \mdot\ from a variable \ha\ line. An estimate of the error in
fitting the \ha\ profile of this star is given nonetheless is Table
\ref{results2}. Unfortunately this approach is not possible for the
other 3 stars since there are no reliable \mdotradio\ values available
in the literature. The problems associated with deriving \mdot\ from
the \ha\ profiles of these stars are discussed in the next section.
  \\

\section{Results}
\label{results}

\subsection{The B supergiant \teff\ scale}
\label{tempscale}


\begin{figure*}
\centering
\includegraphics[scale=0.75,width=12cm]{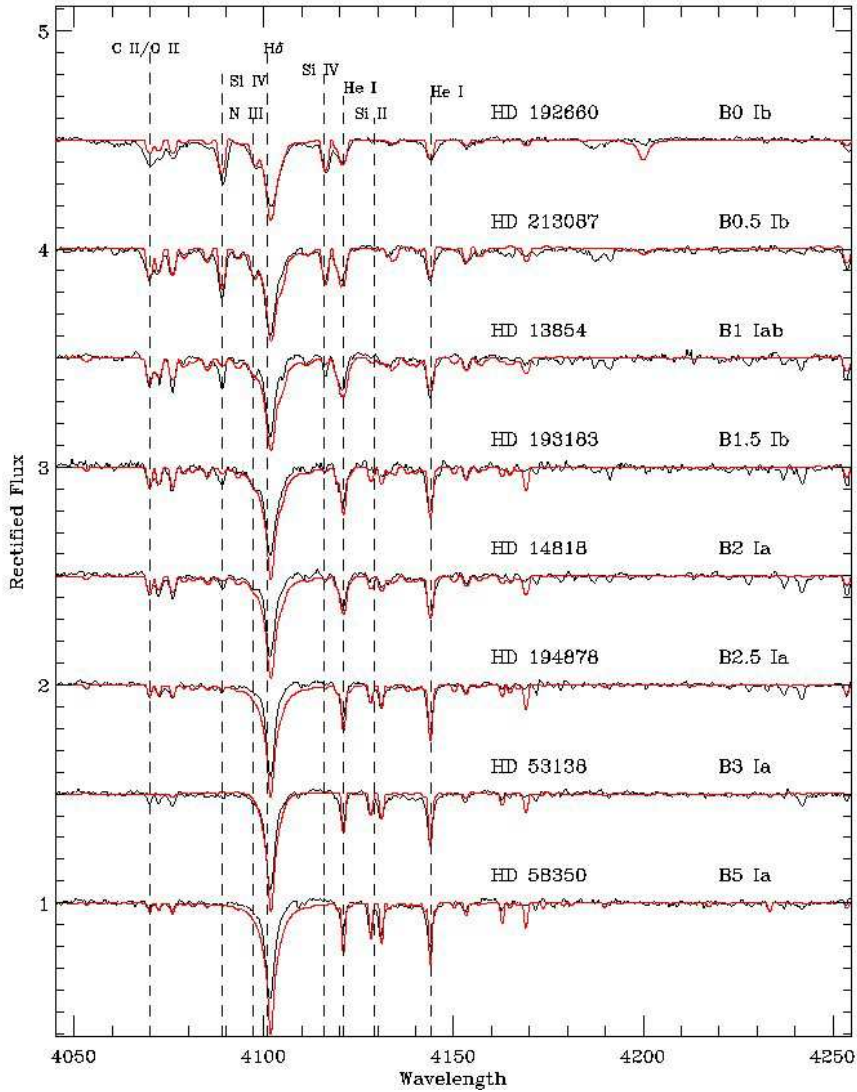}
\caption{Overall {\sc CMFGEN} fit to the optical spectrum of B0 -- B5 supergiants (4050 \AA\ - 4250 \AA). The solid black line is the observed spectrum and the red line denotes the {\sc CMFGEN} model fit.}
\label{op1}
\end{figure*}

\begin{figure*}
\centering
\includegraphics[scale=0.75,width=12cm]{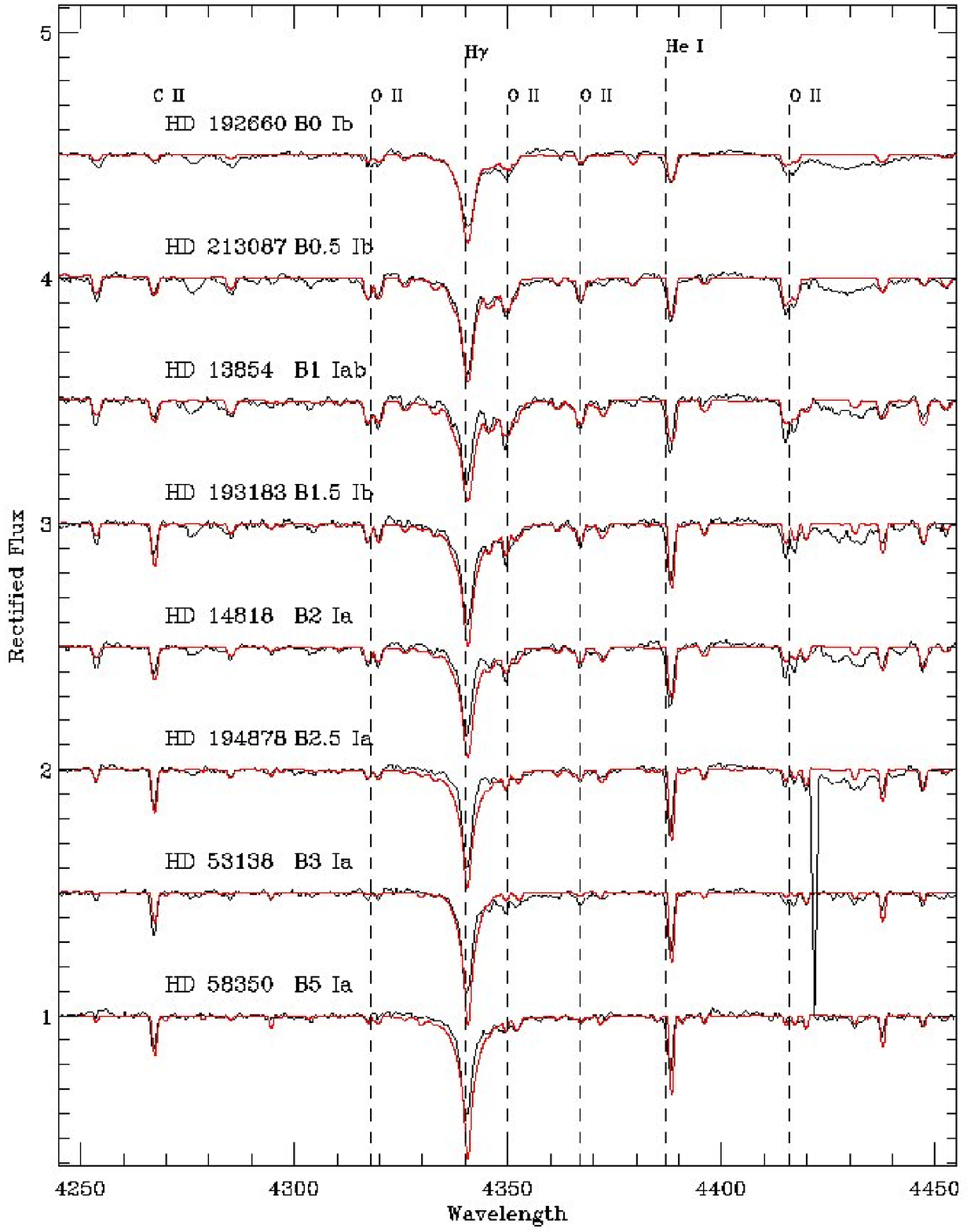}
\caption{Overall {\sc CMFGEN} fit to the optical spectrum of B0 -- B5 supergiants (4200 \AA\ - 4450 \AA). The solid black line is the observed spectrum and the red line denotes the {\sc CMFGEN} model fit.}
\label{op2}
\end{figure*}

\begin{figure*}
\centering
\includegraphics[scale=0.75,width=12cm]{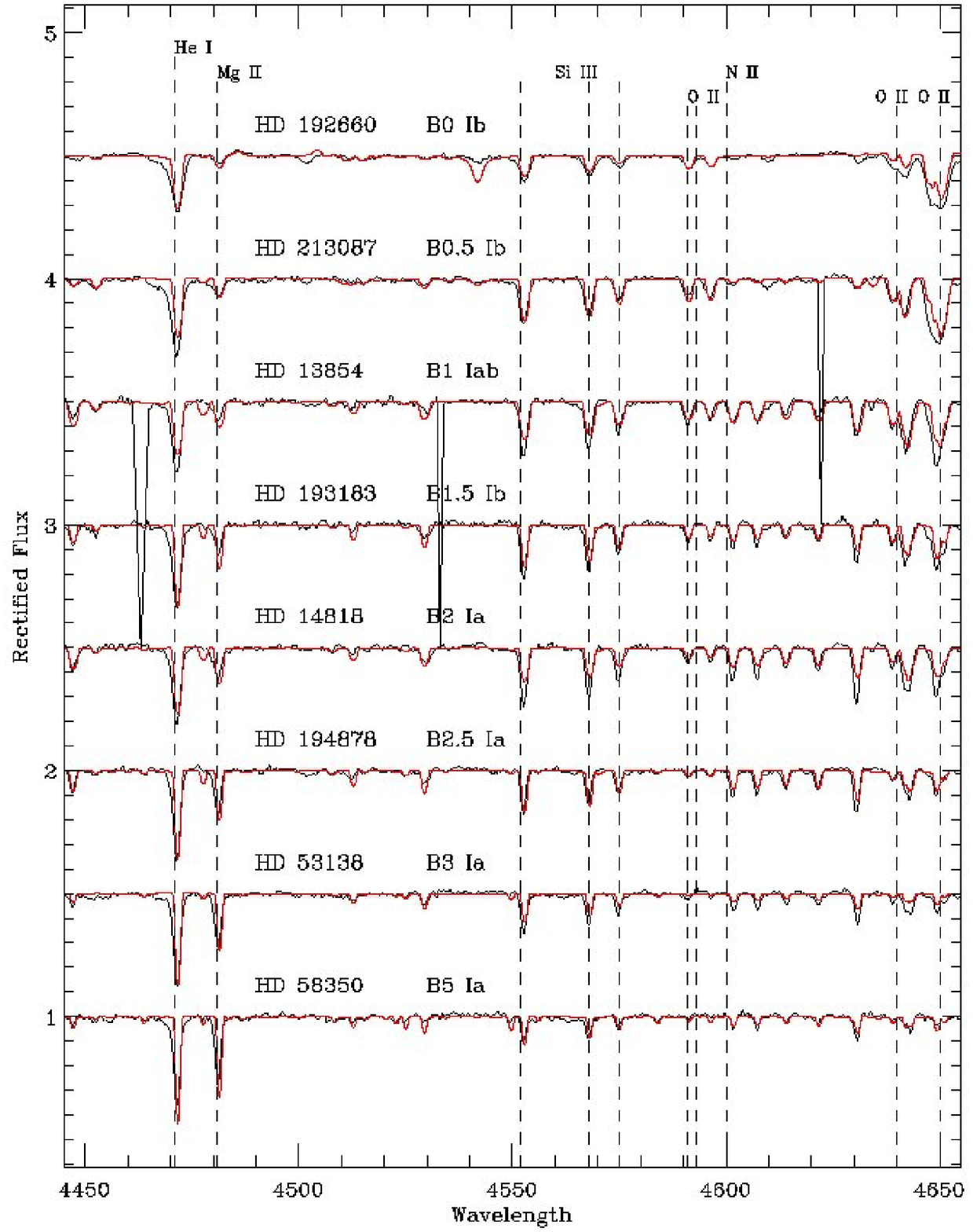}
\caption{Overall {\sc CMFGEN} fit to the optical spectra of B0 -- B5 supergiants (4450 \AA\ - 4650 \AA). The solid black line is the observed spectrum and the red line denotes the {\sc CMFGEN} model fit.}
\label{op3}
\end{figure*}


\begin{table*}
\centering
\begin{normalsize}
\setlength\tabcolsep{5pt}
\caption{Fundamental parameters (\teff, log $g$, \logllsolar, \rstar(\rsolar), E(B-V) \mv and \mevol) derived for the sample of 20 Galactic B supergiants.Values of {\it v$_{e}$} sin {\it i} are taken from \cite{howarth1997} and are expressed in km/s.}
\label{results1}
\begin{tabular}{lllllllllllllll}
\hline\hline\noalign{\smallskip}
HD no. & Sp. type & $\teff\ $ (K) & log $g$ & log (L/L$_{\odot}$) &
\rstar(\rsolar) & E(B-V) & \mv & {\it v$_{e}$} sin {\it i} & \mevol \\
\noalign{\smallskip}
\hline
\noalign{\smallskip}
37128 & B0 Ia        & 27500 $\pm$ 1000 & 3.13 & 5.73 $\pm$ 0.11 & 32.4 $\pm$ 0.75 & 0.08 $\pm$ 0.02 & -6.89 $\pm$ 0.05 & 91 & 40 \\
192660& B0 Ib      & 30000 $\pm$ 1000 & 3.25 & 5.74 $\pm$ 0.13 & 23.4 $\pm$ 1.03 & 0.80 $\pm$ 0.10 & -6.66 $\pm$ 0.10 & 94 & 33 \\
204172& B0.2 Ia   & 28500 $\pm$ 1000 & 3.13 & 5.48 $\pm$ 0.27 & 22.4 $\pm$ 3.23 & 0.12 $\pm$ 0.04 & -6.07 $\pm$ 0.30 & 87 & 41 \\
38771 & B0.5Ia     & 26000 $\pm$ 1000 & 3.00 & 5.48 $\pm$ 0.14 & 27.0 $\pm$ 1.24 & 0.07 $\pm$ 0.01 & -6.48 $\pm$ 0.10 & 91 & 33 \\
185859& B0.5Ia    & 26000 $\pm$ 1000 & 3.13 & 5.54 $\pm$ 0.14 & 29.1 $\pm$ 1.34 & 0.53 $\pm$ 0.02 & -6.54 $\pm$ 0.10 & 74 & 35 \\
213087& B0.5Ib    & 27000 $\pm$ 1000 & 3.13 & 5.69 $\pm$ 0.11& 32.0 $\pm$ 0.01 & 1.30 $\pm$ 0.01 & -6.20 $\pm$ 0.10 & 88 & 40 \\
64760 & B0.5Ib     & 28000 $\pm$ 2000 & 3.38 & 5.48 $\pm$ 0.26 & 23.3 $\pm$ 2.15 & 0.15 $\pm$ 0.05 & -6.36 $\pm$ 0.20 & 265 & 33\\
2905  & BC0.7Ia    & 23500 $\pm$ 1500 & 2.75 & 5.48 $\pm$ 0.22 & 33.0 $\pm$ 1.52 & 0.29 $\pm$ 0.03 & -7.00 $\pm$ 0.10 & 91 & 33 \\
13854 & B1 Iab     & 20000 $\pm$ 2000 & 2.50 & 5.54 $\pm$ 0.57 & 49.2 $\pm$ 1.33 & 0.60 $\pm$ 0.10 & -6.41 $\pm$ 0.50 & 97 & 33\\
190066& B1 Iab    & 21000 $\pm$ 1000 & 2.88 & 5.54 $\pm$ 0.20 & 41.4 $\pm$ 1.89 & 0.55 $\pm$ 0.02 & -6.04 $\pm$ 0.10 & 82 & 33\\
190603& B1.5 Ia+ & 19500 $\pm$ 1000 & 2.38 & 5.41 $\pm$ 0.23 & 44.5 $\pm$ 3.07 & 0.70 $\pm$ 0.05 & -6.85 $\pm$ 0.15 & 79 & 27\\
193183& B1.5 Ib   & 18500 $\pm$ 1000 & 2.63 & 5.00 $\pm$ 0.26 & 30.8 $\pm$ 2.84 & 0.70 $\pm$ 0.06 & -6.43 $\pm$ 0.20 & 68 & 18\\
14818 & B2 Ia         & 18000 $\pm$  500 & 2.38 & 5.40 $\pm$ 0.27 & 51.4 $\pm$ 7.10 & 0.62 $\pm$ 0.10 & -6.70 $\pm$ 0.30 & 82 & 26\\
206165& B2 Ib       & 18000 $\pm$  500 & 2.50 & 5.18 $\pm$ 0.26 & 39.8 $\pm$ 5.50 & 0.56 $\pm$ 0.10 & -6.64 $\pm$ 0.30 & 73 & 21\\
198478& B2.5 Ia    & 17500 $\pm$  500 & 2.25 & 5.26 $\pm$ 0.14 & 46.1 $\pm$ 2.12 & 0.40 $\pm$ 0.05 & -7.26 $\pm$ 0.10 & 61 & 23\\
42087 & B2.5 Ib    & 18000 $\pm$ 1000 & 2.50 & 5.11 $\pm$ 0.24 & 36.6 $\pm$ 1.69 & 0.60 $\pm$ 0.02 & -6.11 $\pm$ 0.10 & 71 & 21\\
53138 & B3 Ia         & 16500 $\pm$  500 & 2.25 & 5.30 $\pm$ 0.27 & 54.7 $\pm$ 7.56 & 0.35 $\pm$ 0.10 & -6.79 $\pm$ 0.30 & 58 & 23\\
58350 & B5 Ia         & 15000 $\pm$  500 & 2.13 & 5.18 $\pm$ 0.17 & 57.3 $\pm$ 2.64 & 0.05 $\pm$ 0.04 & -7.12 $\pm$ 0.10 & 50 & 21\\
164353& B5 Ib     & 15500 $\pm$ 1000 & 2.75 & 4.30 $\pm$ 1.30 & 19.6 $\pm$ 8.05 & 0.71 $\pm$ 0.05 & -6.15 $\pm$ 2.00 & 44 & 10\\
191243& B5 Ib     & 14500 $\pm$ 1000 & 2.75 & 5.30 $\pm$ 0.37 & 70.8 $\pm$ 3.26 & 0.93 $\pm$ 0.03 & -6.59 $\pm$ 0.10 & 38 & 21\\
\hline\hline
\end{tabular}
\end{normalsize}
\end{table*}


\begin{figure}
\centering
\resizebox{\hsize}{!}
{\includegraphics[width=\textwidth]{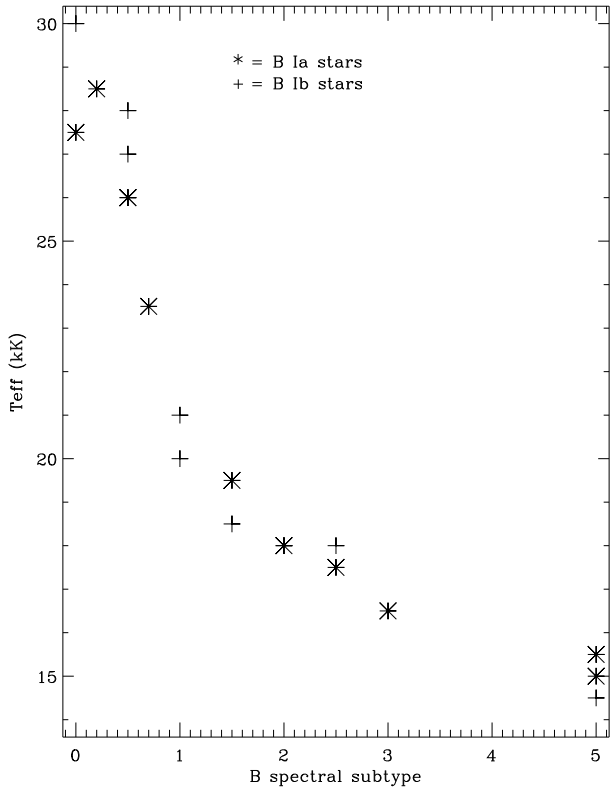}}
\caption{The Galactic B supergiant \teff\ scale as a function of spectral type. B Ia stars are indicated by an asterisk, whilst B Ib stars are marked by a plus sign.}
\label{teffscale}
\end{figure}

\begin{figure}
\centering
\resizebox{\hsize}{!}
{\includegraphics[width=\textwidth]{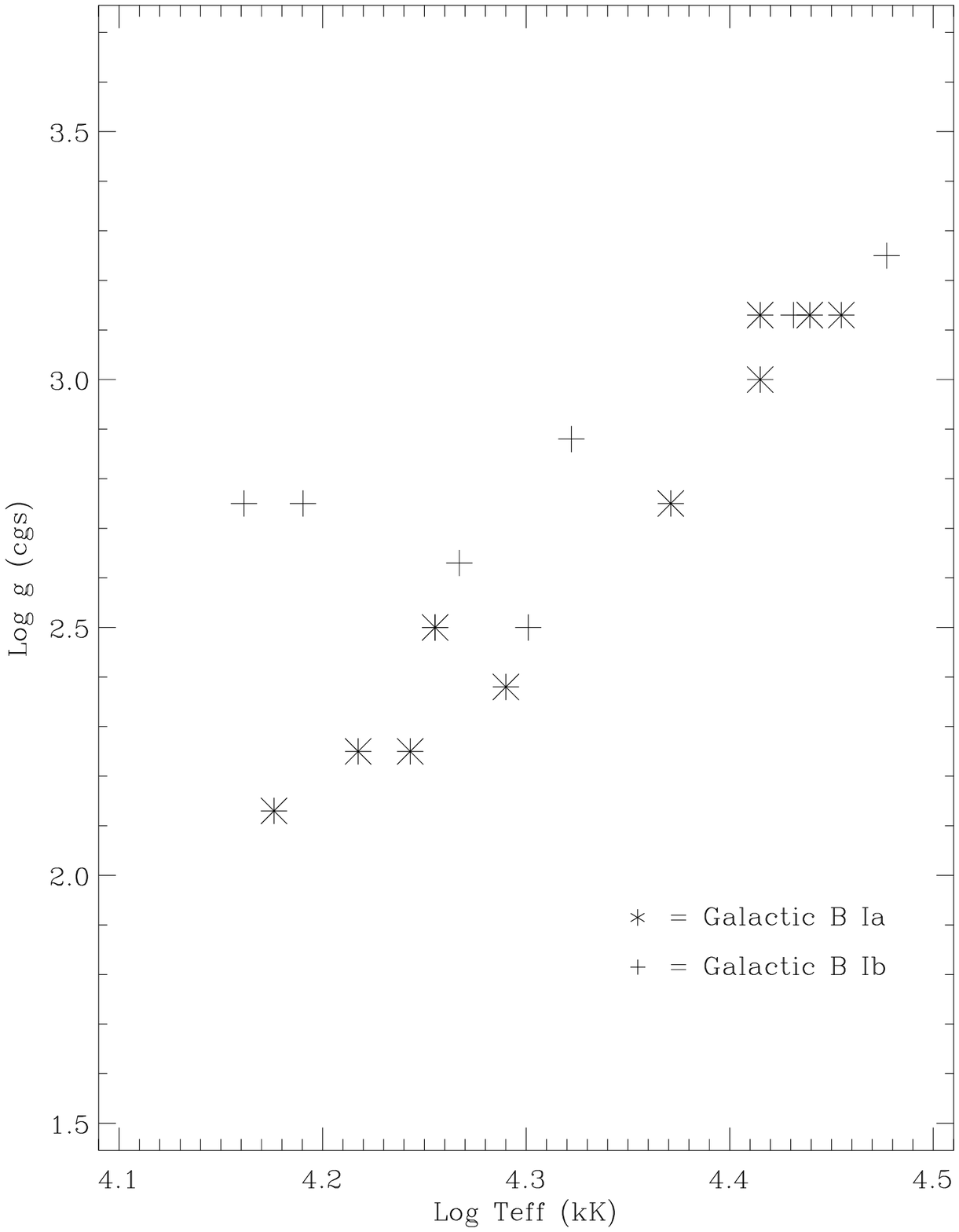}}
\caption{The Galactic B supergiant \teff\ - log $g$ scale.}
\label{tefflogg}
\end{figure}

\noindent
Model fits to the optical spectra of HD 192660 (B0 Ib), HD 213087
(B0.5 Ia), HD 13854 (B1 Iab), HD 193183 (B1.5 Ib), HD 14818 (B2 Ia),
HD 198478 (B2.5 Ia), HD 53138 (B3 Ia) and HD 58350 (B5 Ia) are shown
in Fig. \ref{op1} (4050 -- 4250 \AA), Fig. \ref{op2} (4250 -- 4450 \AA)
and Fig. \ref{op3} (4450 -- 4650 \AA). Overall {\sc CMFGEN} has
succeeded in providing excellent fits to the observed spectrum of each
star. The models succeed in reproducing the H, He, Si and Mg lines
quite accurately. However, some individual spectral lines are more
difficult to model than others. The \ion{Si}{iv} 4089 \AA\ line is
sometimes underestimated in B0~-~B2 supergiants, with the effect being
most pronounced in B1~-~B2 supergiants, which might be partly due to a
blend with \ion{O}{ii}. It is also noticeable that the
model \ion{Si}{iv} line displays a slight sensitivity to mass loss.
\cite{hillier2003} also noted that some model photospheric
  lines can be affected by mass loss and more importantly,
  \cite{dufton2005} noted when using {\sc FASTWIND} that \ion{Si}{iv}
  4116 \AA\ and the \ion{Si}{iii} multiplet 4552, 4568, 4575 \AA\ were
  affected by the stellar wind.  However, whilst using {\sc CMFGEN},
  we have not observed any significant stellar wind effects on the
  \ion{Si}{iii} multiplet. The values of \teff\ derived for these
  stars can still be justified since the model spectrum still fits the
  rest of the spectrum, including the \ion{Si}{iii}, \ion{Mg}{ii} and
  \ion{He}{i} lines, very well. In the cases where the model does
  underestimate the \ion{Si}{iv} 4089 \AA\ line, the use of a model
  with a higher \teff\ that provided a better fit to the \ion{Si}{iv}
  line would provide a worse fit to the rest of the observed spectrum.
  Note that the values of \teff\ obtained in these cases were still
  derived using the silicon ionisation balance and the effect of the
  compromise attained between fitting the \ion{Si}{iv} line
  underestimated by the model and the rest of the spectrum is
  reflected in the value of $\Delta\teff\ $ quoted in Table \ref{results1}. 
It is also intriguing to note that {\sc CMFGEN} predicts two
  absorption lines at 4163 \AA\ (see in B1.5 -- B5 supergiants) and
  4168.5 \AA\ (seen in all B0 -- B5 supergiants) that are not observed
  in any of the sample stars. These predicted lines also appear in the
  {\sc CMFGEN} models of \cite{crowther2006}, where it appears that
  they have identified the line at 4168.5 \AA\ as \ion{He}{i} but no
  explanation is given for the line at 4163 \AA. We can confirm the
  identify of the line at 4168.5 \AA\ and also add that the 
line at 4163 \AA\ is \ion{Fe}{iii}. \\

\begin{table}
\centering
\caption{Values of \teff\ (expressed in terms of 10$^{3}$ K) obtained in this thesis work and from \cite{trundle2004,trundle2005,kudritzki1999,mcerlean1999}. Values marked with an asterisk denote where values from one author have been averaged and are quoted to 1 decimal place. $\dagger$ the B0.5 Ib star HD 64760 has been omitted here because it is a rapid rotator.}
\label{teffsp}
\begin{tabular}{lllllll}
\hline\hline
Sp type & This work & Trundle & McErlean & Crowther \\
\hline
B0 Ia & 27.5 & 27.0* & 28.5 & 27.4* \\ 
B0 Ib & 30.0 & - & - & -  \\
B0.2 Ia & 28.5 & - & 28.5 & - \\
B0.5 Ia & 26.0 & 27.3* & 27.5 & 26.0*  \\
B0.5 Ib & 27.0$\dagger$ & - & 26.5* & -  \\
B0.7 Ia & 23.5 & - & 24.0 & 22.9  \\
B1 Ia & - & 23.8* & - & 22.0  \\
B1 Iab/Ib & 20.5* & - & 23.3* & 21.8*  \\
B1.5 Ia & 19.5 & 21.3* & 21.25 & 18.17  \\
B1.5 Ib & 18.5 & - & 22.3* & -  \\
B2 Ia & 18.0 & 19.0 & 19.83 & 18.6*  \\
B2 Ib & 18.0 & - & 20.8* & -  \\
B2.5 Ia & 17.5 & 16.5 & 18.0 & 16.5  \\
B2.5 Ib & 18.0 & - & 20.5 & -  \\
B3 Ia & 16.5 & 14.0 & 17.9* & 15.8*  \\ 
B4 Iab & - & - & 16.5 & - \\
B5 Ia & 15.0 & 14.5* & 15.4* & -  \\
B5 Ib/II & 15.0* & - & 15.8* & -  \\ 
\hline\hline
\end{tabular}
\end{table}

\noindent
The values of \teff, log $g$, \logllsolar, E(B-V) and \mv\ derived for
each of the 20 B supergiants in the sample are listed in Table
\ref{results1}. These results show that B0~--~B5 supergiants have a
range in \teff\ of 14\,500 -- 30\,000 K, in \logllsolar\ of 4.30 --
5.74 and that their stellar radii vary from about 20 - 71 \rsolar.
They also exhibit a range of $-6.04 \le \mv \le -7.26 $ in brightness,
confirming their status as some of the brightest stars in our Galaxy.
The temperature scale for B supergiants derived here is shown in Fig.
\ref{teffscale}, plotted against spectral type. A drop of up to
10\,000 K in temperature is witnessed between B0~--~B1, whereas at
lower spectral types, the \teff\ scale shows a more gradual decrease
in \teff. The Galactic O star \teff\ scale published by
\cite{repolust2004} ranges from an O2 If star with $\teff =$ 42\,500 K
down to an O9.5 Ia star with $\teff = $ 29\,000 K and an O9.5 Ib star
with \teff $=$ 32\,000 K, meaning that the B supergiant \teff\ scale
presented here carries on smoothly from the Galactic O supergiant
\teff\ scale. Similarly the B supergiant \teff\ scale ends with B5 Ib
stars having $\teff \approx$ 15\,000 K and the Galactic A supergiant
\teff\ scale derived by \cite{venn1995} begins with a \teff\ of 9950
K. A gap between the B and A supergiant \teff\ scales is expected
since none of the recently published B star \teff\ scales include B6-9
stars. The B supergiant temperature scale derived here also
demonstrates the difference in \teff\ between B Ia and B Ib/Iab/II
stars. B0 -- B2 Ib/Iab stars are found to be up to 2\,500 K hotter than
B0 -- B2 Ia stars, with the exception of the stars HD 190603 (B1.5
Ia+) and HD 193183 (B1.5 Ib).
However, a less significant difference of 500 K in \teff\ 
is found between B2 -- B5 Ia and B2 -- B5 Ib/II stars, with the B Ib
stars again being hotter than their more luminous counterparts; this
discrepancy is well within the margins of error in deriving \teff\ as
typically $\Delta$\teff $=$ 500 - 1\,000~K. We have compared our
Galactic B supergiant \teff\ scale to other published values
\citep{trundle2004,trundle2005, mcerlean1999, crowther2006} in Table
\ref{teffsp}. Where each author has several stars with the same
spectral type, the values are averaged and marked with an asterisk in
the table. Note that the results of \cite{mcerlean1999} were obtained
with an unblanketed stellar atmosphere code. If we compare our derived
\teff\ values with those of the unblanketed \cite{mcerlean1999} \teff\ 
scale, the use of a stellar-atmosphere code with a full treatment of
line blanketing has the effect of lowering \teff\ by 1\,000 - 3\,000 K
for Galactic B supergiants. This is not as drastic as the reduction
found for O supergiants, which can be as high as 7\,000 K for extreme
stars \citep{crowther2002}. If we compare our derived \teff's to those
of \cite{mcerlean1999}, with whom we have 10 target stars in common
(HD~37128, HD~38771, HD~2905, HD~13854, HD~193183, HD~14818,
HD~206165, HD~42087 and HD~53138), we find reasonably good agreement
except for B1 Ia/Iabs, where the \cite{mcerlean1999} results imply
that a B1 supergiant is 2500 - 3\,000 K hotter than our values. The
SMC B supergiant temperature scale \citep{trundle2004,trundle2005}
also implies a much hotter B1 supergiant, but it is expected that SMC
stars will be hotter than Galactic stars (see e.g. the O star
temperature scales of \cite{massey2005} (SMC) and \cite{repolust2004}
(Galactic)
where the SMC stars are up to 4\,000 K hotter than the Galactic ones). 

\subsection{Log $g$ estimates}

\begin{figure}
\centering
\resizebox{\hsize}{!}
{\includegraphics[scale=0.6,width=\textwidth]{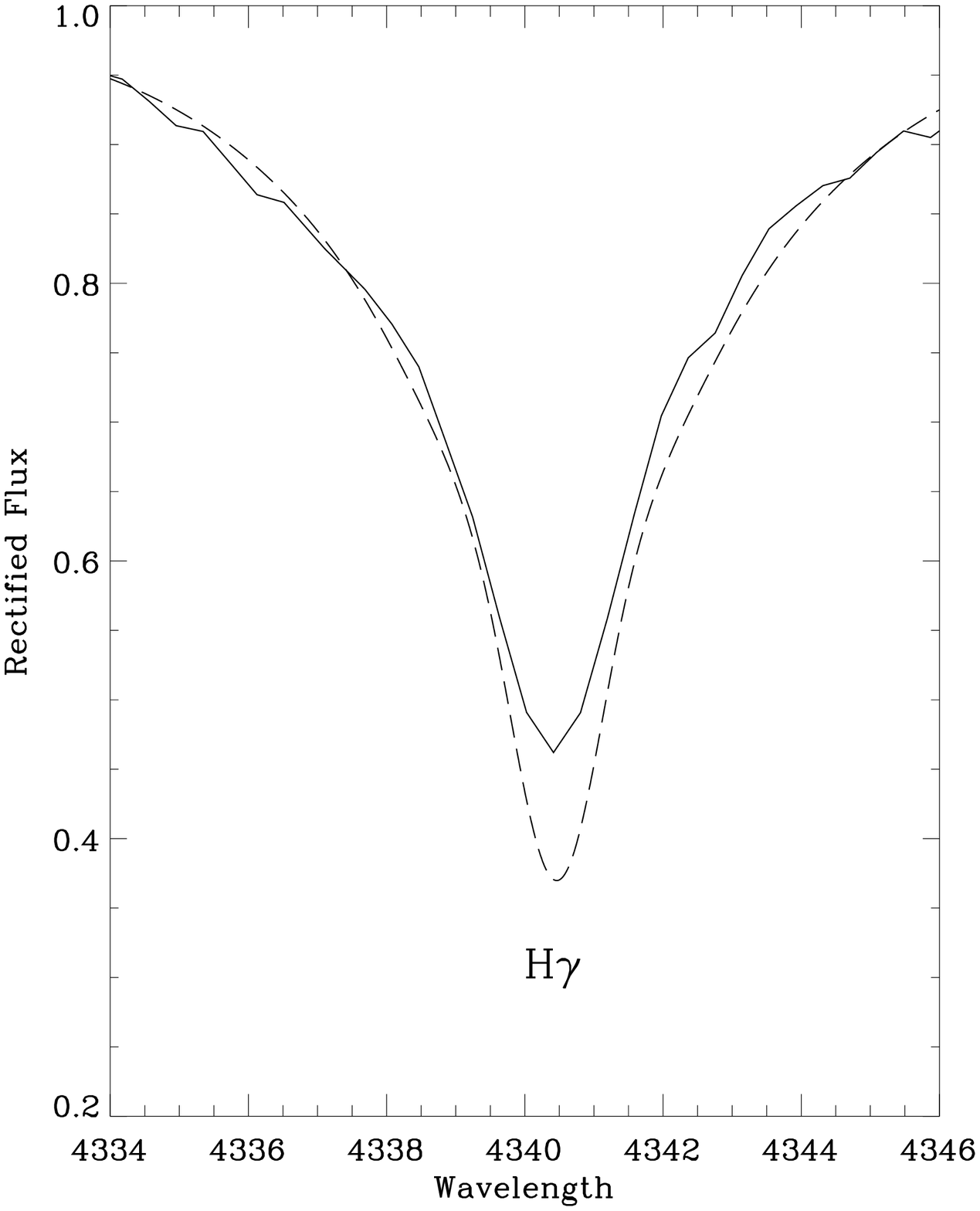}}
\caption{Example of a {\sc TLUSTY} log $g$ fit to \hg\ profile of the B5 Ib/II star HD 164353. A value of log g = 2.75 is used here.}
\label{difflogg}
\end{figure}

\noindent
An example of a {\sc TLUSTY} log $g$ fit to the H$\gamma$ profile of
HD 164353 is shown in Fig. \ref{difflogg}. Fits to the \hg\ 
 and \hd\ profiles of all 20 B supergiants can be found online. {\sc
  TLUSTY} is a purely photospheric code therefore the model Balmer
lines are also photospheric; however in B supergiants the observed
Balmer lines suffer from wind contamination, where hydrogen photons
emitted in the wind at the same wavelengths as H$\gamma$ and H$\delta$
`fill in' the absorption profile. This makes it appear more `shallow'
when compared to a photospheric profile and explains the difference in
depth between the two profiles shown in Fig. \ref{difflogg}.
Values of log $g$ derived from \hg\ and \hd\ were in good
 agreement in general and no discrepancies larger than the margin of
 error on log $g$ were found. Only small discrepancies were found for
B0 -- B0.5 stars where a model with a log $g$ value 0.13 dex higher
than the adopted value might provide a slight better fit to the \hd\ 
profile. However this ambiguity can be attributed to the influence
of a large \ion{N}{iii} blend on the blue wing of \hd\ masking where
the actual wing of the profile should really lie. In these cases a
very good fit is made to \hg\ so the value derived from \hg\ is
taken. Some difficulties were encountered when trying to fit the
\hg\ and \hd\ profiles of HD 192660, HD 64760, HD 190603, HD 13854
\& HD 190066 due to the observed asymmetry of the \hg\ and \hd\ 
profiles. This is particularly evident in the HD 190603, a B1.5 Ia
hypergiant with a strong wind evident from the P Cygni shape of its
\hb\ profile.  The \teff\ -- log $g$ scale derived from this work is
shown in Fig.  \ref{tefflogg}, where higher log $g$ values are found
for B Ib stars.  The log $g$ values derived for B Ia stars are 0.1 -
0.2 dex higher than those obtained by
\cite{kudritzki1999,crowther2006} for a sample of Galactic B
supergiants, whereas the \cite{trundle2004,trundle2005} values for SMC
B supergiants are generally higher than those for
Galactic B supergiants. \\

\subsection{Stellar Masses and the Mass Discrepancy}

\begin{figure}
\centering
\resizebox{\hsize}{!}
{\includegraphics[width=\textwidth]{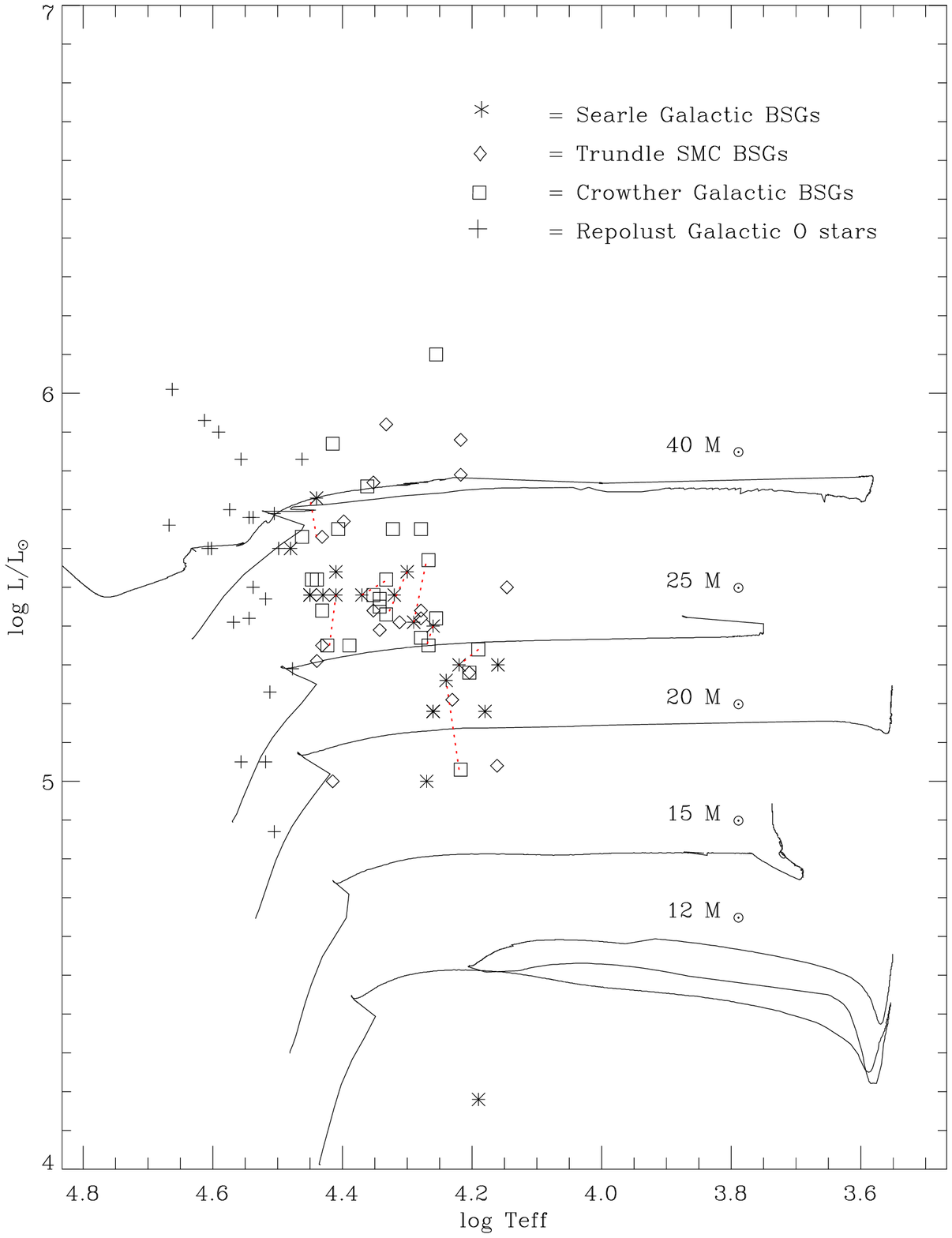}}
\caption{Position of the sample of Galactic B supergiants on the Hertzsprung-Russell diagram, along with other Galactic B supergiants \cite{crowther2006}, SMC B supergiants \cite{trundle2004,trundle2005} and Galactic O stars \cite{repolust2004}. Evolutionary tracks are taken from \cite{meynet2001} and imply 15 \msolar\ $<$ \mevol\ $\le$ 40 \msolar\ for the sample of 20 Galactic B supergiants presented in this work. Galactic B supergiants that are common to both our sample and that of \cite{crowther2006} are joined by a dotted red line to illustrate the effect of different stellar parameters on a star's precise location on the HR diagram. }
\label{obahrd}
\end{figure}

\begin{figure}
\centering
\resizebox{\hsize}{!}
{\includegraphics[width=\textwidth]{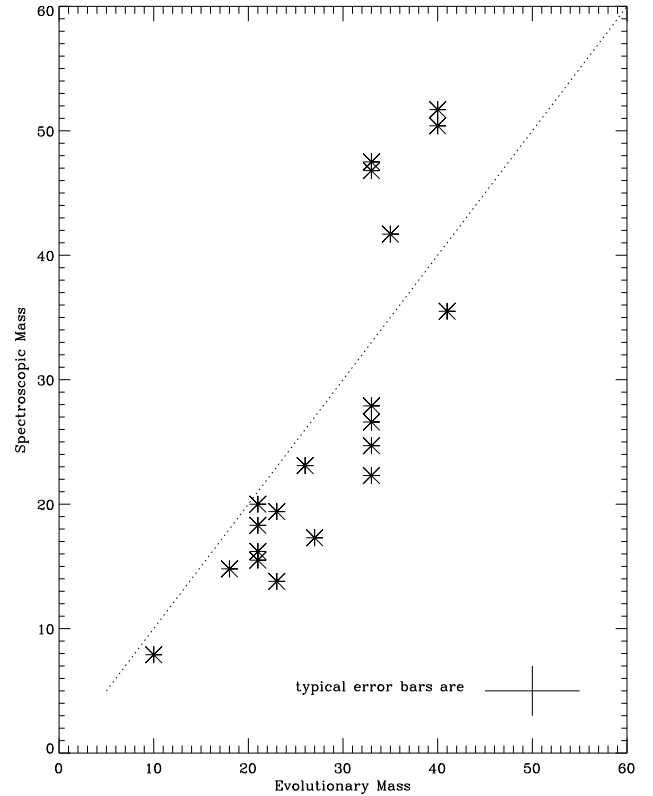}}
\caption{Comparison of evolutionary and spectroscopically-derived stellar masses for the sample of B supergiants. The dotted line indicates 1:1 correspondance.}
\label{mass_cf}
\end{figure}

\begin{figure}
\centering
\resizebox{\hsize}{!}
{\includegraphics[width=\textwidth]{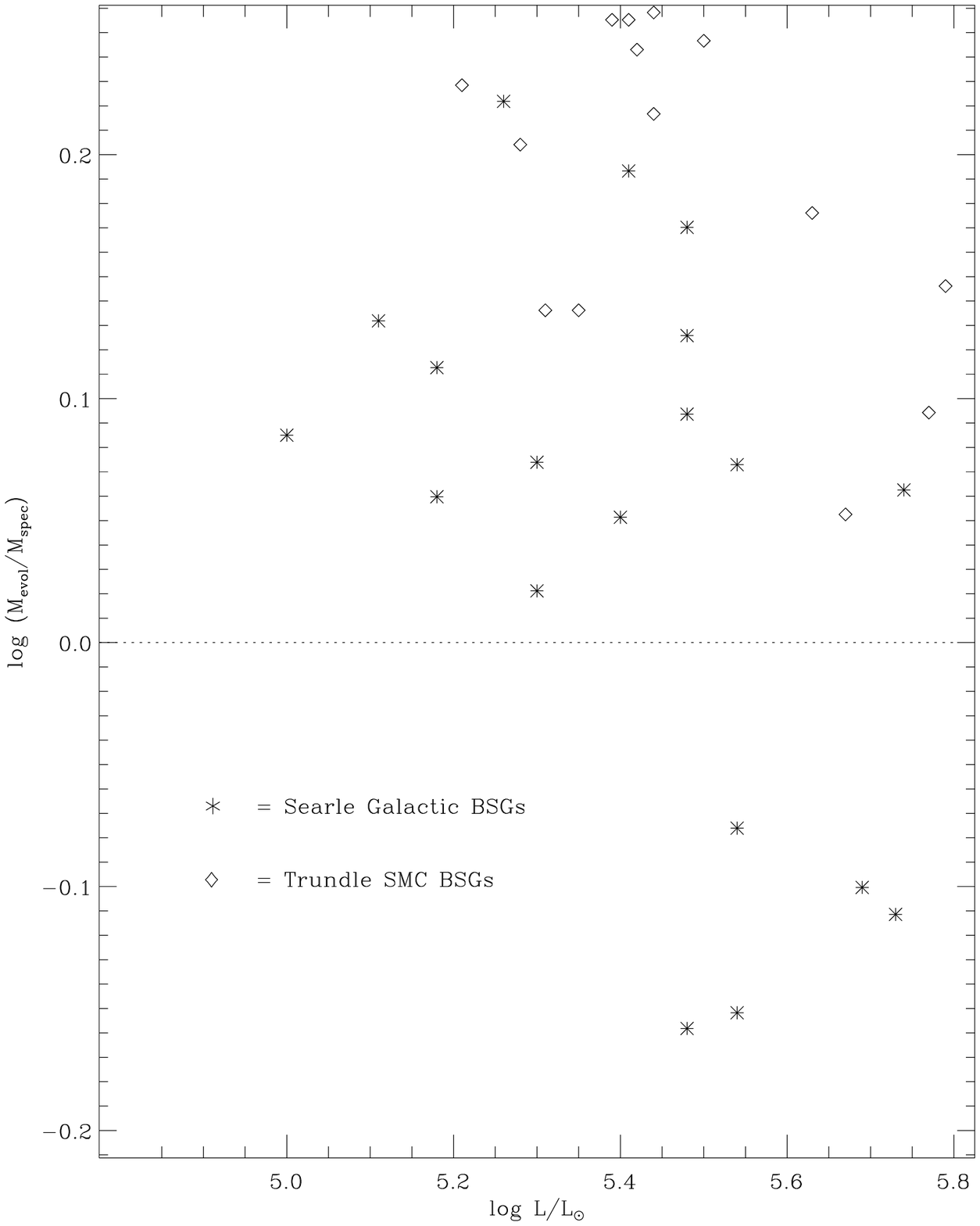}}
\caption{Comparison of $\frac{M_{evol}}{M_{spec}}$ with luminosity (B5 Ib/II stars omitted). Values obtained for the 20 Galactic B supergiants (asterisks) are plotted with those derived for SMC B supergiants (diamonds) \cite{trundle2004,trundle2005}. The dotted line indicates 1:1 correspondance.}
\label{2massratio}
\end{figure}

Using our estimates of log $g$, spectroscopic masses have been derived
for each of the 20 B supergiants and imply a range of $8 \le \mstar
\le 52$. Estimates of the evolutionary mass, \mevol, were then
obtained using our derived stellar parameters and the stellar
evolutionary tracks of \cite{meynet2000}. The positions of our 20
Galactic B supergiants on the Hertzsprung-Russell diagram, along with
other Galactic B supergiants (\citealt{crowther2006}), SMC B
supergiants (\citealt{trundle2004,trundle2005}) and Galactic O stars
(\citealt{repolust2004}), are shown in Fig. \ref{obahrd}. Here, the
\cite{meynet2000} stellar evolutionary tracks have been used, which
include the effects of rotation and are therefore more appropriate for
OB supergiants. In order to demonstrate the effect of different
stellar parameters on a star's precise position on the HR diagram,
Galactic B supergiants common to both our sample and that of
\cite{crowther2006} are joined by a dotted line. A comparison of
both masses is shown in Fig. \ref{mass_cf}. For 14 out of the 20 B
supergiants, \mevol\ $>$ \mspec\ as found by \cite{herrero2002}.
However, for the 5 other stars, which (excluding the rapid rotator HD
64760) have \logllsolar\ $\ge$ 5.54, \mevol\ $<$ \mspec. The
dependence of the mass discrepancy with luminosity is examined further
in Fig.  \ref{2massratio} and compared to the mass discrepancy for SMC
B supergiants investigated by \cite{trundle2004,trundle2005}. Both
data sets exhibit a peak in the mass discrepancy at $ 5.4 \le
\logllsolar\ \le 5.5 $
that drops off quite rapidly. 

\subsection{Calibration of B supergiant fundamental parameters}

\noindent
A calibration of stellar atmosphere parameters (i.e., \teff,
\loglstar, log $g$, \mstar, \rstar) according to spectral type has
been carried out, using the fundamental parameters derived for our
sample and that of \cite{crowther2006}. A linear regression was
applied to the trend of \teff\ with spectral type; once the \teff\
scale had been established, linear regressions were made to the trends
of log \teff\ vs. \loglstar\ and log \teff\ vs.  log $g$, from which
the values of \rstar\ and \mstar\ were then calculated. The resulting
values of \teff, \loglstar, log $g$, \mstar\ and \rstar\ for each
spectral type are shown in Table \ref{spcalib}.

\begin{table}
\begin{center}
\caption{Calibrations of fundamental parameters by spectral type for Galactic B supergiants, based on this work and that of \cite{crowther2006}}
\label{spcalib}
\begin{tabular}{lcccccc}
\hline\hline
Sp. type & \teff & \logllsolar & \rstar (\rsolar) & log $g$ & \mstar (\msolar) \\   
\hline
B0 Ia   & 28.1 & 5.60 & 26.9 & 2.99 & 25 \\ 
B0 Ib   & 29.7 & 5.66 & 23.8 & 3.24 & 37 \\
B0.2 Ia & 26.7 & 5.62 & 30.4 & 3.04 & 36 \\ 
B0.2 Ib & 28.5 & 5.65 & 27.8 & 3.23 & 49 \\
B0.5 Ia & 24.7 & 5.58 & 33.8 & 2.90 & 33 \\
B0.5 Ib & 25.4 & 5.58 & 32.2 & 3.09 & 47 \\
B0.7 Ia & 23.6 & 5.53 & 35.1 & 2.72 & 23 \\
B0.7 Ib & 24.4 & 5.51 & 33.9 & 2.93 & 37 \\
B1 Ia   & 22.0 & 5.44 & 36.5 & 2.41 & 12 \\
B1 Ib   & 21.7 & 5.38 & 34.9 & 2.67 & 22 \\
B1.5 Ia & 19.9 & 5.44 & 44.5 & 2.41 & 18 \\
B1.5 Ib & 19.3 & 5.29 & 39.7 & 2.50 & 19 \\
B2 Ia   & 18.3 & 5.41 & 51.0 & 2.32 & 19 \\
B2 Ib   & 18.1 & 5.27 & 44.4 & 2.46 & 21 \\
B2.5 Ia & 17.2 & 5.39 & 56.5 & 2.24 & 19 \\
B2.5 Ib & 17.6 & 5.25 & 46.2 & 2.43 & 22 \\
B3 Ia   & 16.4 & 5.37 & 60.4 & 2.16 & 19 \\
B3 Ib   & 17.5 & 5.23 & 45.5 & 2.39 & 19 \\
B4 Ia   & 15.8 & 5.34 & 63.5 & 2.06 & 16 \\
B4 Ib   & 17.4 & 5.18 & 43.2 & 2.27 & 13 \\
B5 Ia   & 15.7 & 5.33 & 63.0 & 2.03 & 15 \\
B5 Ib   & 15.2 & 5.09 & 51.7 & 2.11 & 13 \\  
\hline
\end{tabular}
\end{center}
\end{table}

\subsection{Evidence for CNO processing in B supergiants}
\label{cno}

As previously mentioned, it was \cite{walborn1976} who first suggested
that the nitrogen and carbon anomalies found in OB stars can be
explained by their evolutionary status, with OBC stars being the least
evolved. It therefore follows that a typical OB supergiant should
display some partial CNO processing, in the form of nitrogen
enrichment accompanied by CO depletion. Several authors (
\citealt{trundle2004,trundle2005,evans2004b,venn1995}) have
found N enrichments and CO depletions in OBA stars with respect to
solar abundances. All 20 B supergiants in our sample show evidence for
partial CNO processing in their spectra. The details of the CNO
abundances derived for individual stars are given in Table
\ref{results3}. \\


\noindent
The majority of Galactic B supergiants show a modest
nitrogen enrichment, but some stars (HD 37128, HD 192660 and HD
191243) are slightly nitrogen deficient. \cite{walborn1976} observed
that Orion belt stars such as HD 37128 ($\epsilon$ Ori) are nitrogen
deficient due to the weakness of the \ion{N}{iii} 4097 \AA\ and 4640
\AA\ (blend) spectral lines. The largest nitrogen enhancements are
seen for B1 -- B2 stars (HD 13854, HD 190603 and HD 14818). It is of
interest to note that \cite{walborn1976} classed HD 13854 as a
morphologically normal B supergiant (as well as HD 38771), whereas we
have found a modest yet significant N enrichment in this star. \\

\begin{table*}
\centering
\caption{Derived CNO abundances for the sample of Galactic B supergiants (expressed as log $\Big(\frac{N_{x}}{N_{H}}\Big) + 12$). The amount of nitrogen enrichment relative to carbon and oxygen respectively is given in the last two columns, calculated as log $\Big(\frac{N_{x}}{N_{H}}\Big)_{*} -$ log $\Big(\frac{N_{x}}{N_{H}}\Big)_{\odot}$.}
\label{results3}
\begin{tabular}{llllllllll}
\hline\hline
HD no. & Sp. type & \teff\ & {\it v$_{e}$} sin {\it i} & C & N & O & N/C & N/O \\ 
\hline
SUN    & G2 {\sc V} & 5770 & - & 8.39 & 7.78 & 8.66 & -0.61 & -0.88 \\ 
\hline
37128 & B0 Ia     &  27500 &   91  &  7.66 & 7.31 & 8.68 & +0.26 & -0.49 \\
192660& B0 Ib     &  30000 &   94  &  8.02 & 7.51 & 8.73 & +0.10 & -0.40 \\
204172& B0.2 Ia   &  28500 &   87  &  7.66 & 7.71 & 8.66 & +0.66 & -0.07 \\
38771 & B0.5 Ia   &  26000 &   91  &  7.74 & 8.15 & 8.73 & +1.02 & +0.30 \\
185859& B0.5 Ia   &  26000 &   74  &  7.72 & 7.95 & 8.53 & +0.84 & +0.30 \\
213087& B0.5 Ib   &  27000 &   88  &  8.00 & 8.15 & 8.73 & +0.76 & +0.30 \\
64760 & B0.5 Ib   &  28000 &  265  &  7.99 & 8.15 & 8.73 & +0.77 & +0.30 \\
2905  & BC0.7 Ia  &  23500 &   91  &  7.99 & 8.16 & 8.80 & +0.78 & +0.24 \\
13854 & B1 Iab(e) &  20000 &   97  &  7.66 & 8.51 & 8.80 & +1.46 & +0.59 \\
190066& B1 Iab(e) &  21000 &   82  &  7.88 & 8.15 & 8.53 & +0.88 & +0.50 \\
190603& B1.5 Ia+  &  19500 &   79  &  7.48 & 8.76 & 8.73 & +1.89 & +0.91 \\
193183& B1.5 Ib   &  18500 &   68  &  7.66 & 8.15 & 8.73 & +1.10 & +0.30 \\
14818 & B2 Ia     &  18000 &   82  &  7.66 & 8.72 & 8.90 & +1.67 & +0.70 \\
206165& B2 Ib     &  18000 &   73  &  7.96 & 8.15 & 8.43 & +0.80 & +0.60 \\
198478& B2.5 Ia   &  17500 &   61  &  7.86 & 8.29 & 8.45 & +1.04 & +0.72 \\
42087 & B2.5 Ib   &  18000 &   71  &  7.76 & 8.11 & 8.80 & +0.96 & +0.19 \\
53138 & B3 Ia     &  16500 &   58  &  7.78 & 8.32 & 8.60 & +1.15 & +0.60 \\
58350 & B5 Ia     &  15000 &   50  &  7.78 & 8.29 & 8.75 & +1.12 & +0.42 \\
164353& B5 Ib/II  &  15500 &   44  &  7.78 & 7.89 & 8.53 & +0.72 & +0.24 \\
191243& B5 Ib/II  &  14500 &   38  &  7.70 & 7.65 & - & +0.56 & - & \\
\hline\hline
\end{tabular}
\end{table*}

\begin{table*}
\centering
\caption{Comparison of mean published CNO abundances for OBA supergiants (expressed as log $\Big(\frac{N_{x}}{N_{H}}\Big) + 12$).}
\label{meancno}
\begin{tabular}{lllllllll}
\hline\hline
Author & Stellar group & C & N & 0 & N/C & N/0 \\
\hline
This work & Gal BSGs & 7.79 & 8.12 & 8.68 & +0.57 & +0.26 \\
Crowther et al. (2006) & Gal BSGs & 7.87 & 8.33 & 8.47 & +0.68 & +0.59 \\
Trundle et al. (2004-5) & SMC BSGs & 7.27 & 7.71 & 8.13 & +1.38 & 1.04 \\
Evans et al. (2004) & LMC OBSG & 7.49 & 8.55 & 8.02 & +1.28 & +1.26 \\
Evans et al. (2004) & SMC OBSG & 7.30 & 7.94 & 8.01 & +0.86 & +0.66 \\
Venn(1995) & Gal ASGs & 8.14 & 8.05 & - & +0.13 & - \\
Venn(1999) & SMC ASGs & - & 7.33 & 8.14 & - & -0.08 \\
\hline\hline
\end{tabular}
\end{table*}

\noindent
In general, the fits to the CNO diagnostic lines are good. On
 the whole, nitrogen abundances constrained from \ion{N}{ii} 3995
 \AA\ and \ion{N}{iii} 4097 \AA\ are in good agreement; the only
  exceptions being HD 37128 and HD 193183.  It is well-documented in
  the literature that there exists a discrepancy between carbon
  abundances derived from the \ion{C}{ii} 4267.02, 4267.27 \AA\ 
  multiplet line and the \ion{C}{ii} multiplets at 6578 and 6582 \AA,
  due to their strong sensitivity to nLTE effects and the adopted
  stellar parameters (see e.g., \citealt{nieva2006}). A combination of
  high-resolution and high signal-to-noise spectra, along with
  sufficiently- detailed model atoms, are required to attempt to
  resolve this problem. It is unlikely that our data is of a suitable
  resolution and signal-to-noise to attempt to solve this discrepancy,
  but we will nonetheless discuss our findings as appropriate. For our
  sample of B supergiants, the \ion{C}{ii} multiplets at 6578 and 6582
  \AA\ are not prominent for B0 -- B1 supergiants; however for B1 --
  B5 stars the lines are distinguishable. The fits to the \ion{C}{ii}
  4267.02, 4267.27 \AA\ multiplet are very good but {\sc CMFGEN }
  tends to overestimate the \ion{C}{ii} 6578 and 6582 \AA\ multiplets.
  In the case of constraining the nitrogen abundances, in general good
  agreement is found between the abundance implied from the
  \ion{N}{iii} 4097 and \ion{N}{ii} 3995, 4447, 4630 \AA. Some
  exceptions are found for some B0 -- B0.5 supergiants (HD 37128, HD
  204172, HD 38771 HD 192660) where very good fits are made to
  \ion{N}{ii} 3995 and 4447 \AA, but \ion{N}{iii} 4097 and \ion{N}{ii}
  4630 are both underestimated, with the model producing a much weaker
  \ion{N}{ii} 4447 line than observed. Evidently increasing the
  nitrogen abundance would then cause the \ion{N}{ii} 3995 and 4447
  \AA lines to be overestimated.  For HD 185859 and HD 213087, much
  better agreement is found between all four nitrogen diagnostics. \\

\begin{figure}
\centering
\resizebox{\hsize}{!}
{\includegraphics[width=\textwidth]{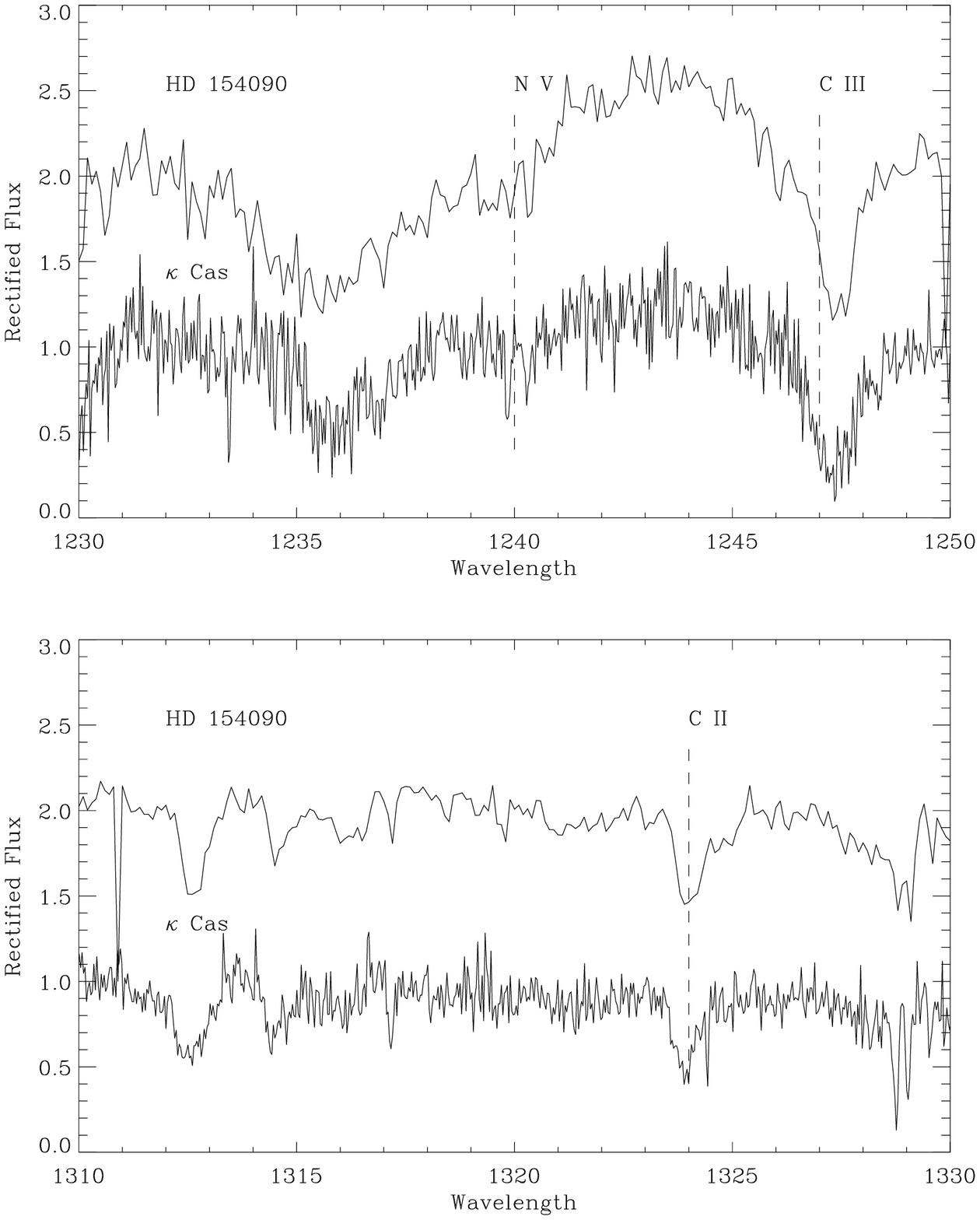}}
\caption{Comparison of the relative strengths of the N and C lines in $\kappa$ Cas and HD 154090}
\label{nc_cf}
\end{figure}

\noindent
There is still an intriguing contradiction that $\kappa$ Cas, which
has been defined as a carbon-rich star (\citealt{walborn1976}) has very
similar CNO abundances to the stars HD 64760, HD 213087 which have not
been noted as carbon rich by any other authors. The original criteria
for classifying $\kappa$ Cas as a carbon-rich star were based on the
weakness of its nitrogen lines as well as the strength of its carbon
lines; this makes sense since (as \citealt{walborn1976} explains) it is
expected that nitrogen deficiency will be accompanied by carbon
enrichment. In order to resolve this discrepancy, the {\it IUE}
spectrum of $\kappa$ Cas has been compared to the {\it IUE} spectrum of
the B0.7 Ia star HD 154090 (see Fig. \ref{nc_cf}). Looking at the
\ion{C}{ii} 1324 \AA\ line, it is certainly no stronger than the same
line in the spectrum of HD 154090. The same is true of the
\ion{C}{iii} line at 1247 \AA.  However, both stars appear to be
nitrogen weak (see e.g., the
\ion{N}{v} wind resonance line around 1240 \AA). Therefore on the
basis of this evidence, it appears that the $\kappa$ Cas should be
defined as a nitrogen weak star, rather than carbon rich.
\cite{crowther2006} found similar results for $\kappa$ Cas, citing
it as having the `least nitrogen enriched abundance' in their sample
as well as the lowest values for the N/C and N/O ratios. \\

\noindent
The CNO abundances derived here for the sample of 20 B supergiants are
compared in Table \ref{meancno} to values obtained by other authors
\citep{trundle2004,trundle2005,evans2004b,crowther2006,venn1995,venn1999}
for OBA supergiants. The results from \citet{trundle2004,trundle2005}
SMC B supergiants have been combined to obtain mean CNO abundances
based on a sample of 18 stars (but only 13 were used for the mean
oxygen abundance since oxygen abundances were not derived for some
B2.5~-~5 stars due to weak, unmeasurable \ion{O}{ii} lines). The data
from \citet{evans2004b} were purely based on CNO abundances derived
from OB supergiants so that the results for nebular and \hii\ regions
included by the authors for comparison were omitted. It is clear from
Table \ref{meancno} that more CNO enrichment occurs in stars belonging
to the Magellanic Clouds than Galactic stars. This is in accordance
with \citet{evans2004b}, who found that OB supergiants in the LMC
display a nitrogen enrichment that is greater than the nitrogen
enrichments in Galactic B supergiants. \citet{evans2004b} conclude
that their sample of Magellanic Cloud stars show significant nitrogen
enrichment due to efficient rotational mixing. The CNO abundances show no
clear trend with effective temperature or {\it v$_{e}$} sin {\it i}. \\

\subsection{Mass loss rates for B supergiants} 
\label{mdot}

\begin{figure}
\centering
\resizebox{\hsize}{!}
{\includegraphics[angle=90,scale=0.25]{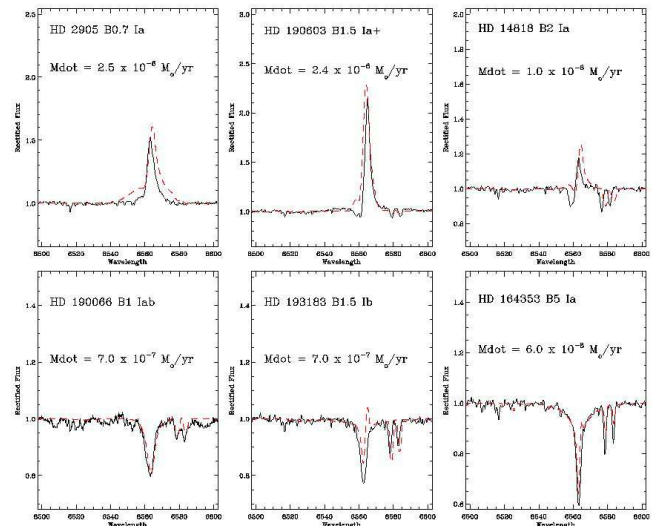}}
\caption{Examples of {\sc CMFGEN} fits to \ha\ profiles of $\kappa$ Cas, HD 190603, HD 14818, HD 190066, HD 193183 and HD 164353, along with the mass loss rate adopted for each star. The red, dotted line represents the {\sc CMFGEN} model fit to each \ha\ profile and the solid, black line indicates the observed \ha\ profile. }\label{eghafits}
\end{figure}

The mass loss rates obtained for this sample of 20 Galactic B
supergiants are based on matches to the \ha\ profile and the resulting
values are listed in Table \ref{results2}. All of these B supergiants
have mass loss rates ranging between $ - 7.22 \le \logmdot \le -
5.30$, except the B5 II/Ib star HD 164353 for which \mdot\ ~$= 6
\times 10^{-8}$ was derived. The errors quoted in Table \ref{results2}
reflect the ambiguity involved in fitting \ha\ `by eye' (and therefore
represent the maximum and minimum values of \mdot\ that fit \ha\ 
reasonably) and are no greater than a factor of 2. In some cases an
upper or lower error limit only is quoted where the model fit over- or
under-estimates the observed \ha\ profile, meaning that a
larger/smaller mass loss rate would not be appropriate. {\sc CMFGEN}
fits to the H$\alpha$ profiles of $\kappa$~Cas, HD~190603, HD~14818,
HD~190066, HD~193183 and HD~164353 are given in Fig.  \ref{eghafits}.
In general, good fits are obtained for each star, but several
difficulties have been encountered in trying to reproduce the observed
\ha\ profiles. It is clear that all the observed profiles are
asymmetric and it is likely that this is caused by the influence of
resonant line scattering that is too weak to produce a `P Cygni'-type
profile so merely results in a slightly asymmetric profile. In some
stars e.g., HD 213087, it appears to be a redward emission component
that partly fills in the profile, to such an extent that in some stars
this red component is visible as a separate emission component (e.g.,
HD 206165) and the H$\alpha$ profile begins to resemble a P Cygni
profile (e.g., HD 14818). In the majority of stars, the peak/trough of
the H$\alpha$ profile has shifted from the line centre as observed in
$\kappa$ Cas. This effect is particularly clear on comparing the
H$\alpha$ profile of $\kappa$ Cas with that of HD 190603, whose peak
is much more central resulting in only a slight asymmetry to the
overall profile. It is also of interest to note that {\sc CMFGEN}
predicts a `bump' in the blueward wing of the H$\alpha$ profile of HD
190603 that is not present in the
observed profile; a similar phenomenon is observed for HD 193183. \\

\noindent
A small, preliminary investigation into the effects of including clumping
on the morphology of the model \ha\ profile was undertaken.
\cite{hillier1999} assume the winds
are clumped with a volume filling factor $f$ and that no inter-clump
medium is present. The volume filling factor, is defined as:

\begin{equation}\label{cleq}
  f = \finf + (1.0 - \finf) e^{-\frac{v}{\vcl}}
\end{equation}

\noindent
where \finf\ is the filling factor at \vinf\ and \vcl\ is the velocity
at which clumping is `switched on' in the wind. However, to carry out
a fair comparison between clumped and homogeneous models involves a
larger parameter space than merely varying \finf\ and \vcl. For
example, it is important to check for consistency in the atmospheric
structure of both models i.e., that they sample the same optical depth
in the photosphere and have the same density structures, which can
include some fine tuning of the velocity law and the point at which
the {\sc TLUSTY} hydrostatic structure joins the {\sc CMFGEN} density
structure. It is also important to check that the same computation
options are selected for both models to ensure that the density
structure is computed using the same methods (e.g, the same number of
$\Lambda$ iterations are specified). This is important as
computational parameters may have been changed for individual models
in order to ease convergence. For these reasons, a rigorous comparison
of
homogeneous and clumped models will be postponed to a later date. \\




\begin{table}
\centering
\caption{Stellar wind parameters (\mdot, $\beta$, \vinf, \vturb) derived for a sample of 20 Galactic B supergiants. The errors given on \mdot\ reflect the errors in fitting each individual \ha\ profile. }
\label{results2}
\renewcommand{\arraystretch}{1.5}
\begin{tabular}{llllllllll}
\hline\hline
HD no. & Sp type & \mdot ($10^{-6}$\msolaryr) & $\beta$ & \vinf(km/s) & \vturb(km/s) \\
\hline
37128 & B0 Ia     & 1.90$^{+0.9}_{-0.0}$ & 1.1 & 1600 & 15 \\
192660& B0 Ib     & 5.00$^{+0.0}_{-3.0}$ & 1.3 & 1850 & 20 \\  
204172& B0.2 Ia   & 0.57$^{+0.7}_{-0.34}$& 1.0 & 1685 & 15 \\
38771 & B0.5 Ia   & 1.20$^{+0.3}_{-0.2}$ & 1.1 & 1390 & 15 \\
185859& B0.5 Ia   & 0.50$^{+0.1}_{-0.1}$ & 1.5 & 1830 & 20 \\
213087& B0.5 Ib   & 0.70$^{+0.4}_{-0.0}$ & 1.5 & 1520 & 20 \\
64760 & B0.5 Ib   & 1.10$^{+1.0}_{-1.0}$ & 1.0 & 1600 & 15 \\
2905  & BC0.7 Ia  & 2.50$^{+0.0}_{-0.5}$ & 1.5 &  850 & 20 \\
13854 & B1 Iab    & 1.50$^{+0.5}_{-0.5}$ & 1.2 &  955 & 10 \\
190066& B1 Iab    & 0.70$^{+0.1}_{-0.1}$ & 1.5 & 1275 & 15 \\
190603& B1.5 Ia+  & 2.40$^{+0.0}_{-0.2}$ & 1.2 &  390 & 15 \\
193183& B1.5 Ib   & 0.23$^{+0.27}_{-0.00}$&1.5 &  565 & 20 \\
14818 & B2 Ia     & 1.00$^{+0.5}_{-0.5}$ & 1.5 &  625 & 15 \\
206165& B2 Ib     & 0.50$^{+0.0}_{-0.2}$ & 1.5 &  640 & 15 \\
198478& B2.5 Ia   & 0.50$^{+0.0}_{-0.3}$ & 1.2 &  550 & 20 \\
42087 & B2.5 Ib   & 0.50$^{+0.0}_{-0.3}$ & 1.2 &  650 & 15 \\
53138 & B3 Ia     & 0.45$^{+0.0}_{-0.3}$ & 1.2 &  500 & 20 \\
58350 & B5 Ia     & 0.70$^{+0.3}_{-0.0}$ & 1.0 &  320 & 20 \\
164353& B5 Ib/II  & 0.06$^{+0.0}_{-0.03}$ & 1.5&  450 & 25 \\
191243& B5 Ib/II  & 0.83$^{+0.0}_{-0.6}$ & 1.0 &  550 & 20 \\
\hline\hline
\end{tabular}
\end{table}

\begin{figure}
\centering
\resizebox{\hsize}{!}{\includegraphics[scale=0.5]{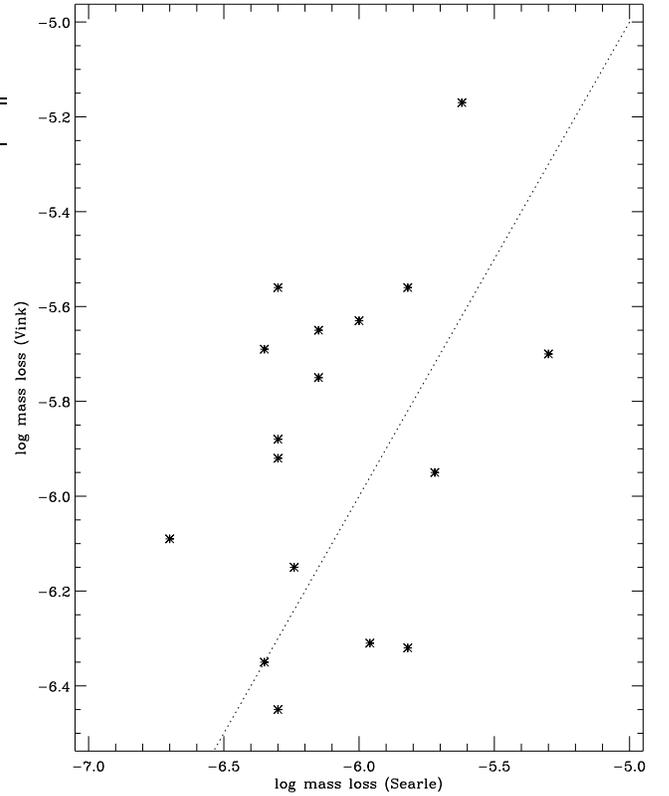}}
\caption{Comparison of CMFGEN derived mass loss rates with theoretical mass loss rates predicted by the \cite{vink2000} mass loss prescription. The dotted line indicates 1:1 correspondance.}
\label{mdotvink}
\end{figure}

\begin{figure}
\centering
\resizebox{\hsize}{!}{\includegraphics[scale=0.5]{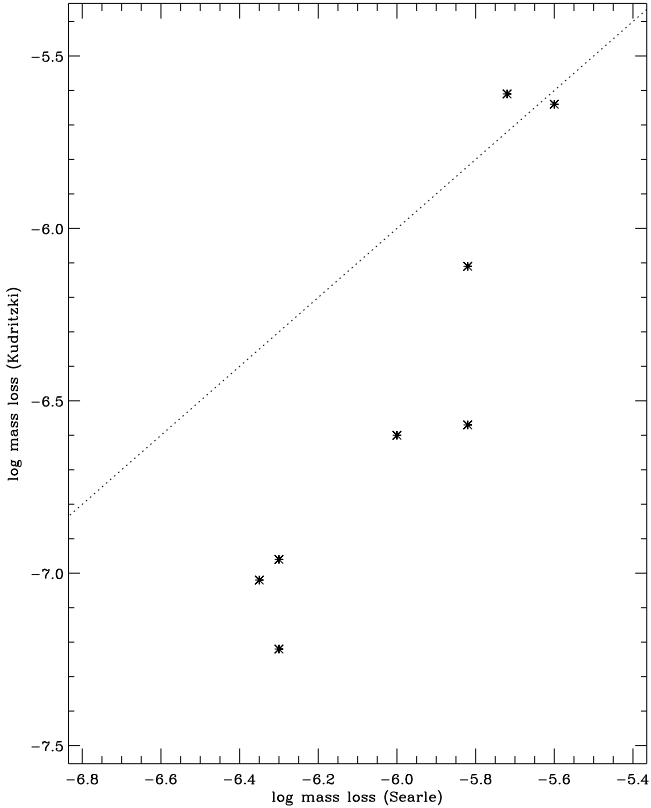}}
\caption{Comparison of CMFGEN derived \mdot\ with those of \cite{kudritzki1999} for 8 stars common to both samples}
\label{mdotkudr}
\end{figure}

\noindent
These mass loss rates have been compared to those predicted by the
theoretical mass loss prescription of \cite{vink2000}, as shown in
Fig.  \ref{mdotvink}, where the values of \teff, \loglstar\ and
\mstar\ derived in the previous section have been input into the
relevant mass loss recipes (equations 12 and 13) quoted in the paper.
We find that the \mdot's derived here are in good agreement with the
\cite{vink2000} predictions, with discrepancies of a factor of 2-3 on
average and the maximum discrepancy a factor of 6.
\cite{trundle2004,trundle2005} found that the values of \mdot$_{vink}$
were a factor of five lower than observed mass loss rates for early B
supergiants, whereas for mid B supergiants \mdot$_{vink}$ was a factor
of seven higher than observed values. No consistent discrepancy is
found in our results but {\em generally} \mdot$_{vink}$ $\le$
\mdothalpha\ for B0 -- B1 supergiants and the reverse is true for B2
-- B3 supergiants. Our values of \mdot\ obtained here were compared to
those of \cite{crowther2006}, with who we have 8 stars in common, and
the mass loss rates are in very good agreement. A comparison has also
been made to the values obtained by \cite{kudritzki1999}, since there
are again 8 stars common to both data sets (Fig. \ref{mdotkudr}). With
the exception of $\epsilon$ Ori, $\kappa$ Cas and HD 206165, all the B
supergiant mass loss rates derived by \cite{kudritzki1999} are smaller
than our values by typically a factor of up to 5. The values derived
for $\epsilon$ Ori and $\kappa$ Cas are well within the errors of our
derived values; however a larger discrepancy of a factor of $\sim$ 10
is found for HD 206165. Initially this is puzzling since in both cases
good fits have been obtained to the observed \ha\ profile of HD 206165
and do not suggest such a large discrepancy in \mdot. However, quite
different stellar parameters have been adopted in terms of \teff\ 
(\teff\ $=$ 18\,000 K in our analysis cf. 20\,000 K from \citealt{kudritzki1999}), 
\logllsolar, \rstar\ and \vinf; more importantly \cite{kudritzki1999} adopt 
a much higher $\beta$ value of 2.5 compared to 1.5 in this work. \\

\subsection{The Wind-Luminosity--Momentum Relation}
\label{wlr}

The concept of a Wind-Luminosity--Momentum Relation (hereafter WLR)
was first proposed by \cite{kudritzki1995}, using the prediction from
the theory of radiatively driven winds that there is a strong
dependence of the total mechanical momentum flow \mdot\vinf\ of the
stellar wind on stellar luminosity \citep[e.g.,][]{cak1975}, which can
be described as

\begin{equation}\label{mom}
  \mdot\vinf \propto \rstar^{-\frac{1}{2}} L^{\frac{1}{\alpha_{eff}}}
\end{equation}

\noindent
The importance of the WLR lies in its potential as an extra-galactic
distance indicator {\em provided that it is reliably calibrated}. The
proportionality shown in Equation \ref{mom} was first confirmed
observationally by \cite{puls1996} for a sample of
Galactic and Magellanic Cloud O stars with $5.5 \le \loglstar\ \le 6.5$.
For $ \loglstar < 5.5 $, a linear fit
was not possible, demonstrating the dependence of the WLR on spectral
type. \cite{kudritzki1999} then showed that a linear fit to the WLR
was also possible for galactic BA supergiants. Since then many authors
(\citealt{repolust2004,
  markova2004,massey2004,massey2005,trundle2004,trundle2005}) have
published values for wind momenta when deriving fundamental parameters
for sets of OBA stars using

\begin{equation}\label{dmom}
  D_{mom} = \mdot\vinf \rstar^{0.5}
\end{equation}

\noindent
where \rstar\ is in solar radii, \mdot\ in g/s and \vinf\ in cm/s
to give $D_{mom}$ in units of cgs. Assuming a WLR of the form

\begin{equation}\label{wlreq}
\mathrm{log} D_{mom} = \mathrm{log} D_{0} + x \logllsolar\
\end{equation}

\noindent
a linear regression can be used to constrain the coefficients $x$ and
log $D_{0}$. The reciprocal of $x$ can be thought of as the {\em
  effective exponent} $\alpha_{eff}$ (see Equation \ref{mom}).
Applying a linear regression to our data gives log $D_{0} = 19.76$ and
$x = 1.61$. Looking at Fig. \ref{dmom}, it can be seen that the hotter
spectral types i.e. O stars have a steeper WLR than cooler B spectral
types.  This is to be expected if, as proposed by \cite{vink1999},
\ion{Fe}{ii} and \ion{Fe}{iii} lines are responsible for driving the
subsonic part of the wind, corresponding to lower ionisation stages.
Fig. \ref{dmom} also illustrates the effect of metallicity on the WLR,
with the more metal poor environment of the Magellanic Cloud hosting
stars with lower values of $D_{mom}$.  This effect is particularly
noticeable between Galactic and SMC B supergiants. Our values of
$D_{mom}$ are compared to those predicted by the theoretical WLR
prescription of \cite{vink2000}, using the parameters derived for our
sample of stars in this chapter. It is found that the observational
values are greater than predicted values for B0~--~B0.7 supergiants
(except for the B0.2 Ia star HD 204172) where \teff\ $\ge$ 23\,000 K,
which is expected for the hotter side of the bi-stability jump.
Conversely, B1~--~B5 supergiants have smaller, observed values of
\dmom\ compared to predicted values.  This is caused by a combination
of several effects. Firstly, just as the predicted \cite{vink2000}
mass loss rates are a factor of 5 larger on the cooler side of the
bi-stability jump, (\teff\ $\sim$ 23\,000 K), the predicted wind
momenta will be greater for B1~--~B5 supergiants, causing a larger
discrepancy between observed and theoretical values of \dmom. In
addition to this, many B1~--~B5 supergiants have \ha\ profiles in
absorption, making it harder to constrain a `true' observed mass loss
rate. Similar results are found by \cite{trundle2005} for their sample
of SMC B supergiants. \cite{repolust2004} suggested that the inclusion
of clumping in the derivation of mass loss rates may help to alleviate
the existing discrepancies between observed and theoretical wind
momenta, which also exist for O stars as well as B stars. Clumping
would reduce the mass loss rate and consequently lower wind momenta,
although the precise amount of clumping present in OB star winds
remains uncertain. Very recently \cite{puls2006} attempted to derive
better constraints on the clumping factor in hot star winds through a
combined optical, infra-red and radio analysis of the wind.  They
found that use of clumped mass loss rates did produce much better
agreement between observed and theoretical wind momenta, since for O
stars the observed wind momenta were originally higher than the
theoretical ones, but the inclusion of clumping reduced the value of
\mdot\ and consequently $D_{mom}$. Unfortunately, for the case of
Galactic B supergiants this may only work for B0 -- B0.7 supergiants;
the use of a lower, clumped \mdot\ would not resolve the discrepancy
for B1 -- B5 supergiants. This may indicate a fundamental difference
between the structure and inhomogeneity of O and B star winds. At
present, it is not possible to obtain a reliable calibration of the
WLR, until the problems associated with its dependence on luminosity
and metallicity, as well
as the effect of clumping on observed mass loss rates, are resolved. \\

\begin{figure}
\centering
\resizebox{\hsize}{!}{\includegraphics{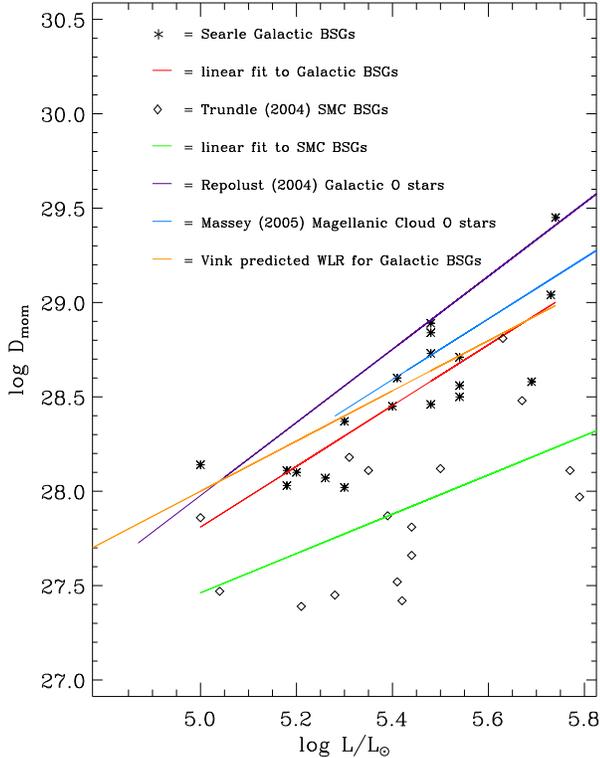}}
\caption{The Wind-Luminosity Momentum Relation for OBA stars. Note the dependence of the WLR on spectral type and metallicity. The theoretical WLR predicted by \cite{vink2000} is calculated for our sample of Galactic B supergiants and is represented by the orange line. }
\label{dmom}
\end{figure}



\section{Testing the UV predictions of CMFGEN}
\label{uv}

\begin{figure}
\centering
\resizebox{\hsize}{!}{\includegraphics{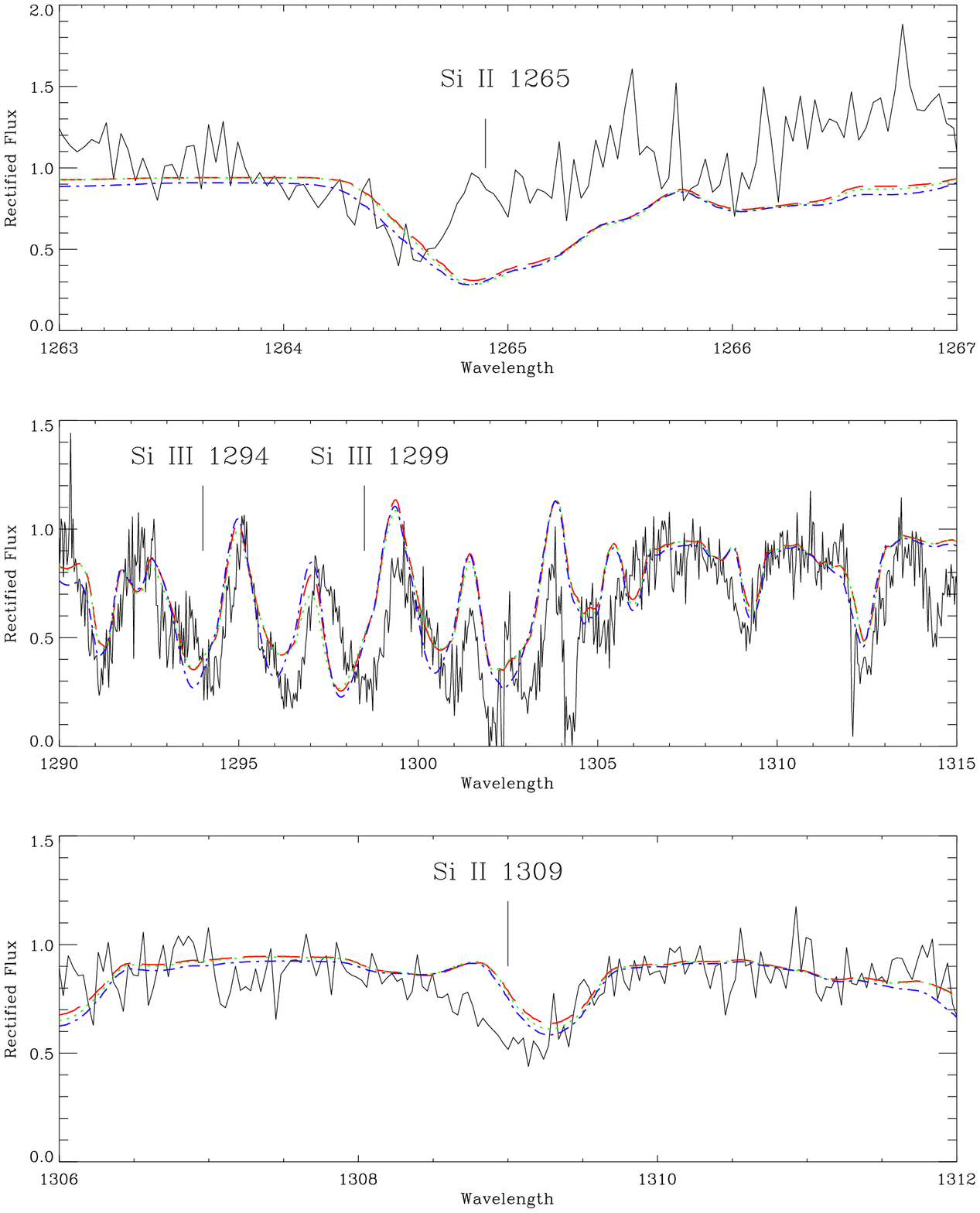}}
\caption[{\sc CMFGEN} model fit to the UV silicon lines of HD 14818 (B2 Ia)]{{\sc CMFGEN} model fit to the {\it IUE} spectrum of HD 14818 (B2 Ia), focusing on the UV silicon \teff\ diagnostics: \ion{Si}{ii} $\lambda$1265 and $\lambda$1309, \ion{Si}{iii} $\lambda$1294, $\lambda$1299 and $\lambda$1417. The `best fit' model derived from the optical has \teff\ $=$ 18\,000 K, L = 2.5 $\times 10^{5}$ \llsolar\ and \mdot\ $= 1.1 \times 10^{-6}$ \msolaryr\ (dashed red line). Others models have \teff\ $=$ 17\,500 K, L = 2.4 $\times 10^{5}$ \llsolar, \mdot\ $=1.2 \times 10^{-6}$ \msolaryr\ (dotted green line) and \teff\ $=$ 18\,500 K, L = 2.5 $\times 10^{5}$ \llsolar, \mdot\ $= 1.8 \times 10^{-6}$ \msolaryr\ (dot-dashed blue line).}
\label{hd14818modeluv}
\end{figure}

The next step of our investigation was to examine if the {\sc CMFGEN}
models presented in the previous section would also provide a good fit
to the UV silicon lines, thus confirming that \teff\ diagnostics at
both optical and UV wavelengths implied the same value of \teff\ for
each star. An example of a {\sc CMFGEN} comparison to \teff-sensitive
silicon lines \ion{Si}{ii} $\lambda$1265, \ion{Si}{iii} $\lambda$1294
and $\lambda$1299 (as noted by \citealt{massa1989}) is given in
Fig.~\ref{hd14818modeluv} for the B2 Ia star HD 14818 and the
corresponding final {\sc CMFGEN} model with \teff\ $=$ 18\,000 K, L =
2.5 $\times 10^{5}$ \llsolar\ and \mdot\ $= 1.1 \times 10^{-6}$
\msolaryr. Two other models with \teff\ $=$ 17\,500~K, L = 2.4 $\times
10^{5}$ \llsolar, \mdot\ $= 1.2 \times 10^{-6}$ \msolaryr\ and \teff\ 
$=$ 18~500 K, L = 2.5 $\times 10^{5}$ \llsolar, \mdot\ $= 1.8 \times
10^{-6}$ \msolaryr\ respectively are also shown to demonstrate the
effects of changing \teff\ on these lines (the slight differences in
the luminosity and mass loss of these models will not significantly
affect the silicon lines). A direct comparison of the observed and
model \ion{Si}{ii} $\lambda$1265 line profiles is difficult since the
continuum is raised about this line, but it is apparent that the model
produces an asymmetric profile (whereas the observed profile is
symmetric) shifted by about 1 \AA\ blue-ward relative to the observed
line profile centre. The model profile is also much broader than the
observed profile and varying \teff\ by $\pm$ 500 K has no significant
effect on this line. In the case of \ion{Si}{ii} $\lambda$1309, the
model line profile is more narrow and shallow than the observed one.
Changes in \teff\ are more apparent on this line, though still make no
significant improvement to the overall line fit. For \ion{Si}{iii}
$\lambda$1294 and $\lambda$1299, the model line profiles are again
asymmetric, unlike the observed profiles, and the blue wings of these
lines are overestimated whilst the absorption troughs are
underestimated. In fact, both observed \ion{Si}{iii} lines appear to
show some evidence of broadening due to the stellar wind despite being
photospheric, which is also evident in the model profiles in the form
of asymmetry.  Additionally change in \teff\ appears to have no affect
on these lines; however the higher value of \mdot\ for the \teff\ $=$
18\,500 K model (blue line) produces a deeper absorption trough for the
profile.  To conclude, varying \teff\ and even \mdot\ has a small
affect on these lines, but will not succeed in reproducing the
observed lines accurately, with the correct
broadness and symmetry. \\

\subsection{The optical/UV discrepancy in wind lines}
\label{opuv}

A large number of UV lines are also strongly affected by mass loss
from the wind, so it is also of interest to investigate whether the
values of \mdot\ derived from \ha\ in \S \ref{mdot} succeed in
reproducing the UV wind resonance lines accurately. This is not the
first time that modelling of hot stars has been extended to the UV and
matching the P Cygni profiles observed there. In the last couple of
years, several authors have begun to consider both optical and UV
stellar properties when deriving fundamental parameters (e.g.,
\citealt{evans2004b,crowther2006}) and \citet{bouret2005} analysed IUE
and FUSE spectra of two Galactic O4 stars with {\sc CMFGEN} and TLUSTY,
presenting one of the first analyses based exclusively on UV
diagnostics that also uses these particular stellar atmosphere codes.
It is evident that an optical analysis provides a much easier way of
obtaining stellar parameters, where diagnostics for e.g. \teff\ and
luminosity are readily available and only depend on abundance, \teff\ 
and/or luminosity. On the other hand, the task of identifying suitable
diagnostic lines is less straightforward, since many UV lines will be
sensitive to mass loss as well as \teff, abundance and
in some cases \vturb.  \\

\begin{figure}
\centering
\resizebox{\hsize}{!}{\includegraphics[angle=90]{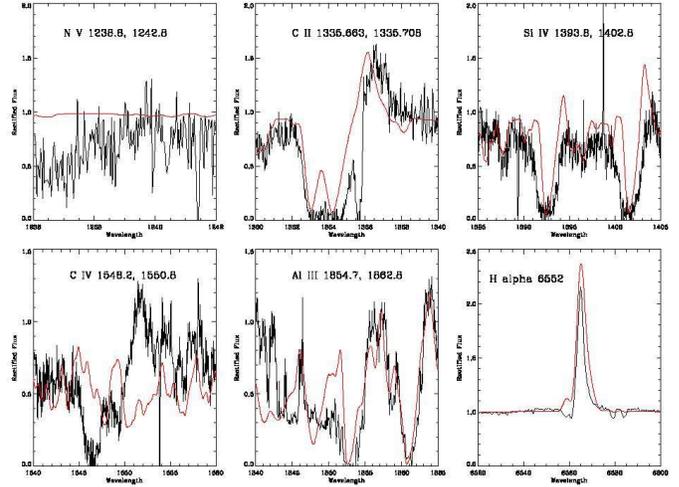}}
\caption[{\sc CMFGEN} UV fit to the IUE spectrum of HD 190603 (B1.5 Ia+)]{{\sc CMFGEN} fit to the IUE spectrum to the \ion{N}{v}, \ion{C}{iv}, \ion{Si}{iv}, \ion{Al}{iii} and \ion{C}{ii} wind resonance lines of HD 190603 (B1.5 Ia+). Note that a good fit to \ha\ does not guarantee the same mass loss rate will provide a good fit to the UV P Cygni profiles. Model parameters are \teff\ = 19\,500 K, \logllsolar\ = 5.41 and \mdot\ = 2.4 $\times 10^{-6}$ \msolar. }
\label{uvfit1}
\end{figure}

\noindent
An example of a {\sc CMFGEN} fit to the IUE spectrum of HD~190603
(B1.5 Ia+) is shown in Fig.~\ref{uvfit1}. Since the mass loss rate has
already been constrained from fits to the \ha\ profile, it is
interesting to see whether the derived value of \mdot\ is confirmed by
reasonable fits to the UV P Cygni profiles, provided that a reasonable
model fit to the \ha\ profile has already been achieved. Looking at
the case of HD~190603 shown in Fig.~\ref{uvfit1}, the fit to the
observed \ha\ profile is good. However, it is clear that {\sc CMFGEN}
does not reproduce any of the observed P~Cygni profiles accurately,
implying that a different value of \mdot\ would be appropriate for the
UV. The model fails to produce sufficient high velocity absorption in
the UV wind resonance lines, to the extent that the predicted
\ion{C}{iv}~$\lambda\lambda$~1548.2,~1550.8 line is only present as a
photospheric line with no evidence of wind contamination. \ion{N}{v}
is not seen as a P~Cygni profile in this star, but the model does not
even produce a distinct, weak photospheric line at 1238 \AA. However,
better fits are achieved at lower ionisation:
\ion{C}{ii}~$\lambda\lambda$~1335.66,~1335.71;
\ion{Si}{iv}~$\lambda\lambda$~1393.8,~1402.8 and
\ion{Al}{iii}~$\lambda\lambda$~1854.7,~1862.8. The observed
\ion{C}{ii}~$\lambda\lambda$~1335.66,~1335.71 line is saturated but
the model produces an unsaturated line, which suggests that either a
model with a higher value of \mdot\ is required or the model
ionisation is incorrect. Adopting a higher value for \mdot\ though
would worsen the effect of the model overestimating the red wings of
the \ion{Si}{iv}~$\lambda\lambda$~1393.8,~1402.8 doublet. It would
have a more positive effect on the
\ion{Al}{iii}~$\lambda\lambda$~1854.7,~1862.8 line, since the observed
blueward doublet is beginning to saturate but the model blueward
doublet is clearly unsaturated, again supporting a higher mass loss
rate. The {\sc CMFGEN} fit to \ha\ would worsen if a higher value of
\mdot\ was adopted, illustrating the discrepancy between the mass loss
rates implied from the optical and UV. It is also noticeable when
comparing the observed and model \ion{Si}{iv} P Cygni profiles that
the model doublet components are narrower than observed. As in the
case of \ion{C}{iv}, this is due to the model predicting to little
absorption at high velocities. For HD 190603, these problems arise in
spite of the fact
that the value adopted for \mdot\ provides a good fit to the \ha\ profile. \\

\begin{figure}
\centering
\resizebox{\hsize}{!}{\includegraphics[angle=90]{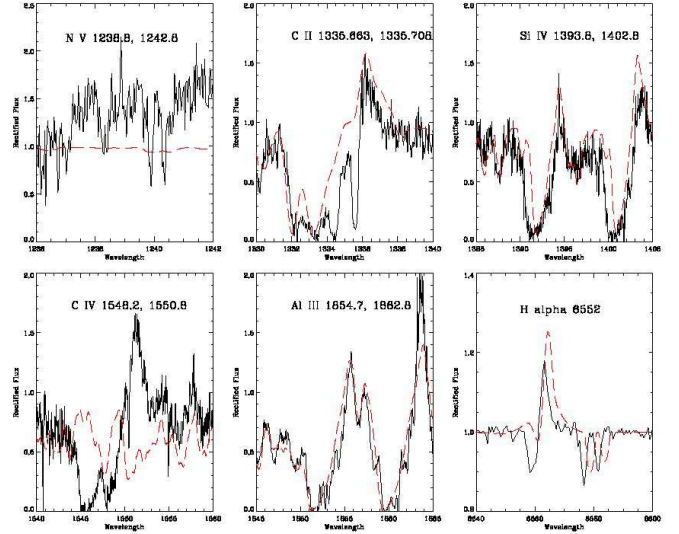}}
\caption[{\sc CMFGEN} UV fit to the IUE spectrum of HD 14818 (B2 Ia)]{{\sc CMFGEN} fit to the IUE spectrum to the \ion{N}{v}, \ion{C}{iv}, \ion{Si}{iv}, \ion{Al}{iii} and \ion{C}{ii} wind resonance lines of HD 14818 (B2 Ia). Even though the fit to \ha\ is not perfect, a reasonable fit is made to the UV P Cygni profiles, particularly \ion{Si}{iv} and \ion{Al}{iii}. Model parameters are \teff\ = 18\,000 K, \logllsolar\ = 5.40 and \mdot\ = 1.0 $\times 10^{-6}$ \msolar. }
\label{uvfit2}
\end{figure}

\begin{figure}
\centering
\resizebox{\hsize}{!}{\includegraphics[angle=90]{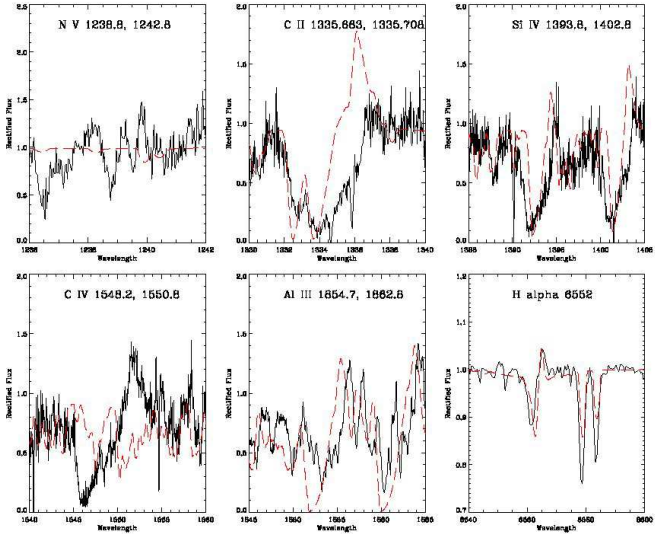}}
\caption[{\sc CMFGEN} UV fit to the IUE spectrum of HD 53138 (B3 Ia)]{{\sc CMFGEN} fit to the IUE spectrum to the \ion{N}{v}, \ion{C}{iv}, \ion{Si}{iv}, \ion{Al}{iii} and \ion{C}{ii} wind resonance lines of HD 53138 (B3 Ia). Although a good fit has been made to \ha\ with the adopted mass loss rate, {\sc CMFGEN} does not reproduce the observed UV P Cygni profiles well. Model parameters are \teff\ = 16\,500 K, \logllsolar\ = 5.30 and \mdot\ = 4.5 $\times 10^{-7}$ \msolar. }
\label{uvfit3}
\end{figure}

\noindent
An example of a better {\sc CMFGEN} fit to the UV wind resonance lines
is given in Fig. \ref{uvfit2} for the B2 Ia star HD 14818. The
observed \ha\ profile displays a P Cygni profile, which has not been
successfully reproduced by the model (as discussed in \S
\ref{mdot}). Despite this, very good fits have been obtained to
\ion{Si}{iv} and \ion{Al}{iii} in comparison to those obtained for HD
190603, though again a lack of high velocity absorption causes the
model to under-estimate the broadness of the absorption trough for
\ion{Si}{iv}. However, the same failure occurs in reproducing the P Cygni
profile of \ion{C}{iv} line, whilst \ion{N}{v} shows no evidence of wind contamination. The fit to \ion{C}{ii} is
reasonable, although the model predicts too much redward emission and
as a result does not match the redward side of the absorption trough.
Conversely an example of a worse fit than either of the previous cases
is shown in Fig. \ref{uvfit3} for HD 53138 (B3 Ia). Its observed \ha\ 
profile is in absorption but shows a small amount of red-ward emission
and is reasonably well matched by {\sc CMFGEN}. On the other hand, the
UV P Cygni profiles are in general poorly matched by the model, with
none of the five wind line profiles being well reproduced. The same
problems seen for HD 190603 and HD 14818 in matching \ion{N}{v},
\ion{C}{iv} and \ion{Si}{iv} also occur here. The red-ward emission
in \ion{C}{ii} is grossly over-estimated and the model produces an
asymmetric \ion{Al}{iii} profile that is not observed. In both cases
the model predicts saturated lines when the observed profiles are not
saturated (though \ion{C}{ii} is beginning to saturate a little). The
high velocity absorption in \ion{Al}{iii} is over-estimated to the
extent that it predicts saturation to occur at a higher velocity than
observed. It is therefore clear from Fig. \ref{uvfit1}, Fig.
\ref{uvfit2} and Fig. \ref{uvfit3} that a discrepancy exists
between the value of \mdot\ required to fit the \ha\ and UV wind
resonance lines (hereafter referred to as the {\em optical/UV discrepancy}). \\

\noindent
In general, {\sc CMFGEN} only succeeds in matching the \ion{C}{iv}
line when it is saturated in early B supergiants, at which point it is
no longer sensitive to \teff\ and \mdot\ so a reliable fit cannot be
obtained as altering these parameters will have no affect on the model
line profile. Otherwise, {\sc CMFGEN} manages to reproduce most of
the observed P Cygni profile for \ion{C}{ii}, \ion{Al}{iii} and
\ion{Si}{iv}, but fails to produce enough high velocity absorption to
reproduce the full extent of the observed absorption trough. As a
result, the model often under-estimates the blueward absorption as
well as over-estimating the redward emission, especially in the case
of \ion{Si}{iv}. This can sometimes lead to the model giving an
asymmetry to the P Cygni profile that is certainly not observed in the
spectrum. Additionally, {\sc CMFGEN} never succeeds in producing the
\ion{N}{v} P Cygni profile when present in B0 -- B1 supergiants and
even when a weak, photospheric profile is observed, the model fails to
produce a discernible spectral line at the correct wavelength for
\ion{N}{v}. In the hotter B supergiants, the model grossly
underestimates the photospheric \ion{Al}{iii} and \ion{C}{ii} lines.
However when the same resonance lines are seen as P Cygni profiles,
the model has a tendency to reproduce them as saturated when they are
observed to be unsaturated. All these discrepancies suggest that the
problem lies within the predicted ionisation structure of the models.
{\sc CMFGEN} fits to the overall IUE spectra of 10 B0 -- B5
supergiants are available as online material (Fig.s \ref{uv1},
\ref{uv2}, \ref{uv3}). Very similar problems in matching the UV P
Cygni profiles have also been encountered by \cite{evans2004b} and
\cite{crowther2006} when modelling O and early B supergiants with {\sc
  CMFGEN}.

\subsection{Modelling the UV exclusively}
\label{modeluv}

The {\sc CMFGEN} models examined in the last section demonstrate a
clear discrepancy between \mdothalpha\ and the value of \mdot\ implied
by the P Cygni profiles of the wind resonance lines. It is hardly
surprising that they are unsuccessful in reproducing the observed UV
wind diagnostics accurately. In this section, the possibility of
modelling a star solely from its UV spectra will be investigated
(ignoring any prior knowledge of values of parameters from the
optical) to see if the UV can be reproduced more accurately.
In order to do this, we must first identify suitable UV
 diagnostic lines by which values of \teff, \logllsolar, \mdot, \vinf, $\beta$ and abundances could be constrained. \\

\noindent
Looking back to the problems mentioned in the previous section, one
potential difficulty is immediately apparent. {\sc CMFGEN} is unable
to reproduce the \ion{C}{iv} line accurately, which makes it hard to
constrain \vinf\ and $\beta$ from this line. Suitable UV \teff\ 
diagnostics also need to be found besides the photospheric
\ion{Si}{ii} and \ion{Si}{iii} lines discussed in \S \ref{uv}.
\ion{Si}{iv}~$\lambda\lambda$~1393.8,~1402.8 could be a good candidate
but it is also very sensitive to luminosity and mass loss; moreover it
is often saturated, reducing its sensitivity to both parameters, and
{\sc CMFGEN} rarely reproduces it accurately. Other potential \teff\ 
diagnostics are \ion{Al}{iii} and \ion{C}{ii} which also show some
sensitivity to mass loss and are therefore not ideal. Another possible
\teff\ diagnostic is the photospheric \ion{Si}{ii}~1526.7,~1533.4~\AA\ 
line, but {\sc CMFGEN} does not model these lines well either, often
completely failing to reproduce the blue-ward part of the doublet.
More importantly, the 1533.4~\AA\ doublet becomes blended with
\ion{C}{iv}~$\lambda\lambda$~1548.2,~1550.8~\AA\ at high value of
\vinf.  At this stage, we have no photospheric lines to use as 
reliable \teff\ diagnostics, since they are not well matched by {\sc CMFGEN}.
The best we can do is look at the UV lines
best reproduced by {\sc CMFGEN} (i.e., \ion{Si}{iv}, \ion{Al}{iii} and
\ion{C}{ii}) and analyse their sensitivity to the main stellar parameters.  \\

\begin{figure}
\centering
\resizebox{\hsize}{!}{\includegraphics{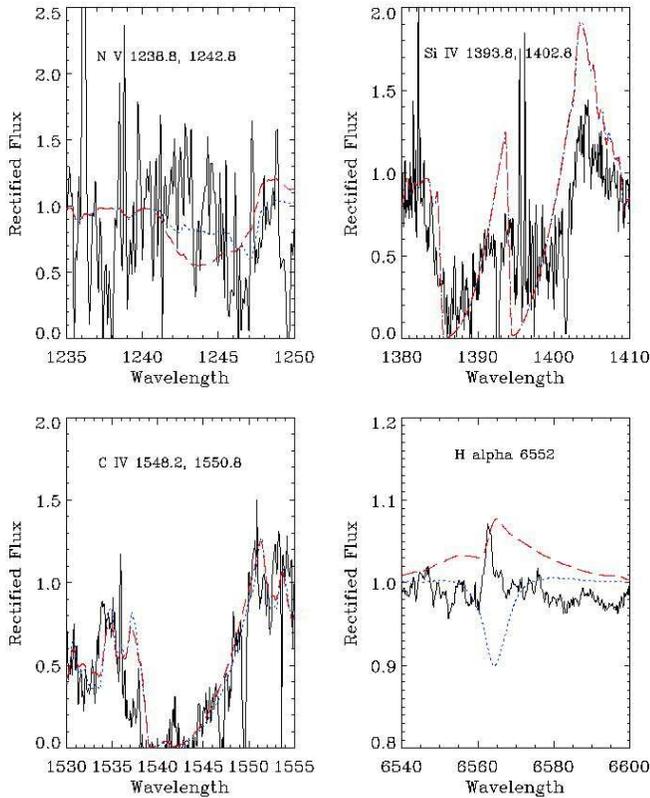}}
\caption[Comparison of UV wind resonance lines of HD 192660 for models with different mass loss rates]{Comparison of UV wind resonance lines of HD 192660 for models with \mdot\ $=$ $5.0 \times 10^{-6}$ \msolar\ (red dashed line) and $2.6 \times 10^{-6}$ \msolar\ (blue dotted line) for HD 192660 (B0 Ib).}
\label{hd192660_diffMdot}
\end{figure}

\begin{figure}
\centering
\resizebox{\hsize}{!}{\includegraphics{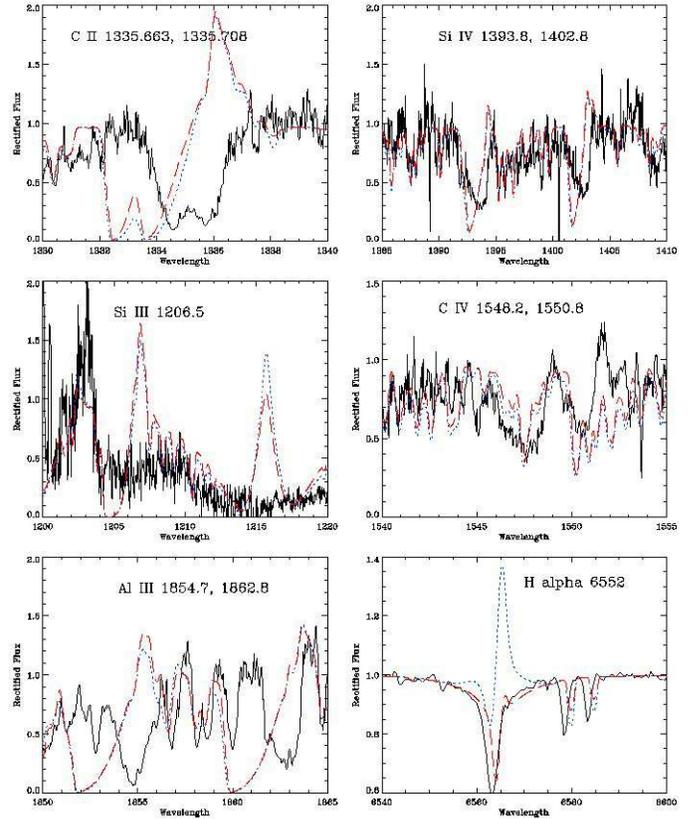}}
\caption[Comparison of UV wind resonance lines of HD 164353 for models with different mass loss rates]{Comparison of UV wind resonance lines of HD 164353 for models with \mdot\ $= 6 \times 10^{-8}$ \msolaryr\ (red dashed line) and $= 1.8 \times 10^{-7}$ \msolaryr\ (blue dotted line) respectively}
\label{hd164353_diffMdot}
\end{figure}

\noindent
In practice, another major problem materialises. It is difficult to
disentangle the effects of \teff\ and \mdot\ on \ion{Si}{iv},
\ion{Al}{iii} and \ion{C}{ii}, plus they are often too saturated to be
sensitive enough to these parameters. When \ion{Si}{iv} is not
observed to be saturated, {\sc CMFGEN} still predicts a saturated
profile that is virtually insensitive to \teff\ and \mdot, making it
difficult to use as a \teff\ and \mdot\ diagnostic. In fact,
 the lack of a significant difference between model P Cygni profiles
 when varying mass loss presents a serious obstacle to any attempt to
 derive parameters from the UV, as we will now show. For B0-B1
supergiants, the model often produces a saturated \ion{C}{iv} P Cygni
profile and over-estimates the \ion{Si}{iv} P Cygni profile. It may
appear logical that adopting a model with a lower mass loss rate would
provide a better fit to the observed \ion{C}{iv} and \ion{Si}{iv}
lines. However, the lack of sensitivity of this line to mass loss
becomes apparent when the \mdot\ adopted by the model is altered. This
is illustrated in Fig.  \ref{hd192660_diffMdot}, where it can be seen
that lowering the value of \mdot\ from $5.0 \times 10^{-6}$ to $2.6
\times 10^{-6}$ has no affect on the wind resonance lines (implying
that they are still optically thick), despite producing model \ha\ 
profiles in emission and absorption respectively (note that the broad
feature seen in the model between 1242 - 1247 \AA\ is {\em not}
\ion{N}{v} but \ion{C}{iii}, which interestingly enough {\em does}
show some sensitivity to mass loss). It could still be argued that a
larger decrease in mass loss is required to fit these lines. However
Fig.  \ref{hd164353_diffMdot} disproves this idea as yet again no
difference is seen between P Cygni profiles for models with \mdot\ $=6
\times 10^{-8}$ \msolaryr\ (red dashed line) and $= 1.8 \times
10^{-7}$ \msolaryr\ (blue dotted line) respectively. This is in spite
of the fact that this difference in mass loss again results in model
\ha\ profiles in emission and absorption, as well as having a
significant difference on the amplitude of Ly $\alpha$ (1216 \AA). HD
164353 presents an interesting case study for how {\sc CMFGEN} deals
with the ionisation in the stellar wind, as it is a B5 Ib/II star that
possesses a very weak wind with $ \mdot\ = 6 \times
10^{-8}$~\msolaryr\ and can be thought of as a star with negligible
mass loss and stellar wind contamination. This is confirmed by looking
at the observed \ion{C}{ii} and \ion{Al}{iii} resonance lines (Fig.
\ref{hd164353_diffMdot}), which show some asymmetric broadening.
However the model predicts strongly saturated profiles for both lines
despite the low mass loss rate adopted for the model, again suggesting
that the predicted ionisation structure is at fault. This
 highlights another significant problem that, given their lack of
 sensitivity to significant changes in mass loss, the \ion{C}{iv} and
  \ion{Si}{iv} P Cygni profiles would not make suitable mass loss
  diagnostics. It appears that the root of the problem lies in {\sc
  CMFGEN} predicting ionisation fractions for \ion{C}{ii} and
\ion{Al}{iii} that are too high, resulting in a large optical depth
that produces too many absorbers at too high a velocity. The observed
profiles on the other hand show us that absorption is only occurring
around the rest velocity of the line. The model over-estimation of
\ion{C}{ii} and \ion{Al}{iii} may therefore only be resolvable by
lowering the ionisation fraction of these two elements and cannot be
resolved by altering the mass loss rate of the model in question.
From this, we conclude that the UV wind resonance lines are
 not suitable candidates for deriving \teff\ and \mdot. Even if the
 ionisation structure was correctly predicted, more diagnostic lines
 would be required to determine all the necessary stellar parameters
 other than \teff\ and \mdot, as well as ensuring
 that an accurate analysis had been carried out.  \\

\noindent
In addition to the wind resonance lines, the UV subordinate lines can
potentially be used to provide additional constraints on the mass loss
adopted for the model. An example is the \ion{Si}{iv}~1122,~1128~\AA\ 
line in the FUV, whose upper energy level is coincident with the lower
energy level of \ion{Si}{iv}~1400~\AA. This means that if the model
over-populates the lower level of \ion{Si}{iv}~1400~\AA, the upper
level of \ion{Si}{iv}~1122,~1128~\AA\ will also be over-populated,
pushing the line into emission when it is observed to be in
absorption. If this predicted line is seen to be in emission in a
model when the observed line is in absorption, this is a direct
indication that the adopted mass loss rate of the model is too high.
There are no examples of \ion{Si}{iv}~1128~\AA\ being in emission in
the models used for the sample of 20 Galactic B supergiants, so this
would suggest that the adopted mass loss rates are within reason. 
This also means that we could not have used
this line as a mass loss diagnostic in this analysis. \\

\begin{figure}
\centering
\resizebox{\hsize}{!}{\includegraphics{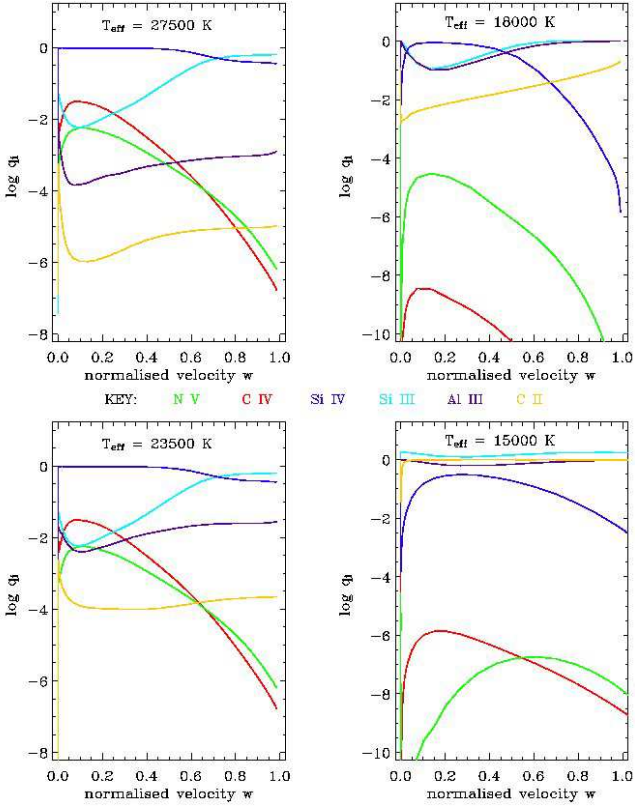}}
\caption[CMFGEN predicted ionisation structure at different \teff]{CMFGEN predicted ionisation structure at different \teff, plotted against normalised veocity $w$. The ions are colour-coded as follows: \ion{N}{v} $=$ green, \ion{C}{iv} $=$ red, \ion{Si}{iv} $=$ dark blue, \ion{Si}{iii} $=$ light blue, \ion{Al}{iii} $=$ purple and \ion{C}{ii} $=$ yellow. \ion{Si}{iv} is predicted to be dominant for B0-B2 supergiants ($30\,000 K \le \teff \le 18\,000 K$), after which \ion{Si}{iii}, \ion{Al}{iii} and \ion{C}{ii} take over as the dominant ions in the wind for \teff\ $\le$ 18\,000 K. }
\label{qvteff}
\end{figure}

\noindent
It is possible to provide alternative perspectives on the UV behaviour
by comparing the ionisation stages present in any given star, rather
than trying to reproduce individual line profiles. Fig. \ref{qvteff}
shows the predicted ionisation structure against $w$ at four different
\teff; 27\,500 K, 23\,500 K, 18\,000 K and 15\,000 K for the six ions
(\ion{N}{v}, \ion{C}{iv}, \ion{Si}{iv}, \ion{Si}{iii}, \ion{Al}{iii}
and \ion{C}{ii}). {\sc CMFGEN} predicts that \ion{Si}{iv} will be
dominant as expected for \teff\ $=$ 27\,500 K, but shows very low
levels of \ion{N}{v} and \ion{C}{iv}. This is hardly surprising since
it explains the complete absence of a \ion{N}{v} P Cygni profile (when
present observationally), as well as the difficulties in generating a
P Cygni profile for \ion{C}{iv} when it is unsaturated. It is also
interesting to note that the levels of ionisation drop off rapidly in
the model as $w$ increases, contradicting the empirical determinations
of \cite{prinja2005} where winds became more highly ionised at high
$w$. This is direct evidence of the model failing to generate enough
high-velocity absorption to sustain the same level of ionisation
further out in the wind. This is the reason for the `narrowness' of
the model \ion{C}{iv} and \ion{Si}{iv} P~Cygni profiles compared to
the broad absorption troughs of the observed P~Cygni profiles. If the
model cannot sustain enough ionisation in the inner and outer parts of
the wind, then it will be unable to fully reproduce the blue-ward part
of the profile.  {\sc CMFGEN} predicts \ion{Si}{iv} to be dominant
down to \teff\ $=$ 18\,000~K, at which point \ion{Si}{iii} and
\ion{Al}{iii} take over as the dominant ions in the wind. At this
\teff, \ion{C}{ii} has also increased in
strength, becoming a dominant ion at \teff\ $=$~15\,000~K. 
Whereas this approach has provided us with valuable insight into
why {\sc CMFGEN } struggles to predict the P Cygni profiles correctly,
it too fails to provide us with an alternative means of constraining
parameters due to the incorrectly-predicted ionisation structure. \\
 
\noindent
Given all these problems with {\sc CMFGEN} mismatching the
observed UV P Cygni profiles, a investigation into the effects of
clumping on these lines would not be worthwhile at present. In
addition, mass loss in the UV is only sensitive to $\rho$, rather than
$\rho^{2}$ as in \ha\ and radio-dominated regions of the wind, so it
is not a particularly sensitive indicator of clumping. First the
models need to predict the correction ionisation structure for B
supergiants. Secondly, the problems associated with investigating the
effects of clumping on \ha\ (as discussed in \S~\ref{mdot}) need to be
sorted as \ha\ is an important diagnostic of clumping and can provide
important insight into its behaviour, which would aid a subsequent
analysis of clumping in the UV. Furthermore, in comparison to O stars
which possess strong indicators of UV clumping e.g.,
\ion{P}{v}~$\lambda\lambda$~1118,~1128 (see
\citealt{fullerton2006,bouret2005}), B supergiants do not possess an
equally convenient UV diagnostic. A tentative examination of the
effect of clumping on the UV profiles has shown that it does not
improve the fits to the observed P Cygni profiles, as expected, but
can alleviate the over-estimation of emission seen in the red-ward
part of the model \ion{Si}{iii} and \ion{Al}{iii} profiles.  In the
case of the photospheric \ion{Si}{iii} lines around $\sim$ 1300 \AA,
some broadening of these lines due to the stellar wind is seen
observationally, and the models also show some sensitivity to clumping
in these lines. The models exaggerate the effect of the stellar wind
on these lines by producing slightly asymmetric profiles, but the
inclusion of clumping can help to lessen this asymmetry. This is
logical since the inclusion of clumping will increase the wind density
locally, providing more absorption at the point at which the line
forms in the wind, helping to reduce the excess red-ward emission seen
in many of the model P Cygni profiles. The inclusion of clumping will
have no affect on saturated lines in the model as they are no longer
sensitive to density changes in the wind. All the afore-mentioned issues associated with investigating clumping effects need to be addressed before any truly meaningful analysis of clumping in the UV can be carried out. \\


\section{Conclusions}
\label{conc}

A quantitative study of the optical and UV properties of B0~--~B5 Ia,
Iab, Ib/II supergiants has been carried out, using the nLTE,
line-blanketed stellar atmosphere code of \cite{hillier1998}. A
revised B supergiant \teff\ scale (derived using a stellar atmosphere
code that includes the effects of line blanketing) has been presented,
giving a range of 14.5\,000~K $\le \teff \le$ 30\,000~K for these
stars. This scale shows a drop of up to 10\,000K from B0 Ia/b to B1
Iab and a difference of up to 2\,500 K between Ia and Ib stars. It
also shows that on average the effect of including line blanketing in
the model produces a modest reduction of up to 1\,000 K for B0~--~B0.7
and B3~--~B5 supergiants, whereas a larger reduction of up to 3\,000 K
is seen for B1 -- B2 supergiants (see Table \ref{teffsp}). The 20
Galactic B supergiants also displayed a range of 2.1 $\le$ log $g \le$
3.4 in surface gravity. These results, together with those of
\cite{crowther2006}, have been used to construct a new set of averaged
fundamental parameters for B0~--~B5 supergiants, according to spectral
type. Mass loss rates derived from \ha\ proved B supergiant winds to
be generally weaker than those of O supergiants (as expected since
they are lower-luminosity objects) with \mdot\ ranging from $-7.22 \le
\logmdot \le -5.30$. All 20 B supergiants also shown signs of CNO
processing, with the largest nitrogen enrichments being seen for B1-B2
supergiants. Evidence for a mass discrepancy is found between
estimates of \mspec\ and \mevol, with the largest differences peaking
at a value of \logllsolar\ $\sim$ 5.4. \\

\noindent
A Wind-Momentum--Luminosity relation has also been derived for our
sample, which is lower in value for B1 -- B5 supergiants than that
predicted by \cite{vink2000}, but greater than predicted values for B0
-- B0.7 supergiants. For this reason it is not possible to reconcile
this difference in observed and theoretical WLRs over the whole B
supergiant spectral range by adopting clumped \mdot\ as is the case
for O stars. A severe problem exists in the form of the optical-UV
discrepancy, where the model fails to reproduce some of the P Cygni
profiles accurately. This highlights a failure in the model to
generate enough high-velocity absorption to succeed in reproducing the
observed P Cygni profile and more crucially highlights that the models
are not predicting the correct ionisation structure. Given that B
supergiants, along with other massive stars, have their peak flux in
the UV, it is imperative that this discrepancy is resolved if we are
to have confidence that fundamental parameters derived by this method
are a true representation of the star's properties. Furthermore it
underlines the incompleteness of our current understanding of the
physics of massive star winds and the necessity to review the standard
model. A more thorough analysis of the ionisation structure of early B
supergiant winds will be presented in Paper II. \\


\begin{acknowledgements}
    SCS would like to acknowledge PPARC and BFWG for financial support. 
Thanks go to John Hillier for assistance in using {\sc CMFGEN} as well 
as Callum Wright and Jeremy Yates for computing support. We also thank 
the referee for his comments.  
\end{acknowledgements}

\begin{appendix}

\section{Error Analysis}

In this section, the errors affecting each derived parameter are
discussed. The error on \teff\ is estimated from the quality of the
{\sc CMFGEN} model fit to the diagnostic silicon, helium and magnesium
lines and therefore represents the range in \teff\ over which a
satisfactory fit to the observed spectrum of the star could be
obtained. Luminosity is primarily constrained through dereddening the
observed spectra with respect to the model spectral distribution, its
error depends on $\Delta$\mv, whose errors are estimated from
dereddening the observed spectrum with respect to the model spectral
energy distribution. $\Delta$\loglstar\ also depends on
$\Delta$\mbol, since $ \mbol = \mv + B.C. $, therefore  $\Delta$\loglstar\ is calculated as 

\begin{equation}\label{errlum}
\Delta \loglstar = \loglstar \frac{2.5 \Delta\mbol}{(\mbol - 4.72)}
\end{equation}

\noindent
The error on \rstar\ depends on the square root of the sum of $(\Delta\teff)^{2}$ and $
(\Delta\lstar)^{2}$, with $\Delta\lstar$ having the greatest influence
on $\Delta\rstar$.  For 15 out of the 20 B supergiants, $\Delta\rstar$
is within 10\% of the absolute value of \rstar; those stars with
larger $\Delta\rstar$ are discussed separately in this section.
The error on log $g$ for 16 of the 20 stars star is estimated to be 0.25 dex, 
based on the accuracy of line fits and the effect of $\Delta\teff$ in
determining log $g$. For four of the stars in our sample, we adopted
$\Delta$ log $g$ $=$ 0.38 dex. In the cases of HD 190603 and HD 14818,
this was due to asymmetric nature of the \hg\ and \hd\ profiles,
whereas for HD 64760 and HD 13854 it reflected the larger error in
\teff\ of 2000 K (cf. to 500 - 1500 K for other stars in the sample.
The resulting uncertainty on the spectroscopic mass,
\mspec, due to errors in constraining log $g$ and \rstar\ range from
0.08 $\le \Delta\mstar \le$ 14.52. In comparison, $\Delta M_{evol}$ is typically 5 \msolar
as shown in Fig. \ref{obahrd}, which demonstrates the effect of assuming different values of \teff\ and \logllsolar\
on a star's position in the HR diagram. \\


\noindent
Determining the error in constraining the mass loss rate is more
complicated, since it depends on both $\Delta\rstar$ and the error in
fitting the \ha\ profile by varying \mdot\ and $\beta$. However, the
errors incurred from uncertainties in deriving \rstar\ are negligible
compared to those arising from fitting the \ha\ profile, so we are
justified in defining $\Delta \mdot\ $ as solely the error in fitting
the \ha\ profile, accounting for the degeneracy in varying $\beta$ to
fit \ha\ profiles in emission. Values of \vinf\ are taken from SEI
analysis of UV wind resonance lines (the result of which will be
presented in Paper II) and are accurate to $\pm$ 50 km/s.
The values for \vturb\ are constrained with an uncertainty of $\pm$ 5
km/s, as dictated by sensitivity of fitting the \ion{Si}{iii} lines by eye. \\

\noindent
Some uncertainties exist in our analysis that warrant further
discussion.  Although for the majority of stars it was possible to
constrain \teff\ within $\pm$ 1\,000 K, this was not possible for the
stars HD~64760 and HD~13854. HD~64760 is a rapid rotator and the large
width of its spectral lines makes it harder to make an accurate
distinction between model fits whose \teff\ differ by 1\,000~K,
resulting in $\Delta$~\teff~$=$~$\pm$~2\,000~K for this star. In the
case of HD~13854, if the adopted \teff\ of 20\,000~K is increased to
22\,000~K (keeping the same luminosity) then a much better fit is made
to the silicon lines (i.e., \ion{Si}{iv}~4089 and \ion{Si}{iii}~4552,
4568 and 4575) but at the expense of grossly over-estimating the
hydrogen and helium lines (i.e., \hb, \hg, \hd, \ion{He}{i} 4121,
4144, 4387 and 4471). Normally in our analysis, the fitting of silicon
lines would be given priority, but given the weakness of
\ion{Si}{iv}~4089 at this spectral type (B1 Iab), it is reasonable to
assign greater importance to fitting the helium lines. Furthermore,
adopting a model that only fits the silicon lines well and largely
over-estimates the hydrogen and helium lines will give a misleading
indication of the value of \teff. HD~13854 also has quite a large
error in \mv, which consequently propagates into significant
uncertainties in \logllsolar\ and, combined with a larger
$\Delta$\teff, leads to a very large $\Delta$\rstar. This arises from
a noisy IUE SWP spectrum and the absence of a LWR spectrum leading to
a large dereddening error; the same is true for the considerable
errors on the values of \mv\ and \rstar\ obtained for HD~204172. Low
quality IUE spectra generating higher $\Delta$ E(B-V) also explain the
$\Delta$\logllsolar\ found for HD~14818 and HD~206165. However, for
the B5 II (Ib) star HD~164353, it is the value of \mv\ $=$ -4.2 that
poses a problem; {\sc CMFGEN} simply fails to calculate a
succesfully-converged model at the required luminosity. This
 explains the large values of $\Delta$\logllsolar, $\Delta$\rstar and
 $\Delta$\mv. The adopted value of \mv\ has been independently
 confirmed by several different sources in the literature so we
 believe it to be correct. Furthermore, four stars (HD 37128,
 HD192660, HD198478 and HD 42087) have larger errors in the observed
 value of \mv\ than the value of \mv\ derived from dereddening, so in
 practise the quoted value of $\Delta$\mv\ could be up to 0.2 mag
  larger. \\

\end{appendix}

\bibliographystyle{aa}
\bibliography{7125}

\begin{thebibliography}{70}
\expandafter\ifx\csname natexlab\endcsname\relax\def\natexlab#1{#1}\fi

\bibitem[{Anderson(1989)}]{anderson1989}
Anderson, L.~S. 1989, ApJ, 339, 588

\bibitem[{Bianchi(2002)}]{bianchi2002}
Bianchi, L.~Garcia, M. 2002, ApJ, 581, 610

\bibitem[{{Bidelman}(1988)}]{bidelmann1988}
{Bidelman}, W.~P. 1988, \pasp, 100, 1084

\bibitem[{{Blomme} {et~al.}(2002){Blomme}, {Prinja}, {Runacres}, \&
  {Colley}}]{blomme2002}
{Blomme}, R., {Prinja}, R.~K., {Runacres}, M.~C., \& {Colley}, S. 2002, \aap,
  382, 921

\bibitem[{{Bouret} {et~al.}(2005){Bouret}, {Lanz}, \& {Hillier}}]{bouret2005}
{Bouret}, J.-C., {Lanz}, T., \& {Hillier}, D.~J. 2005, \aap, 438, 301

\bibitem[{{Bouret} {et~al.}(2003){Bouret}, {Lanz}, {Hillier}, {Heap}, {Hubeny},
  {Lennon}, {Smith}, \& {Evans}}]{bouret2003}
{Bouret}, J.-C., {Lanz}, T., {Hillier}, D.~J., {et~al.} 2003, \apj, 595, 1182

\bibitem[{{Brown} {et~al.}(1994){Brown}, {de Geus}, \& {de Zeeuw}}]{brown1994}
{Brown}, A.~G.~A., {de Geus}, E.~J., \& {de Zeeuw}, P.~T. 1994, \aap, 289, 101

\bibitem[{Castor {et~al.}(1975)Castor, Abbott, \& Klein}]{cak1975}
Castor, J.~I., Abbott, D.~C., \& Klein, R.~I. 1975, ApJ, 195, 157

\bibitem[{{Cranmer} \& {Owocki}(1996)}]{cranmer1996}
{Cranmer}, S.~R. \& {Owocki}, S.~P. 1996, \apj, 462, 469

\bibitem[{{Crowther} {et~al.}(2002){Crowther}, {Hillier}, {Evans}, {Fullerton},
  {De Marco}, \& {Willis}}]{crowther2002}
{Crowther}, P.~A., {Hillier}, D.~J., {Evans}, C.~J., {et~al.} 2002, \apj, 579,
  774

\bibitem[{{Crowther} {et~al.}(2006){Crowther}, {Lennon}, \&
  {Walborn}}]{crowther2006}
{Crowther}, P.~A., {Lennon}, D.~J., \& {Walborn}, N.~R. 2006, \aap, 446, 279

\bibitem[{{de Zeeuw} {et~al.}(1999){de Zeeuw}, {Hoogerwerf}, {de Bruijne},
  {Brown}, \& {Blaauw}}]{dezeeuw1999}
{de Zeeuw}, P.~T., {Hoogerwerf}, R., {de Bruijne}, J.~H.~J., {Brown}, A.~G.~A.,
  \& {Blaauw}, A. 1999, \aj, 117, 354

\bibitem[{{Dufton} {et~al.}(2005){Dufton}, {Ryans}, {Trundle}, {Lennon},
  {Hubeny}, {Lanz}, \& {Allende Prieto}}]{dufton2005}
{Dufton}, P.~L., {Ryans}, R.~S.~I., {Trundle}, C., {et~al.} 2005, A\&A, 434,
  1125

\bibitem[{{Egret}(1978)}]{egret1978}
{Egret}, D. 1978, \aap, 66, 275

\bibitem[{{Evans} {et~al.}(2004{\natexlab{a}}){Evans}, {Crowther}, {Fullerton},
  \& {Hillier}}]{evans2004b}
{Evans}, C.~J., {Crowther}, P.~A., {Fullerton}, A.~W., \& {Hillier}, D.~J.
  2004{\natexlab{a}}, \apj, 610, 1021

\bibitem[{{Evans} {et~al.}(2004{\natexlab{b}}){Evans}, {Lennon}, {Trundle},
  {Heap}, \& {Lindler}}]{evans2004a}
{Evans}, C.~J., {Lennon}, D.~J., {Trundle}, C., {Heap}, S.~R., \& {Lindler},
  D.~J. 2004{\natexlab{b}}, \apj, 607, 451

\bibitem[{{Fernie}(1983)}]{fernie1983}
{Fernie}, J.~D. 1983, \apjs, 52, 7

\bibitem[{{Fitzgerald}(1970)}]{fp1970}
{Fitzgerald}, M.~P. 1970, \aap, 4, 234

\bibitem[{{Fullerton} {et~al.}(2006){Fullerton}, {Massa}, \&
  {Prinja}}]{fullerton2006}
{Fullerton}, A.~W., {Massa}, D.~L., \& {Prinja}, R.~K. 2006, \apj, 637, 1025

\bibitem[{{Fullerton} {et~al.}(1997){Fullerton}, {Massa}, {Prinja}, {Owocki},
  \& {Cranmer}}]{fullerton1997}
{Fullerton}, A.~W., {Massa}, D.~L., {Prinja}, R.~K., {Owocki}, S.~P., \&
  {Cranmer}, S.~R. 1997, \aap, 327, 699

\bibitem[{{Garmany} \& {Stencel}(1992)}]{garmany1992}
{Garmany}, C.~D. \& {Stencel}, R.~E. 1992, \aaps, 94, 211

\bibitem[{{Herrero} {et~al.}(2002){Herrero}, {Puls}, \&
  {Najarro}}]{herrero2002}
{Herrero}, A., {Puls}, J., \& {Najarro}, F. 2002, \aap, 396, 949

\bibitem[{{Hillier} {et~al.}(2003){Hillier}, {Lanz}, {Heap}, {Hubeny}, {Smith},
  {Evans}, {Lennon}, \& {Bouret}}]{hillier2003}
{Hillier}, D.~J., {Lanz}, T., {Heap}, S.~R., {et~al.} 2003, \apj, 588, 1039

\bibitem[{Hillier \& Miller(1998)}]{hillier1998}
Hillier, D.~J. \& Miller, D.~L. 1998, ApJ, 496, 407

\bibitem[{{Hillier} \& {Miller}(1999)}]{hillier1999}
{Hillier}, D.~J. \& {Miller}, D.~L. 1999, \apj, 519, 354

\bibitem[{{Hiltner}(1956)}]{hiltner1956}
{Hiltner}, W.~A. 1956, \apjs, 2, 389

\bibitem[{{Hoffleit} \& {Jaschek}(1982)}]{hj1982}
{Hoffleit}, D. \& {Jaschek}, C. 1982, {The Bright Star Catalogue} (The Bright
  Star Catalogue, New Haven: Yale University Observatory (4th edition), 1982)

\bibitem[{{Howarth} {et~al.}(1997){Howarth}, {Siebert}, {Hussain}, \&
  {Prinja}}]{howarth1997}
{Howarth}, I.~D., {Siebert}, K.~W., {Hussain}, G.~A.~J., \& {Prinja}, R.~K.
  1997, \mnras, 284, 265

\bibitem[{Hubeny \& Lanz(1995)}]{hubeny1995}
Hubeny, I. \& Lanz, T. 1995, ApJ, 439, 875

\bibitem[{{Humphreys}(1978)}]{humphreys1978}
{Humphreys}, R.~M. 1978, \apjs, 38, 309

\bibitem[{{Kaufer} {et~al.}(2006){Kaufer}, {Stahl}, {Prinja}, \&
  {Witherick}}]{kaufer2006}
{Kaufer}, A., {Stahl}, O., {Prinja}, R.~K., \& {Witherick}, D. 2006, \aap, 447,
  325

\bibitem[{Kaufer \& Stahl(2002)}]{kaufer2002}
Kaufer, A.~Prinja, R.~K. \& Stahl, O. 2002, A\&A, 382, 1032

\bibitem[{{Kudritzki} {et~al.}(1995){Kudritzki}, {Lennon}, \&
  {Puls}}]{kudritzki1995}
{Kudritzki}, R.-P., {Lennon}, D.~J., \& {Puls}, J. 1995, in Science with the
  VLT, ed. J.~R. {Walsh} \& I.~J. {Danziger}, 246

\bibitem[{{Kudritzki} \& {Puls}(2000)}]{kudritzki2000}
{Kudritzki}, R.-P. \& {Puls}, J. 2000, \araa, 38, 613

\bibitem[{Kudritzki {et~al.}(1999)Kudritzki, Puls, Lennon, Venn, Reetz,
  Najarro, McCarthy, \& Herrero}]{kudritzki1999}
Kudritzki, R.~P., Puls, J., Lennon, D.~J., {et~al.} 1999, A\&A, 350, 970

\bibitem[{Lanz(2003)}]{lanz2003}
Lanz, T.and~Hubeny, I. 2003, ApJS, 146, 417

\bibitem[{{Lennon} {et~al.}(1992){Lennon}, {Dufton}, \&
  {Fitzsimmons}}]{lennon1992}
{Lennon}, D.~J., {Dufton}, P.~L., \& {Fitzsimmons}, A. 1992, \aaps, 94, 569

\bibitem[{{Lesh}(1968)}]{lesh1968}
{Lesh}, J.~R. 1968, \apjs, 17, 371

\bibitem[{{Maeder} \& {Meynet}(2001)}]{meynet2001}
{Maeder}, A. \& {Meynet}, G. 2001, \aap, 373, 555

\bibitem[{{Markova} {et~al.}(2004){Markova}, {Puls}, {Repolust}, \&
  {Markov}}]{markova2004}
{Markova}, N., {Puls}, J., {Repolust}, T., \& {Markov}, H. 2004, \aap, 413, 693

\bibitem[{Martins {et~al.}(2005)Martins, Schaerer, \& Hillier}]{martins2005}
Martins, F., Schaerer, D., \& Hillier, D.~J. 2005, A\&A, 436, 1049

\bibitem[{{Massa}(1989)}]{massa1989}
{Massa}, D. 1989, \aap, 224, 131

\bibitem[{{Massa} {et~al.}(2003){Massa}, {Fullerton}, {Sonneborn}, \&
  {Hutchings}}]{massa2003}
{Massa}, D., {Fullerton}, A.~W., {Sonneborn}, G., \& {Hutchings}, J.~B. 2003,
  \apj, 586, 996

\bibitem[{{Massa} {et~al.}(1995){Massa}, {Prinja}, \& {Fullerton}}]{prinja1995}
{Massa}, D., {Prinja}, R.~K., \& {Fullerton}, A.~W. 1995, \apj, 452, 842

\bibitem[{{Massey} {et~al.}(2004){Massey}, {Bresolin}, {Kudritzki}, {Puls}, \&
  {Pauldrach}}]{massey2004}
{Massey}, P., {Bresolin}, F., {Kudritzki}, R.~P., {Puls}, J., \& {Pauldrach},
  A.~W.~A. 2004, \apj, 608, 1001

\bibitem[{Massey {et~al.}(2005)Massey, Puls, Pauldrach, Bresolin, Kudritzki, \&
  Simon}]{massey2005}
Massey, P., Puls, J., Pauldrach, A. W.~A., {et~al.} 2005, ApJ, 627, 477

\bibitem[{McErLean {et~al.}(1999)McErLean, Lennon, \& Dufton}]{mcerlean1999}
McErLean, N.~D., Lennon, D., \& Dufton, P. 1999, A\&A, 349, 553

\bibitem[{{Meynet} \& {Maeder}(2000)}]{meynet2000}
{Meynet}, G. \& {Maeder}, A. 2000, \aap, 361, 101

\bibitem[{{Morel} {et~al.}(2004){Morel}, {Marchenko}, {Pati}, {Kuppuswamy},
  {Carini}, {Wood}, \& {Zimmerman}}]{morel2004}
{Morel}, T., {Marchenko}, S.~V., {Pati}, A.~K., {et~al.} 2004, \mnras, 351, 552

\bibitem[{{Nieva} \& {Przybilla}(2006)}]{nieva2006}
{Nieva}, M.~F. \& {Przybilla}, N. 2006, \apjl, 639, L39

\bibitem[{{Owocki} {et~al.}(1988){Owocki}, {Castor}, \& {Rybicki}}]{owocki1988}
{Owocki}, S.~P., {Castor}, J.~I., \& {Rybicki}, G.~B. 1988, \apj, 335, 914

\bibitem[{{Prinja} {et~al.}(2002){Prinja}, {Massa}, \&
  {Fullerton}}]{prinja2002}
{Prinja}, R.~K., {Massa}, D., \& {Fullerton}, A.~W. 2002, \aap, 388, 587

\bibitem[{{Prinja} {et~al.}(2005){Prinja}, {Massa}, \& {Searle}}]{prinja2005}
{Prinja}, R.~K., {Massa}, D., \& {Searle}, S.~C. 2005, \aap, 430, L41

\bibitem[{{Prinja} {et~al.}(2004){Prinja}, {Rivinius}, {Stahl}, {Kaufer},
  {Foing}, {Cami}, \& {Orlando}}]{prinja2004}
{Prinja}, R.~K., {Rivinius}, T., {Stahl}, O., {et~al.} 2004, \aap, 418, 727

\bibitem[{{Puls} {et~al.}(1996){Puls}, {Kudritzki}, {Herrero}, {Pauldrach},
  {Haser}, {Lennon}, {Gabler}, {Voels}, {Vilchez}, {Wachter}, \&
  {Feldmeier}}]{puls1996}
{Puls}, J., {Kudritzki}, R.-P., {Herrero}, A., {et~al.} 1996, \aap, 305, 171

\bibitem[{Puls {et~al.}(2006)Puls, Markova, Scuderi, Stanghellini, Taranova,
  Burnley, \& Howarth}]{puls2006}
Puls, J., Markova, N., Scuderi, S., {et~al.} 2006, \aap, submitted

\bibitem[{Repolust {et~al.}(2004)Repolust, Puls, \& Herrero}]{repolust2004}
Repolust, T., Puls, J., \& Herrero, A. 2004, A\&A, 415, 349

\bibitem[{{Runacres} \& {Owocki}(2002)}]{owocki2002}
{Runacres}, M.~C. \& {Owocki}, S.~P. 2002, \aap, 381, 1015

\bibitem[{{Rusconi} {et~al.}(1980){Rusconi}, {Sedmak}, {Stalio}, \&
  {Arpigny}}]{rusconi1980}
{Rusconi}, L., {Sedmak}, G., {Stalio}, R., \& {Arpigny}, C. 1980, \aaps, 42,
  347

\bibitem[{{Schild}(1985)}]{schild1985}
{Schild}, H. 1985, \aap, 146, 113

\bibitem[{{Townsend}(1997)}]{townsend1997}
{Townsend}, R.~H.~D. 1997, \mnras, 284, 839

\bibitem[{{Trundle} \& {Lennon}(2005)}]{trundle2005}
{Trundle}, C. \& {Lennon}, D.~J. 2005, \aap, 434, 677

\bibitem[{{Trundle} {et~al.}(2004){Trundle}, {Lennon}, {Puls}, \&
  {Dufton}}]{trundle2004}
{Trundle}, C., {Lennon}, D.~J., {Puls}, J., \& {Dufton}, P.~L. 2004, A\&A, 417,
  217

\bibitem[{{Venn}(1995)}]{venn1995}
{Venn}, K.~A. 1995, \apj, 449, 839

\bibitem[{{Venn}(1999)}]{venn1999}
{Venn}, K.~A. 1999, \apj, 518, 405

\bibitem[{{Vink} {et~al.}(1999){Vink}, {de Koter}, \& {Lamers}}]{vink1999}
{Vink}, J.~S., {de Koter}, A., \& {Lamers}, H.~J.~G.~L.~M. 1999, \aap, 350, 181

\bibitem[{{Vink} {et~al.}(2000){Vink}, {de Koter}, \& {Lamers}}]{vink2000}
{Vink}, J.~S., {de Koter}, A., \& {Lamers}, H.~J.~G.~L.~M. 2000, \aap, 362, 295

\bibitem[{{Walborn}(1971)}]{walborn1971}
{Walborn}, N.~R. 1971, ApJS, 23, 257

\bibitem[{{Walborn}(1972)}]{walborn1972a}
{Walborn}, N.~R. 1972, \aj, 77, 312

\bibitem[{Walborn(1976)}]{walborn1976}
Walborn, Nolan, R. 1976, ApJ, 205, 419

\end{thebibliography}






   
  






\onlfig{1}{
\begin{figure*}
\centering
\includegraphics[width=12cm]{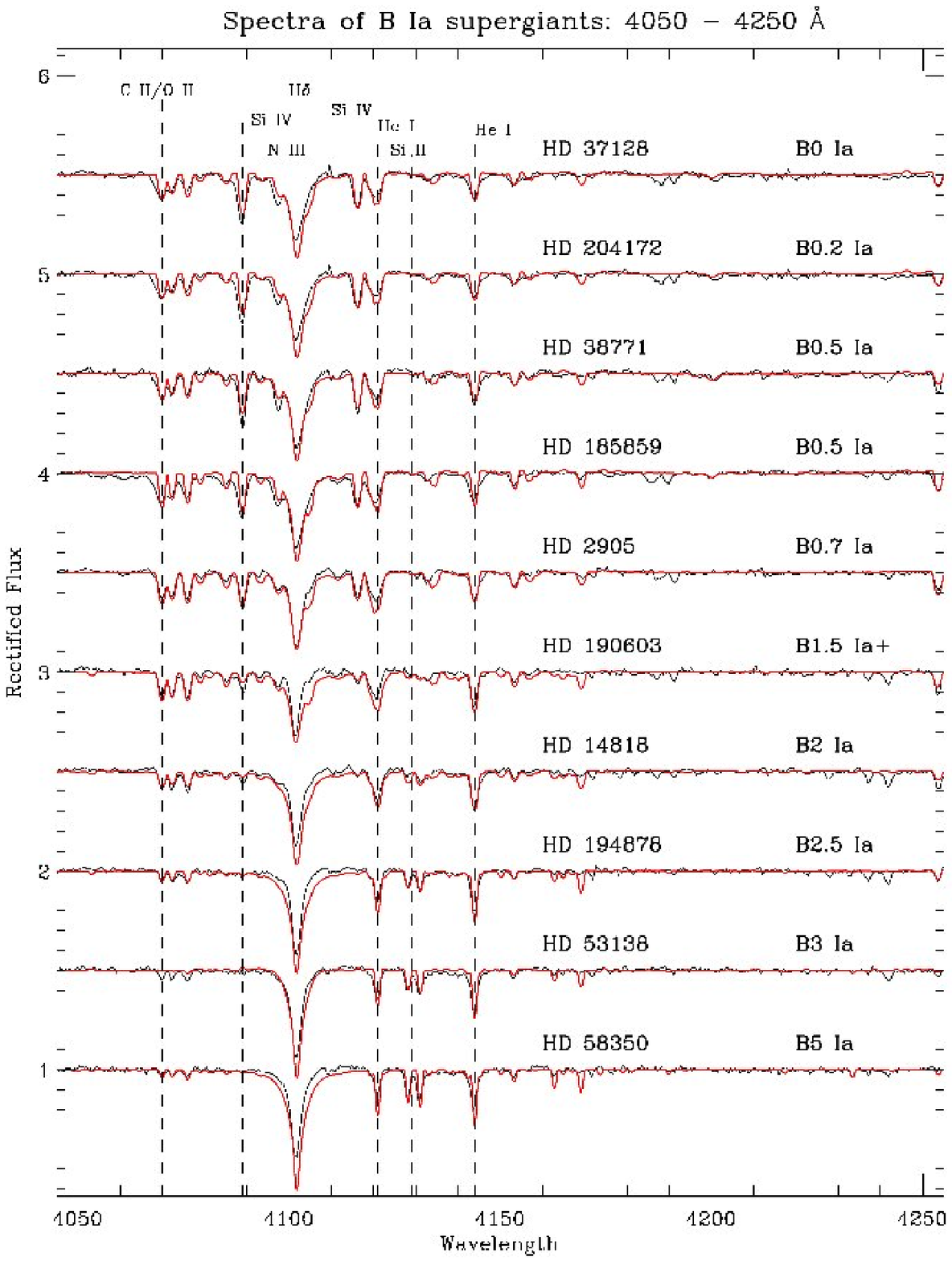}
\caption{{\sc CMFGEN} model fits (4050 -- 4250 \AA) to the optical spectra of 10 B Ia supergiants, with the \teff, luminosity and CNO diagnostic lines marked as shown. Optical spectrum is in black, {\sc CMFGEN} model fit is shown in red.}
\label{op1b}
\end{figure*}}

\onlfig{2}{
\begin{figure*}
\centering
\includegraphics[width=12cm]{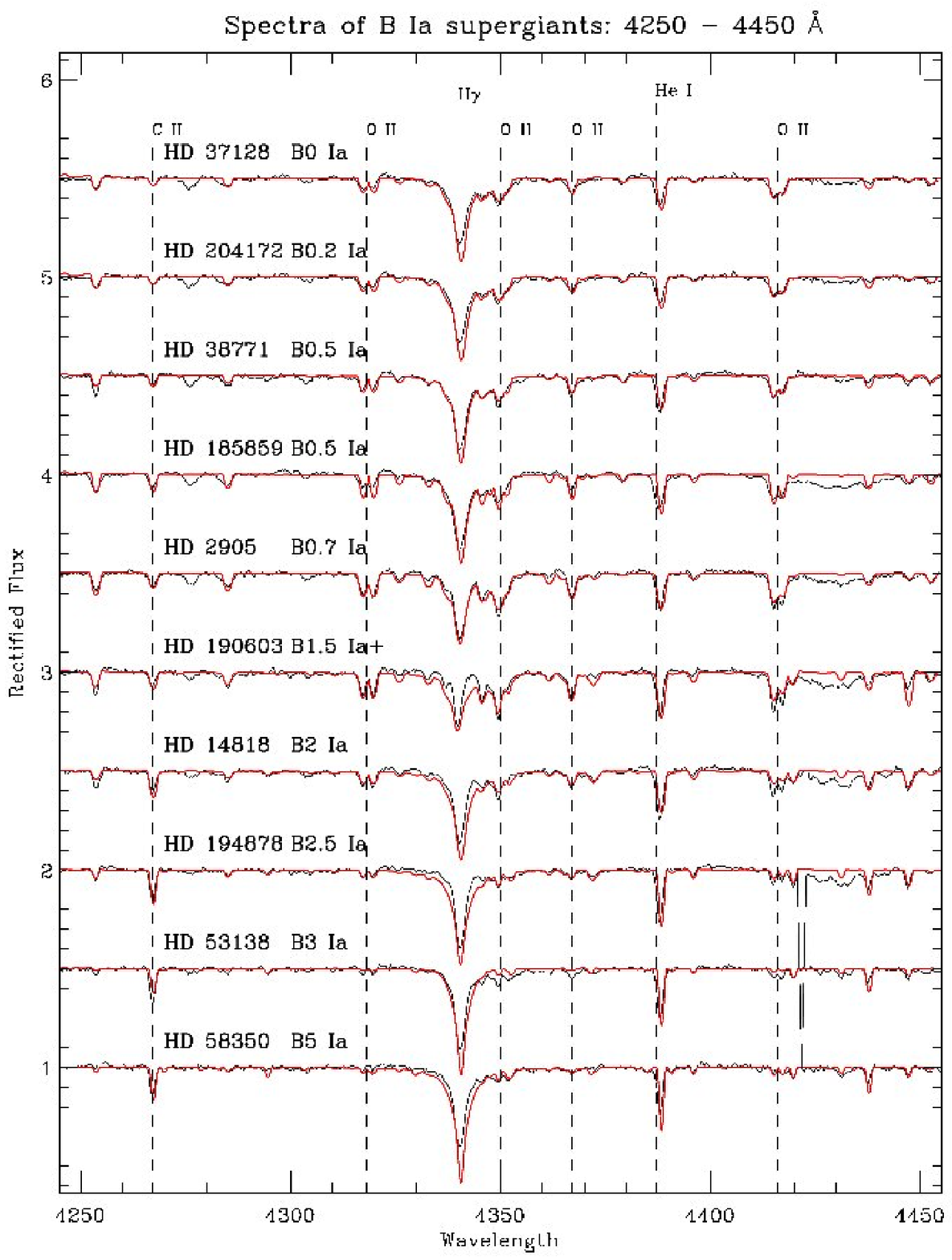}
\caption{{\sc CMFGEN} model fits (4250 -- 4450 \AA) to the optical spectra of 10 B Ia supergiants, with the \teff, luminosity and CNO diagnostic lines marked as shown. Optical spectrum is in black, {\sc CMFGEN} model fit is shown in red.}
\label{op2b}
\end{figure*}}

\onlfig{3}{
\begin{figure*}
\centering
\includegraphics[width=12cm]{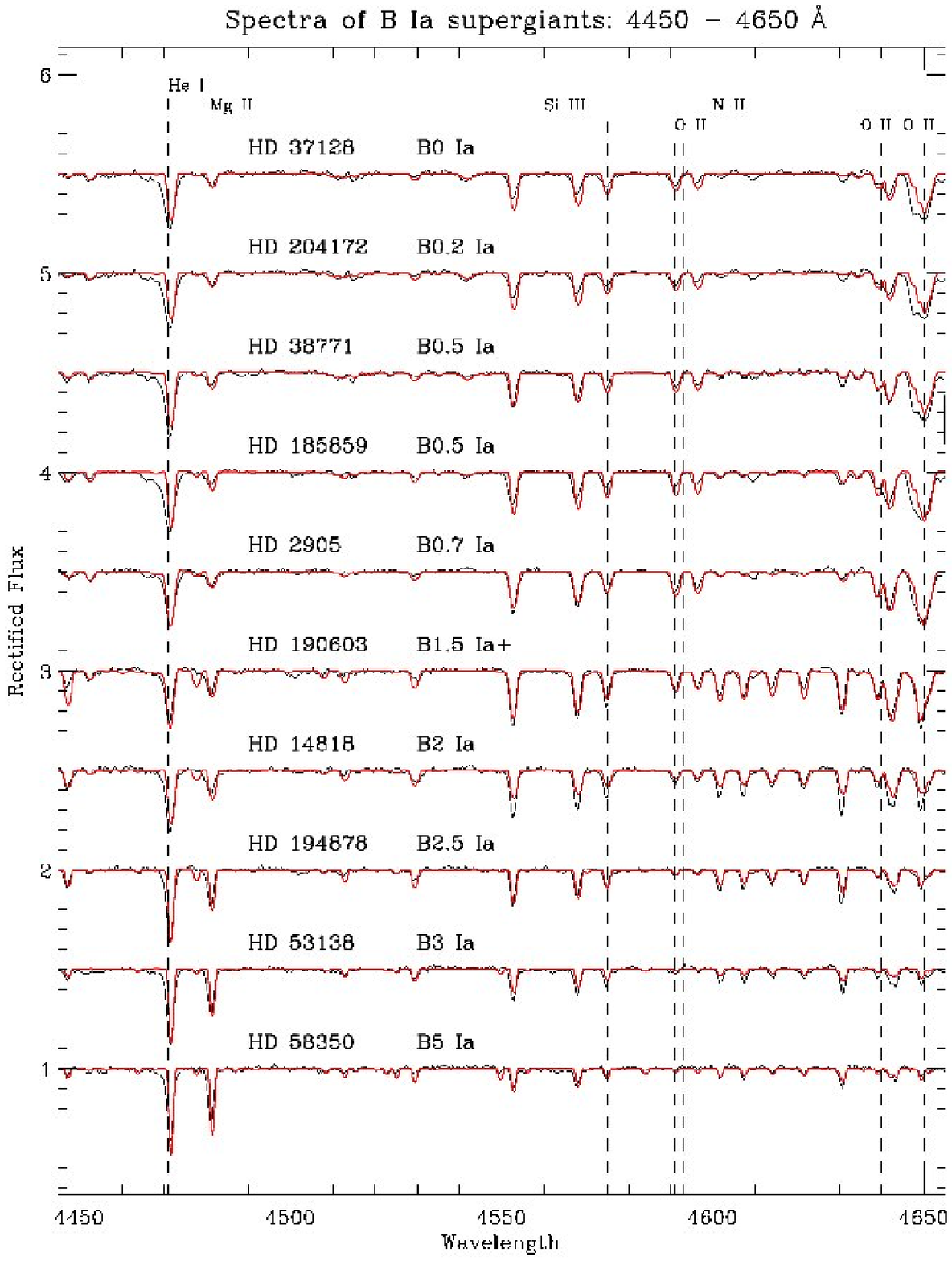}
\caption{{\sc CMFGEN} model fits (4450 -- 4650 \AA) to the optical spectra of 10 B Ia supergiants, with the \teff, luminosity and CNO diagnostic lines marked as shown. Optical spectrum is in black, {\sc CMFGEN} model fit is shown in red.}
\label{op3b}
\end{figure*}}

\onlfig{4}{
\begin{figure*}
\centering
\includegraphics[width=12cm]{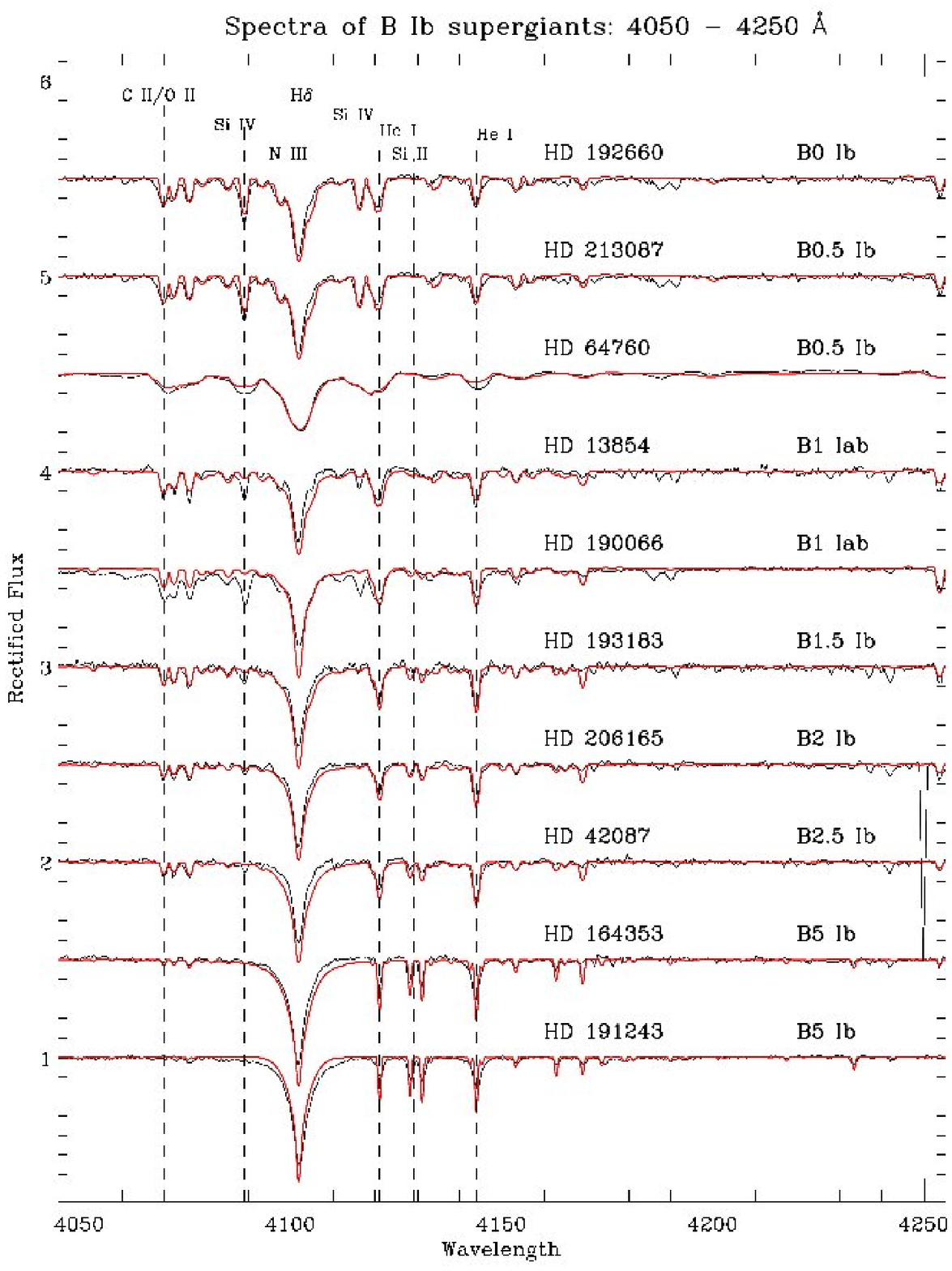}
\caption{{\sc CMFGEN} model fits (4050 -- 4250 \AA) to the optical spectra of 10 B Ib supergiants, with the \teff, luminosity and CNO diagnostic lines marked as shown. Optical spectrum is in black, {\sc CMFGEN} model fit is shown in red.}
\label{op4}
\end{figure*}}

\onlfig{5}{
\begin{figure*}
\centering
\includegraphics[width=12cm]{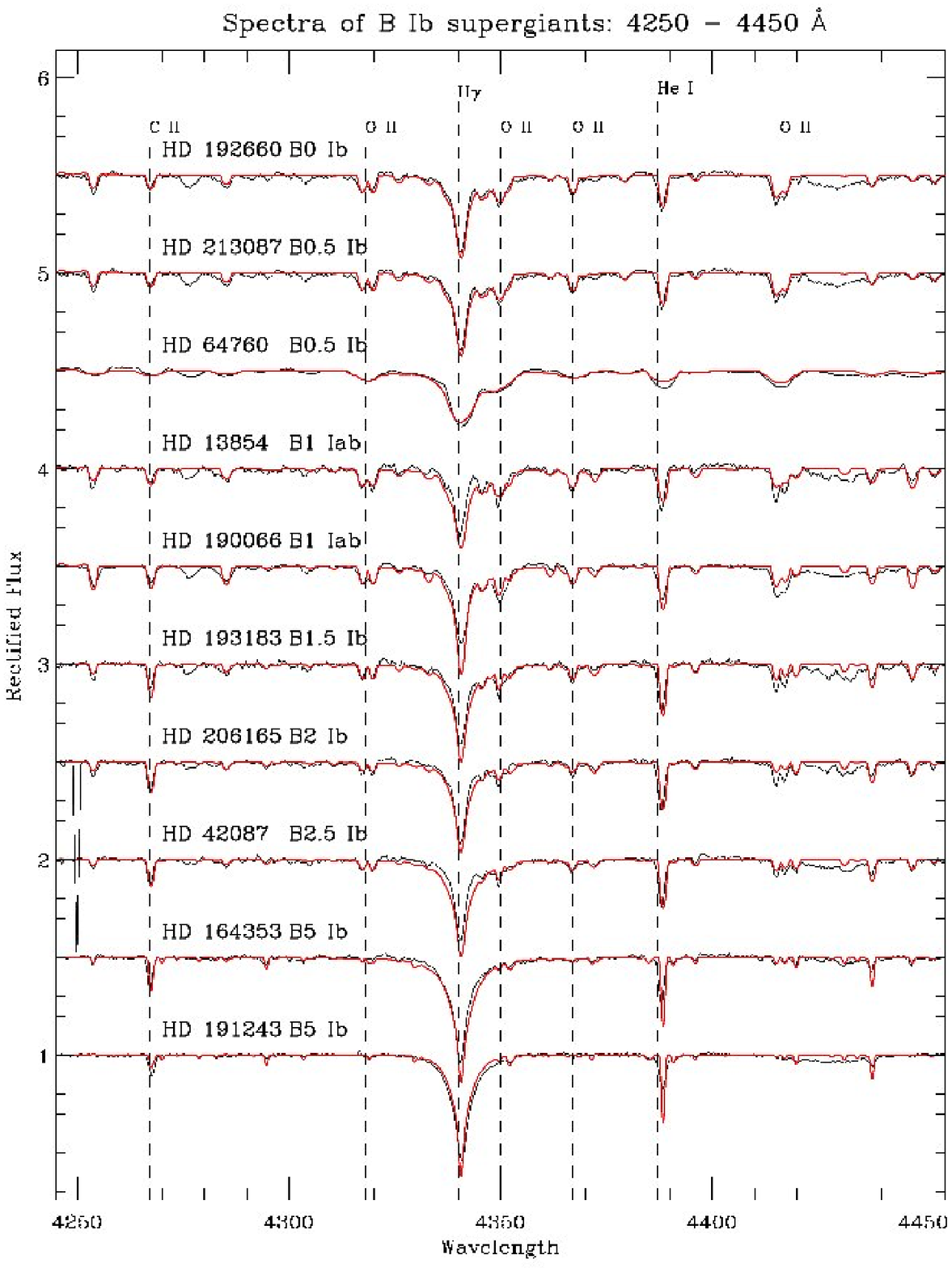}
\caption{{\sc CMFGEN} model fits (4250 -- 4450 \AA) to the optical spectra of 10 B Ib supergiants, with the \teff, luminosity and CNO diagnostic lines marked as shown. Optical spectrum is in black, {\sc CMFGEN} model fit is shown in red.}
\label{op5}
\end{figure*}}

\onlfig{6}{
\begin{figure*}
\centering
\includegraphics[width=12cm]{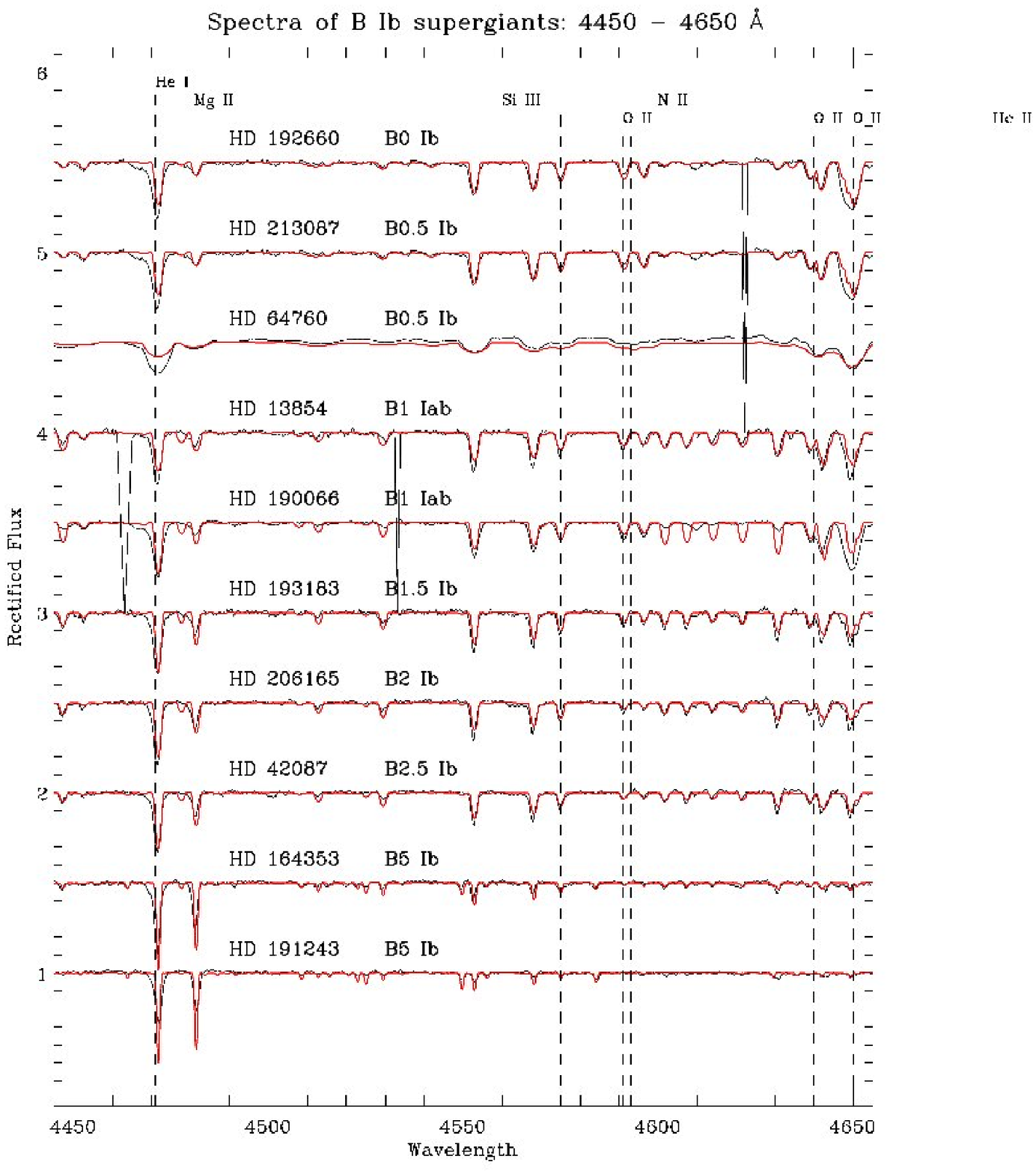}
\caption{{\sc CMFGEN} model fits (4450 -- 4650 \AA) to the optical spectra of 10 B Ib supergiants, with the \teff, luminosity and CNO diagnostic lines marked as shown. Optical spectrum is in black, {\sc CMFGEN} model fit is shown in red.}
\label{op6}
\end{figure*}}


\onlfig{7}{
\begin{figure*}
\centering
\includegraphics[width=12cm]{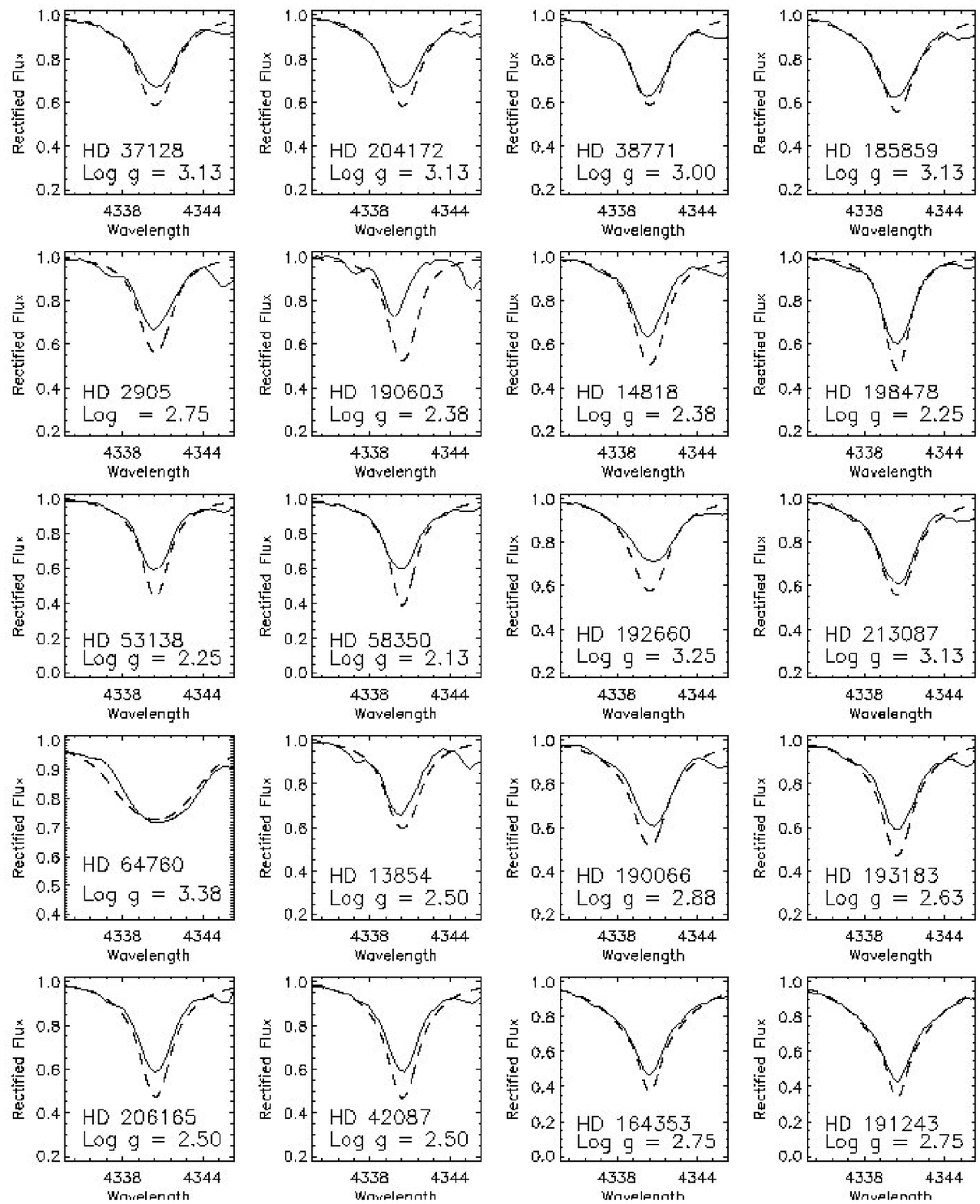}
\caption{{\sc TLUSTY} model fits to the \hg\ profile of all 20 supergiants. Optical spectrum is represented by a solid, black line; {\sc TLUSTY} model fit is shown as a dotted black line.}
\label{hg}
\end{figure*}}

\onlfig{8}{
\begin{figure*}
\centering
\includegraphics[width=12cm]{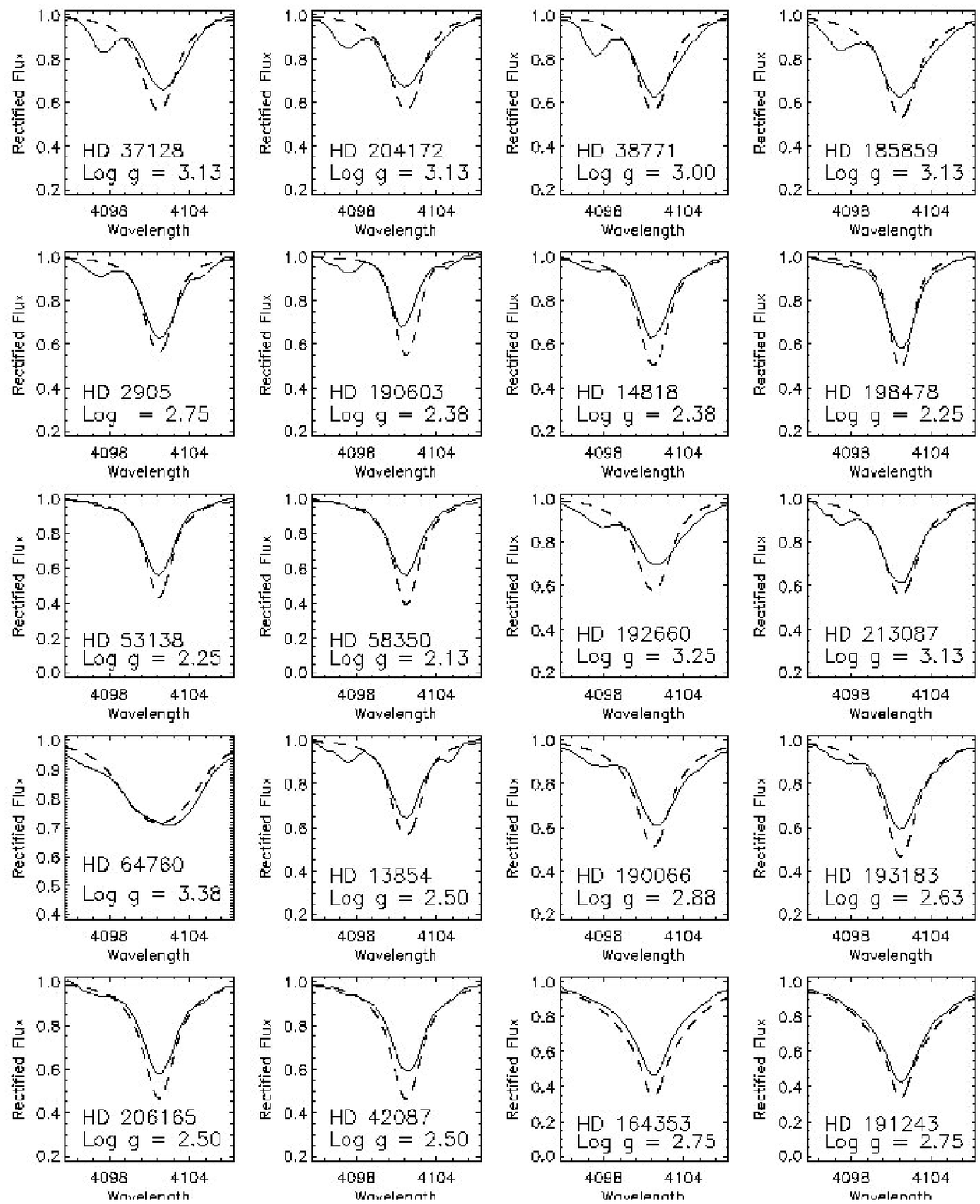}
\caption{{\sc TLUSTY} model fits to the \hd\ profile of all 20 supergiants. Optical spectrum is represented by a solid, black line; {\sc TLUSTY} model fit is shown as a dotted black line.}
\label{hd}
\end{figure*}}


\onlfig{9}{
\begin{figure*}
\centering
\includegraphics[width=12cm]{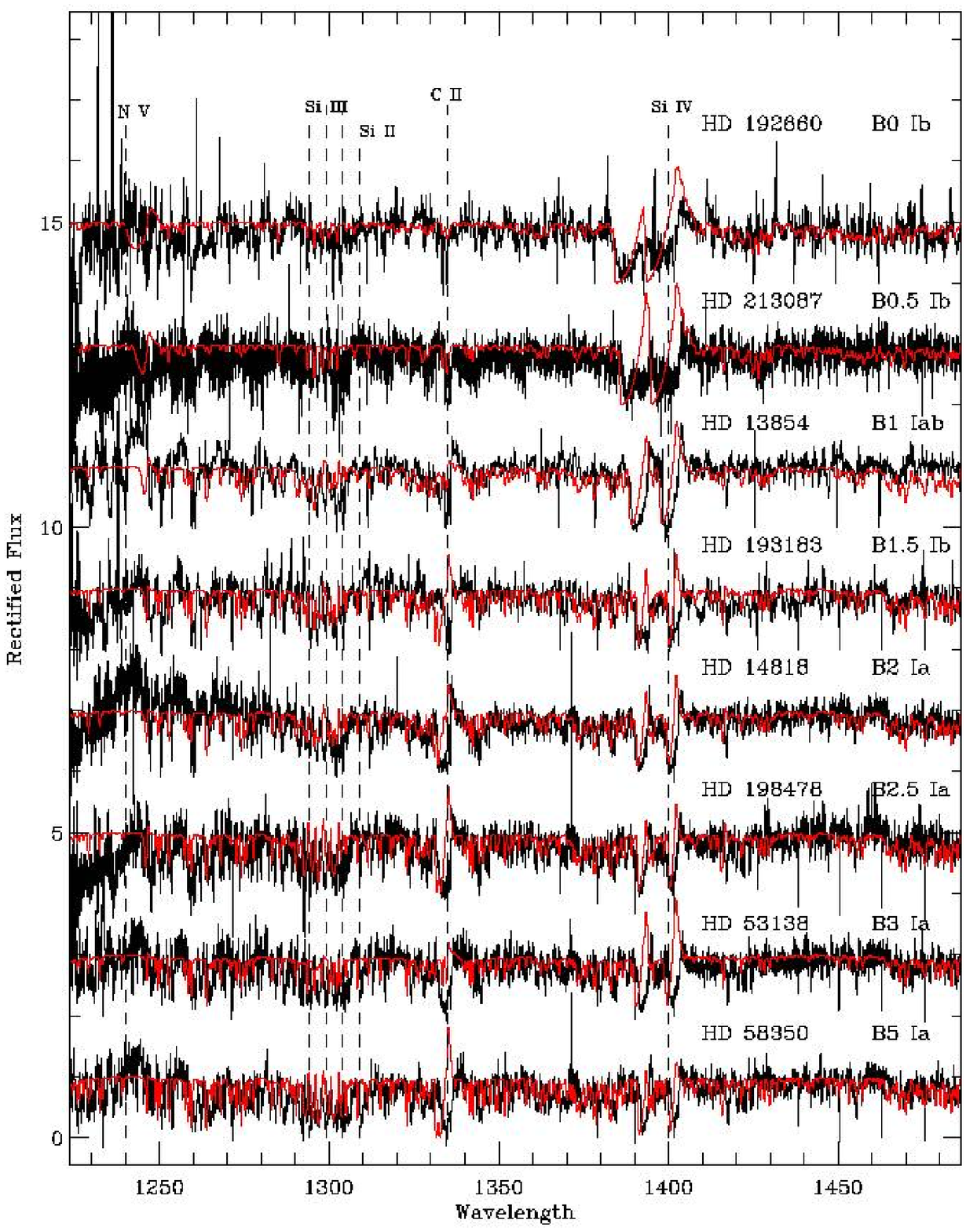}
\caption{{\sc CMFGEN} model fit to the IUE spectra of 10 B0 -- B5 supergiants (1230 \AA\ -- 1480 \AA). The solid red line represents the model fits whereas the solid black line is the IUE spectrum. }\label{uv1}
\end{figure*}}

\onlfig{10}{
\begin{figure*}
\centering
\includegraphics[width=12cm]{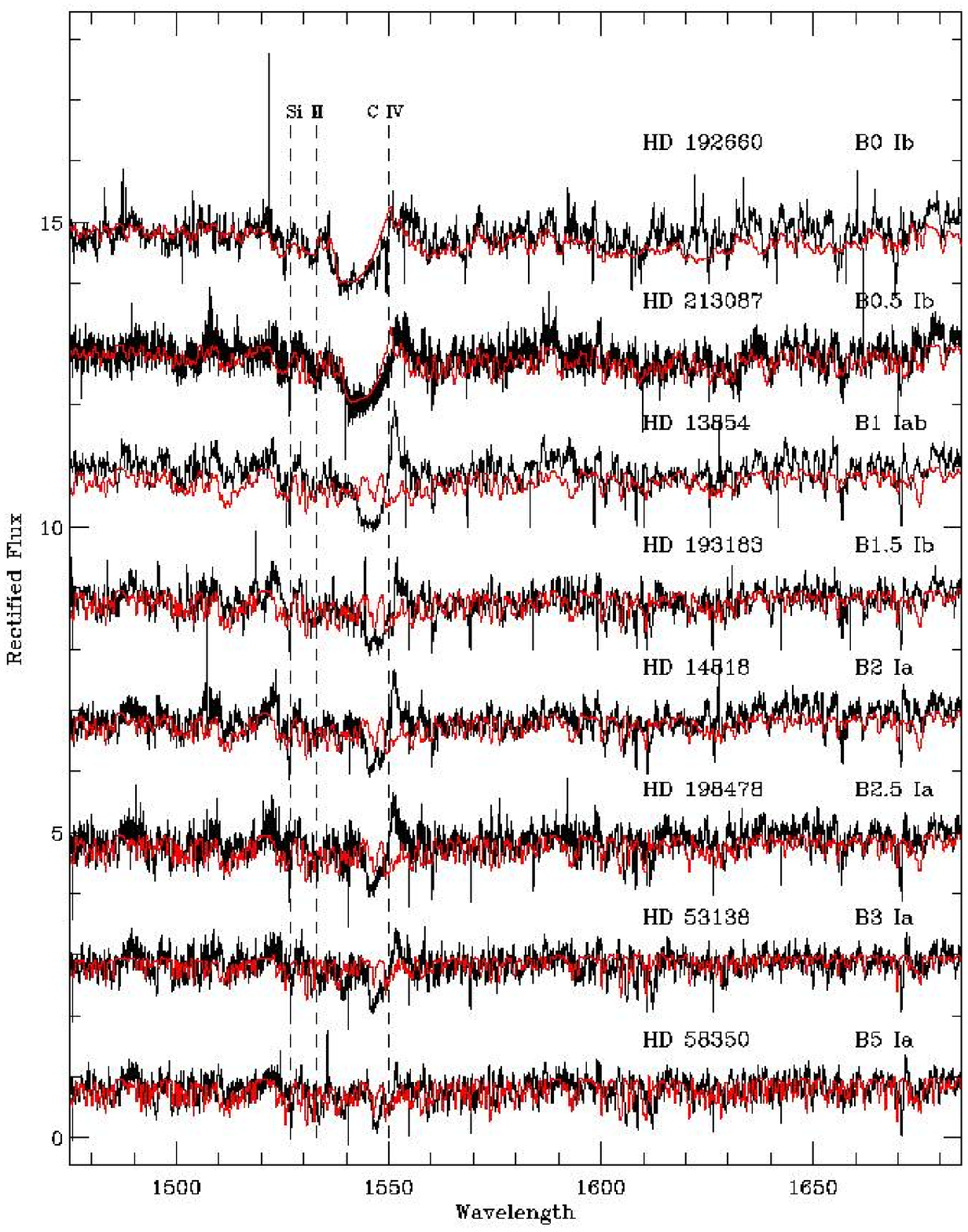}
\caption{{\sc CMFGEN} model fit to the IUE spectra of 10 B0 -- B5 supergiants (1480 \AA\ -- 1680 \AA). The solid red line represents the model fits whereas the solid black line is the IUE spectrum. }\label{uv2}
\end{figure*}}

\onlfig{11}{
\begin{figure*}
\centering
\includegraphics[width=12cm]{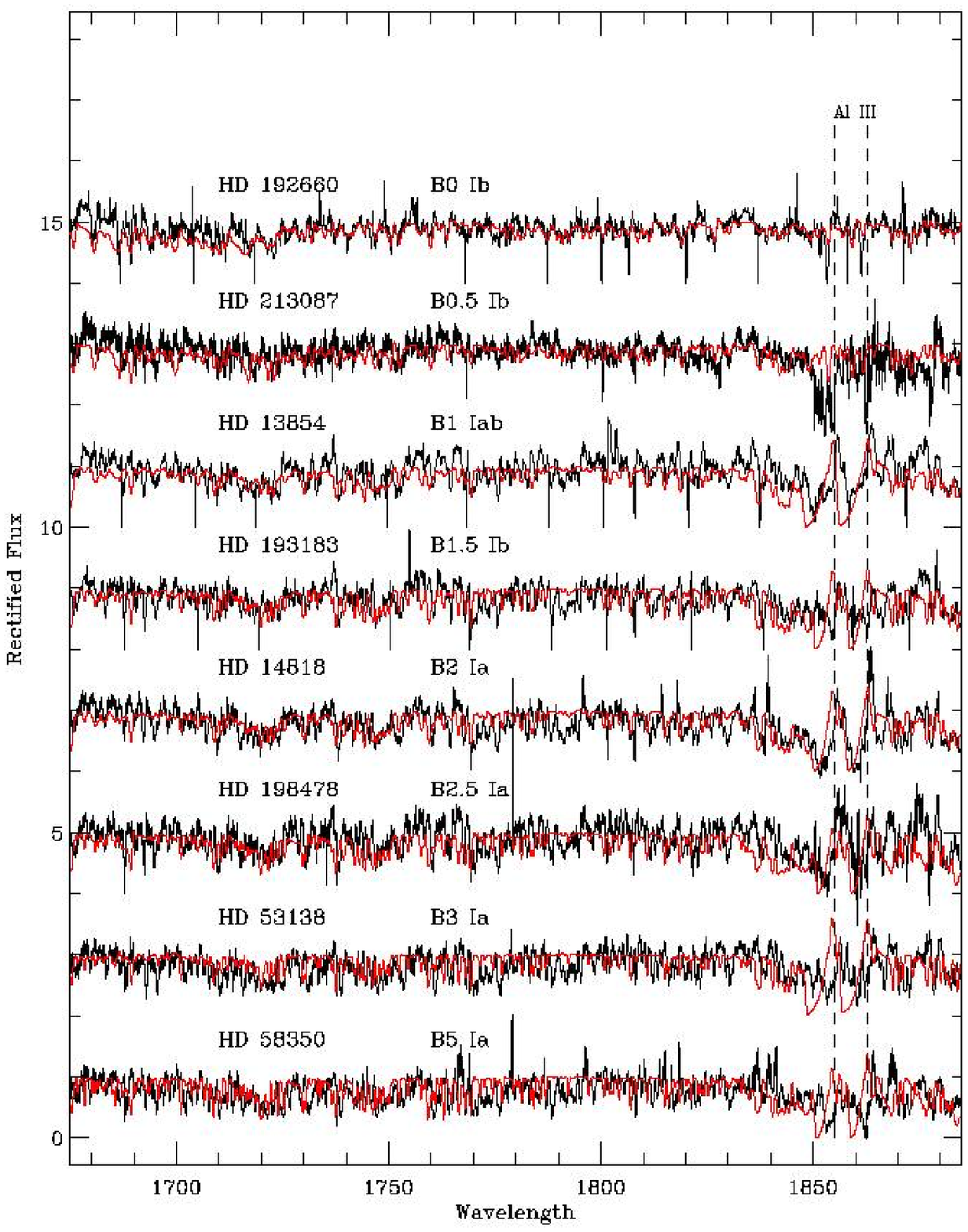}
\caption{{\sc CMFGEN} model fit to the IUE spectra of 10 B0 -- B5 supergiants (1680 \AA\ -- 1880 \AA). The solid red line represents the model fits whereas the solid black line is the IUE spectrum. }\label{uv3}
\end{figure*}}


\end{document}